\documentclass[12pt,a4paper]{article}
\usepackage{graphicx,epsfig}
\usepackage{hhline}

\usepackage{amsmath,amssymb}
\usepackage{times}
\usepackage[varg]{txfonts}
\DeclareMathAlphabet{\mathbold}{OML}{txr}{b}{it}

\usepackage{array,multirow,dcolumn}
\usepackage[mathlines,displaymath]{lineno}
\usepackage{rotating}
\usepackage{captcont} 
\usepackage{verbatimbox}
\usepackage[numbers,square,comma,sort&compress]{natbib}
\usepackage{textcomp}
\makeatletter
\g@addto@macro\bfseries{\boldmath}
\makeatother

\newcolumntype{.}{D{.}{.}{-1}}
\newcolumntype{-}{D{-}{-}{-1}}

\usepackage[dvipsnames]{color}
\definecolor{rltred}{rgb}{0.75,0,0}
\definecolor{rltgreen}{rgb}{0,0.5,0}
\definecolor{rltblue}{rgb}{0,0,0.5}
\usepackage{xcolor}
\usepackage[T1]{fontenc} 

\usepackage[hyperindex,bookmarks,bookmarksnumbered,breaklinks,a4paper,unicode]{hyperref}
\hypersetup{%
  pdftitle        = {HERAPDF2.0Jets NNLO},
  urlcolor        = rltblue,       
  urlbordercolor  = 0 0 0.5,
  filecolor       = rltblue,       
  filebordercolor = 0 0 0.5,
  linkcolor       = rltred,        
  linkbordercolor = 0.75 0 0,
  citecolor       = rltgreen,      
  citebordercolor = 0 0.5 0,
  pagecolor       = rltgreen,      
  pagebordercolor = 0 0.5 0,
  menucolor       = rltgreen,      
  menubordercolor = 0 0.5 0,
  colorlinks    = true,
  pdfauthor     = {H1 plus ZEUS Collaborations},
  pdfsubject    = { },
  pdfkeywords   = {High-Energy Physics, Particle Physics, Proton Structure, DIS}
}

\newlength{\dinwidth}
\newlength{\dinmargin}
\newcommand{\asmz}{\alpha_s(M_Z^2)}
\newcommand{\msbar}{\mbox{$\overline{\rm{MS}}$}\ }
\begin{document}

\makeatletter \def\NAT@space{} \makeatother

\begin{titlepage}
 
\noindent
DESY-21-206 \\
December 1, 2021 \\

\vspace*{1.0cm}

\begin{center}
\begin{large}

{\bfseries Impact of jet-production data on the next-to-next-to-leading-order determination of HERAPDF2.0 parton distributions}

\end{large}
\end{center}

\vspace*{1cm}
\begin{abstract} \noindent
  The HERAPDF2.0 ensemble
  of parton distribution functions (PDFs) was
  introduced in 2015.
  The final stage is presented,
  a next-to-next-to-leading-order (NNLO)
  analysis of the HERA data on inclusive deep
  inelastic $ep$ scattering together with jet data as published
  by the H1 and ZEUS collaborations.
  A perturbative QCD fit, simultaneously of $\asmz$ and
  and the PDFs,
  was performed with the result
  $\asmz = 0.1156 \pm 0.0011~{\rm (exp)}~ ^{+0.0001}_{-0.0002}~
  {\rm (model}$ ${\rm + parameterisation)}~ \pm 0.0029~{\rm (scale)}$.  
  The PDF sets of 
  HERAPDF2.0Jets NNLO were determined 
  with separate fits using two fixed values of $\asmz$, $\asmz=0.1155$
  and $0.118$, since the latter value was already chosen
  for the published HERAPDF2.0 NNLO analysis based on HERA inclusive
  DIS data only.
  The different sets of PDFs are presented, evaluated and compared. 
  The consistency of the PDFs determined with and without the jet data
  demonstrates the consistency of HERA inclusive and jet-production
  cross-section data.
  The inclusion of the jet data reduced the uncertainty on the gluon PDF.
  Predictions based on the PDFs of HERAPDF2.0Jets NNLO give an excellent
  description of the jet-production data used as input.

\end{abstract}

\vspace*{0.5cm}

\begin{center}
  
  
\end{center}
\vspace*{0.5cm}

\end{titlepage}

%
%
\noindent
I.\,Abt$^{48}$,
R.\,Aggarwal$^{62}$,
V.\,Andreev$^{45}$,
M.\,Arratia$^{64}$,
V.\,Aushev$^{34}$,
A.\,Baghdasaryan$^{83}$,
A.\,Baty$^{26}$,
K.\,Begzsuren$^{74}$,
O.\,Behnke$^{23}$,
A.\,Belousov$^{45\,\dagger}$
% ...
%
%
\par\medskip
{\small \setlength{\parindent}{0pt}
\par$^{1}$ I. Physikalisches Institut der RWTH, Aachen, Germany
\par$^{2}$ LAPP, Universit\'e de Savoie, CNRS/IN2P3, Annecy-le-Vieux, France
\par$^{3}$ University of Michigan, Ann Arbor, MI 48109, USA$^{a16}$
\par$^{4}$ Inter-University Institute for High Energies ULB-VUB, Brussels and Universiteit Antwerpen, Antwerp, Belgium$^{a2}$
\par$^{5}$ University of Bath, Bath, United Kingdom
\par$^{6}$ Lawrence Berkeley National Laboratory, Berkeley, CA 94720, USA$^{a16}$
\par$^{7}$ School of Physics and Astronomy, University of Birmingham, Birmingham, United Kingdom$^{a3}$
\par$^{8}$ INFN Bologna, Bologna, Italy$^{a4}$
\par$^{9}$ Physikalisches Institut der Universit\"at Bonn, Bonn, Germany$^{a5}$
\par$^{10}$ Department of Physics and Astronomy, The University of Sussex, Brighton, BN1 9RH, United Kingdom
\par$^{11}$ Horia Hulubei National Institute for R\&D in Physics and Nuclear Engineering (IFIN-HH) , Bucharest, Romania$^{a6}$
\par$^{12}$ GSI Helmholtzzentrum f\"{u}r Schwerionenforschung GmbH, Darmstadt, Germany
\par$^{13}$ STFC, Rutherford Appleton Laboratory, Didcot, Oxfordshire, United Kingdom$^{a3}$
\par$^{14}$ Institut f\"ur Physik, TU Dortmund, Dortmund, Germany$^{a7}$
\par$^{15}$ Joint Institute for Nuclear Research, Dubna, Russia
\par$^{16}$ Institute for Particle Physics Phenomenology, Durham University, Durham, DH1 3LE, United Kingdom
\par$^{17}$ Institut f\"ur Kernphysik, Goethe Universit\"at, Frankfurt am Main, Germany
\par$^{18}$ CERN, Geneva, Switzerland
\par$^{19}$ Irfu/SPP, CE Saclay, Gif-sur-Yvette, France
\par$^{20}$ School of Physics and Astronomy, University of Glasgow, Glasgow, United Kingdom$^{a3}$
\par$^{21}$ II. Physikalisches Institut, Universit\"at G\"ottingen, G\"ottingen, Germany
\par$^{22}$ Hamburg University, Institute of Experimental Physics, Hamburg, Germany$^{a8}$
\par$^{23}$ Deutsches Elektronen-Synchrotron DESY, Notkestr. 85, 22607 Hamburg, Germany
\par$^{24}$ European X-ray Free-Electron Laser facility GmbH, Hamburg, Germany
\par$^{25}$ Physikalisches Institut, Universit\"at Heidelberg, Heidelberg, Germany$^{a7}$
\par$^{26}$ Rice University, Houston, TX 77005-1827, USA
\par$^{27}$ University of Houston, Houston, TX 77004, USA
\par$^{28}$ Institute of Technology (KIT), Karlsruhe, Germany
\par$^{29}$ Department of Physics, Kobe University, Kobe, Japan$^{a9}$
\par$^{30}$ Institute of Nuclear Physics, Polish Academy of Sciences, Krakow, Poland$^{a10}$
\par$^{31}$ AGH University of Science and Technology, Faculty of Physics and Applied Computer Science, Krakow, Poland
\par$^{32}$ Department of Physics, Jagellonian University, Krakow, Poland$^{a11}$
\par$^{33}$ National Centre for Particle Physics, Universiti Malaya, 50603 Kuala Lumpur, Malaysia$^{a12}$
\par$^{34}$ Department of Nuclear Physics, National Taras Shevchenko University of Kyiv, Kyiv, Ukraine
\par$^{35}$ Department of Physics, University of Lancaster, Lancaster, United Kingdom$^{a3}$
\par$^{36}$ Faculdade de Ci\^encias, Universidade de Lisboa, Lisboa, Portugal
\par$^{37}$ Laboratory of Instrumentation and Experimental Particles Physics (LIP), Lisbon, Portugal
\par$^{38}$ Department of Physics, University of Liverpool, Liverpool, United Kingdom$^{a3}$
\par$^{39}$ {\L}\'od\'z University, {\L}\'od\'z, Poland
\par$^{40}$ School of Physics and Astronomy, Queen Mary, University of London, London, United Kingdom$^{a3}$
\par$^{41}$ Physics and Astronomy Department, University College London, London, United Kingdom$^{a3}$
\par$^{42}$ Sant Longowal Institute of Engineering and Technology, Longowal, Punjab, India
\par$^{43}$ Aix Marseille Univ, CNRS/IN2P3, CPPM, Marseille, France
\par$^{44}$ Institute for Theoretical and Experimental Physics, Moscow, Russia$^{a13}$
\par$^{45}$ Lebedev Physical Institute, Moscow, Russia
\par$^{46}$ Institute for Information Transmission Problems RAS, Moscow, Russia$^{a14}$
\par$^{47}$ Lomonosov Moscow State University, Skobeltsyn Institute of Nuclear Physics, Moscow, Russia
\par$^{48}$ Max-Planck-Institut f\"ur Physik, M\"unchen, Germany
\par$^{49}$ Rockefeller University, New York, NY 10065, USA
\par$^{50}$ Oak Ridge National Laboratory, Oak Ridge, TN 37831, USA
\par$^{51}$ Joint Laboratory of Optics, Palack\`y University, Olomouc, Czech Republic
\par$^{52}$ Department of Physics, York University, Ontario, M3J 1P3, Canada$^{a15}$
\par$^{53}$ IJCLab, Universit\'e Paris-Saclay, CNRS/IN2P3, Orsay, France
\par$^{54}$ Department of Physics, University of Oxford, Oxford, United Kingdom$^{a3}$
\par$^{55}$ INFN Padova, Padova, Italy$^{a4}$
\par$^{56}$ Dipartimento di Fisica e Astronomia dell' Universit\`a and INFN, Padova, Italy$^{a4}$
\par$^{57}$ LLR, Ecole Polytechnique, CNRS/IN2P3, Palaiseau, France
\par$^{58}$ Department of Physics, Temple University, Philadelphia, PA 19122, USA$^{a16}$
\par$^{59}$ Faculty of Science, University of Montenegro, Podgorica, Montenegro$^{a17}$
\par$^{60}$ Institute of Physics, Academy of Sciences of the Czech Republic, Praha, Czech Republic$^{a18}$
\par$^{61}$ Faculty of Mathematics and Physics, Charles University, Praha, Czech Republic$^{a18}$
\par$^{62}$ DST-Inspire Faculty, Department of Technology, SPPU, Pune, Maharashtra, India
\par$^{63}$ Department of Particle Physics and Astrophysics, Weizmann Institute, Rehovot, Israel
\par$^{64}$ University of California, Riverside, CA 92521, USA
\par$^{65}$ Dipartimento di Fisica Universit\`a di Roma Tre and INFN Roma 3, Roma, Italy
\par$^{66}$ INFN Roma, Roma, Italy
\par$^{67}$ Shandong University, Shandong, P.R.China
\par$^{68}$ Stony Brook University, Stony Brook, NY 11794, USA$^{a16}$
\par$^{69}$ Raymond and Beverly Sackler Faculty of Exact Sciences, School of Physics, Tel Aviv University, Tel Aviv, Israel$^{a19}$
\par$^{70}$ Department of Physics, Tokyo Institute of Technology, Tokyo, Japan$^{a9}$
\par$^{71}$ Universit\`a del Piemonte Orientale, Novara, and INFN, Torino, Italy$^{a4}$
\par$^{72}$ Universit\`a di Torino and INFN, Torino, Italy$^{a4}$
\par$^{73}$ Institute of Particle and Nuclear Studies, KEK, Tsukuba, Japan$^{a9}$
\par$^{74}$ Institute of Physics and Technology of the Mongolian Academy of Sciences, Ulaanbaatar, Mongolia
\par$^{75}$ Ulaanbaatar University, Ulaanbaatar, Mongolia
\par$^{76}$ Brookhaven National Laboratory, Upton, NY 11973, USA
\par$^{77}$ Universit\'e Claude Bernard Lyon 1, CNRS/IN2P3, Villeurbanne, France
\par$^{78}$ Paul Scherrer Institut, Villigen, Switzerland
\par$^{79}$ Faculty of Physics, University of Warsaw, Warsaw, Poland
\par$^{80}$ National Centre for Nuclear Research, Warsaw, Poland
\par$^{81}$ Department of Physics and Astronomy, Purdue University, West Lafayette, IN 47907, USA
\par$^{82}$ Fachbereich C, Universit\"at Wuppertal, Wuppertal, Germany
\par$^{83}$ Yerevan Physics Institute, Yerevan, Armenia
\par$^{84}$ Departamento de Fisica Aplicada, CINVESTAV, M\'erida, Yucat\'an, M\'exico$^{a20}$
\par$^{85}$ Deutsches Elektronen-Synchrotron DESY, Platanenallee 6, 15738 Zeuthen, Germany
\par$^{86}$ Institut f\"ur Teilchenphysik, ETH, Z\"urich, Switzerland$^{a21}$
\par$^{87}$ Physik-Institut der Universit\"at Z\"urich, Z\"urich, Switzerland$^{a21}$
\par$^{88}$ Institut f\"ur Theoretische Physik, ETH, Z\"urich, Switzerland
\par$^\dagger$ deceased
%
%
\par\smallskip
\par$^{a1}$ also supported by DESY, Hamburg, Germany
\par$^{a2}$ supported by FNRS-FWO-Vlaanderen, IISN-IIKW and IWT and by Interuniversity Attraction Poles Programme, Belgian Science Policy
\par$^{a3}$ supported by the UK Science and Technology Facilities Council, and formerly by the UK Particle Physics and Astronomy Research Council
\par$^{a4}$ supported by the Italian National Institute for Nuclear Physics (INFN)
\par$^{a5}$ supported by the German Federal Ministry for Education and Research (BMBF), under contract No. 05 H09PDF
\par$^{a6}$ supported by the Romanian National Authority for Scientific Research under the contract PN 09370101
\par$^{a7}$ supported by the Bundesministerium f\"ur Bildung und Forschung, FRG, under contract numbers 05H09GUF, 05H09VHC, 05H09VHF, 05H16PEA
\par$^{a8}$ supported by the German Federal Ministry for Education and Research (BMBF), under contract No. 05h09GUF, and the SFB 676 of the Deutsche Forschungsgemeinschaft (DFG)
\par$^{a9}$ supported by the Japanese Ministry of Education, Culture, Sports, Science and Technology (MEXT) and its grants for Scientific Research
\par$^{a10}$ partially supported by Polish Ministry of Science and Higher Education, grant DPN/N168/DESY/2009
\par$^{a11}$ supported by the Polish National Science Centre (NCN) grant no. DEC-2014/13/B/ST2/02486
\par$^{a12}$ supported by HIR grant UM.C/625/1/HIR/149 and UMRG grants RU006-2013, RP012A-13AFR and RP012B-13AFR from Universiti Malaya, and ERGS grant ER004-2012A from the Ministry of Education, Malaysia
\par$^{a13}$ Russian Foundation for Basic Research (RFBR), grant no 1329.2008.2 and Rosatom
\par$^{a14}$ Russian Foundation for Sciences, project no 14-50-00150
\par$^{a15}$ supported by the Natural Sciences and Engineering Research Council of Canada (NSERC)
\par$^{a16}$ supported by the U.S.\ DOE Office of Science
\par$^{a17}$ partially supported by Ministry of Science of Montenegro, no. 05-1/3-3352
\par$^{a18}$ supported by the Ministry of Education of the Czech Republic under the project INGO-LG14033
\par$^{a19}$ supported by the Israel Science Foundation
\par$^{a20}$ supported by CONACYT, M\'exico, grant 48778-F
\par$^{a21}$ supported by the Swiss National Science Foundation
}

\newpage

\newpage


\section{Introduction \label{sec:int}}

Data from deep inelastic scattering (DIS) of
electrons\footnote{From here on, 
the word ``electron'' refers to both electrons and positrons.} 
on protons, $ep$, at centre-of-mass energies of
up to $\sqrt{s} \approx 320\,$GeV recorded
at HERA, have been central to the exploration
of proton structure and quark--gluon dynamics as
described by perturbative Quantum Chromodynamics (pQCD)~\cite{saturation}. 
The combination of H1 and ZEUS data on inclusive $ep$ scattering
and the subsequent pQCD analysis, introducing the ensemble of 
parton density functions (PDFs) known as HERAPDF2.0,
were milestones in the exploitation~\cite{HERAPDF20} of the HERA data.
These analyses are based on pQCD fits to the HERA DIS data in the
DGLAP~\cite{Gribov:1972ri,Gribov:1972rt,Lipatov:1974qm,Dokshitzer:1977sg,Altarelli:1977zs}
formalism using the \msbar scheme~\cite{MSbar}. 

The sets of PDFs presented in this work
complete the
HERAPDF2.0 ensemble~\cite{HERAPDF20} of PDFs.
They were determined with a next-to-next-to-leading-order (NNLO) analysis
of HERA inclusive DIS data~\cite{HERAPDF20} and selected jet-production data
as published separately by the H1 and ZEUS
collaborations~\cite{h1highq2oldjets,h1lowq2jets,zeus9697jets,zeusdijets,h1highq2newjets,h1lowq2newjets}.
An analysis of jet data at NNLO was not feasible at the time of the
introduction of the HERAPDF2.0 ensemble but
has become possible by the recent provision of 
jet cross-section predictions for $ep$ scattering at
NNLO~\cite{nnlojet1,nnlojet2,nnlojet:general,fastnlo1-claire,fastnlo2-claire,fastnlo3-claire,applgrid1,applgrid2,applfast:dis}.

The strategy chosen for the analysis presented here follows that of the
previous HERAPDF2.0 Jets NLO  analysis~\cite{HERAPDF20}.
First, the jet cross-section data were included in the pQCD analysis to
constrain the gluon PDF.
Since the gluon PDF is correlated with the
value of the strong coupling constant, $\asmz$,
a simultanous fit of the PDFs and 
$\asmz$ was performed.
Subsequently, the resulting $\asmz$ was used
to refit the PDFs with $\asmz$ fixed to this value.
In this way, the 
uncertainties of the PDFs 
at this value of $\asmz$ were determined.
The PDFs were also determined for the conventional fixed value of 
$\asmz=0.118$.

The calculation of jet cross sections at NNLO is based on    
jets constructed from massless partons.
The inclusive data, on the other hand, are treated within the
Variable Flavour Number Scheme (VFNS)
RTOPT~\cite{Thorne:1997ga,Thorne:2006qt,Thorne:RTopt},
which requires values
of the parameters for the charm- and beauty-quark masses, $M_c$ and $M_b$,
as input.
These parameters were optimised 
 via QCD fits using both the
 inclusive data and the
 cross sections for
 charm and beauty production
 that were published as combined data by
 the H1 and ZEUS collaborations~\cite{combi-cb}. 
 However,
 the heavy-quark data were not explicitly included in the pQCD fits
 that included jet data.

The results presented here are based entirely on HERA data,
i.e.\ inclusive DIS and jet-production data.
The HERA inclusive data are a single, consistent data set,
taking all systematic uncertainties into account.
The jet 
and inclusive
data have been found to be consistent
in the framework of an NLO~\cite{HERAPDF20} and an NNLO~\cite{nnlojetalphas}
analysis.
The analysis presented here also tests this consistency at NNLO.
The HERAPDF2.0 ensemble of PDFs provides a benchmark to which
PDFs including data from the LHC collider may be compared.
Such comparison is sensitive to Beyond Standard Model effects or the need for
an extension of the QCD analyses for some processes.

\section{Data}
Data taken by the H1 and ZEUS collaborations from 1993 to 2007 were combined
to form a consistent set of inclusive HERA $ep$ DIS
cross sections~\cite{HERAPDF20} taking all systematic uncertainties
into account. 
This set of data was already used as input to the determinations of all previous
members of the HERAPDF2.0 ensemble.
The HERAPDF2.0Jets analysis at NLO, in addition, used selected
data~\cite{h1highq2oldjets,h1lowq2jets,h1highq2newjets,zeus9697jets,zeusdijets}
on inclusive jet and dijet production
from H1 and ZEUS. These data were also used for the present analysis at NNLO.
In addition, new data published
by the H1~collaboration on jet
production~\cite{h1lowq2newjets}
were added as input to the present NNLO analysis.
These data reach to lower~$Q^2$, 
where $Q^2$ is the squared four-momentum-transfer in the DIS process,
and also provide six new high-$Q^2$ points at low $p_{\rm T}$,
where $p_{\rm T}$ is the transverse momentum of the jet.
For all data sets used in the analysis, massless jets were
identified with the $k_{\rm T}$ algorithm with the $R$ parameter set to one.
A summary of these data sets 
is provided in Table~\ref{tab:jet-data}.

The predictions for inclusive jet and dijet production at NNLO were
used for a slightly reduced phase space compared to 
HERAPDF2.0Jets NLO in order to
limit the NNLO scale, $\mu$, uncertainties of the theoretical predictions
to below 10\,\%. 
Jets from the inclusive-jet data with
$\mu^2 = (        p_{\rm T}^2         + Q^2) \le (10.0$\,GeV$)^2$ 
and dijets with 
$\mu^2 = (\langle p_{\rm T} \rangle_2^2 + Q^2) \le (10.0$\,GeV$)^2$,
where $\langle p_{\rm T} \rangle_2 $ is the average of the transverse
momenta of the two jets,
were excluded.
These requirements on $\mu$  also ensure that $\mu$ is 
larger than the $b$-quark mass, which is necessary
because the jets
are built from massless partons in
the calculation of the NNLO predictions.
In addition, for each $Q^2$ interval, the six data points with the lowest
$\langle p_{\rm T}\rangle_2$,
were excluded from the ZEUS dijet data set.
Due to the kinematic cuts applied for the selection of dijet events,
the Born-level dijet contribution vanishes in these bins.
Consequently, the NNLO theory predictions for dijet production amount
only to NLO accuracy here.
The resulting reduction of data points is detailed in
Table~\ref{tab:jet-data}.
Furthermore, the trijet data~\cite{h1highq2newjets}, which were used
for HERAPDF2.0Jets NLO,
were excluded as NNLO theory predictions for trijet production were not available.

Since complete NNLO predictions were not available 
for heavy quarks,
the inclusive charm data~\cite{HERAccombi}, which were included
in the analysis at NLO~\cite{HERAPDF20}, were not explicitly used
in the PDF fits of the analysis presented here.
Heavy-quark data~\cite{combi-cb} were used only to optimise
the mass parameter values for charm, $M_c$, and beauty, $M_b$,
which are required as input to the adopted 
RTOPT~\cite{Thorne:RTopt}
NNLO approach to the fitting of the inclusive data.

\section{QCD analysis}
The present analysis was performed in the same way
as all previous HERAPDF2.0 analyses~\cite{HERAPDF20}.
Only cross sections for $Q^2 \ge Q^2_{\rm min}$,
with $Q^2_{\rm min} = 3.5$\,GeV$^2$, 
were used in the analysis. 
The $\chi^2$ definition was taken from equation\,(32) 
of the previous paper~\cite{HERAPDF20}.
The value of the starting scale
for the DGLAP evolution was taken as 
$\mu_{\rm f0}^2 = 1.9$\,GeV$^2$.
The parameterisation of the PDFs and the choice of free parameters
also followed the prescription for the HERAPDF2.0Jets NLO analysis,
see Section~\ref{sec:modpar}.

All fits were performed using the program 
QCDNUM~\cite{QCDNUM} within the xFitter (formerly HERAFitter)
framework~\cite{HERAfitter} and were cross-checked with an independent
program, which was already used for cross-checks in the
HERAPDF2.0 analysis.
The results obtained using the two programs
were in excellent agreement.
All numbers presented here were obtained using xFitter.
The light-quark coefficient functions were calculated in QCDNUM.
The heavy-quark coefficient functions were calculated in the 
VFNS RTOPT~\cite{Thorne:1997ga}, with recent 
modifications~\cite{Thorne:2006qt,Thorne:RTopt}, see Section~\ref{sec:mcmb}.

The present analysis  was made possible by
the newly available calculation of jet-production cross sections at
NNLO~\cite{nnlojet1,nnlojet2,nnlojet:general,fastnlo1-claire,fastnlo2-claire,fastnlo3-claire,applgrid1,applgrid2,applfast:dis}  using the
zero-mass scheme.
This is expected
to be a reasonable approximation when the relevant QCD scales
are significantly above the charm- and beauty-quark masses.
The jet data were included in the fits at full NNLO  using predictions
for the jet cross sections calculated using
NNLOJET~\cite{nnlojet1,nnlojet2,nnlojet:general},
which was interfaced to the fast
grid-interpolation codes,
fastNLO~\cite{fastnlo1-claire,fastnlo2-claire,fastnlo3-claire} and
APPLgrid~\cite{applgrid1,applgrid2} using the APPLfast
framework~\cite{applfast:dis}, in order
to achieve the required speed for the convolutions needed in an
iterative PDF fit.
The NNLO jet predictions
were provided in the massless scheme
and were corrected
for hadronisation and 
$Z^0$ exchange before they were used in the fits.
A running electromagnetic $\alpha$ as implemented in the 2012 version of 
the programme EPRC~\cite{Spiesberger:95} was used in the treatment
of the jet cross sections.
The predictions included estimates of the numerical precision, 
which were taken into account in all fits
as 50\,\% correlated and 50\,\% 
uncorrelated between processes and bins.

The choice of scales for the jet data had to be adjusted for the NNLO analysis.
At NLO, the factorisation scale was chosen as for the inclusive data,
i.e.\ $\mu_{\rm f}^2 = Q^2$,
while the renormalisation scale was linked to the transverse
momenta, $p_{\rm T}$, of the jets as $\mu_{\rm r}^2 = (Q^2 + p_{\rm T}^2)/2$.
For the NNLO analysis, $\mu_{\rm f}^2 =\mu_{\rm r}^2= Q^2 + p_{\rm T}^2$
was used for inclusive jets and
$\mu_{\rm f}^2 =\mu_{\rm r}^2= Q^2 + \langle p_{\rm T} \rangle_2^2 $
for dijets.
These changes resulted in improved $\chi^2$ values for the fits,
confirming previously published studies~\cite{H1lovesDaniel}.
Scale variations were also considered  and are discussed in
Sections~\ref{sec:as} and~\ref{sec:PDF}.
In general, scale variations are used to estimate
the uncertainties due to missing higher-order contributions.

\subsection{Choice of PDF parameterisation and model parameters}
\label{sec:modpar}
The choice of parameterisation follows the original concept of
HERAPDF2.0, for which all details have been previously
published~\cite{HERAPDF20}. The parameterisation is an effective way to
store the information derived from many data points in a limited set 
of numbers.
The parameterised PDFs, $xf(x)$, are the gluon distribution $xg$, 
the valence-quark distributions $xu_v$, $xd_v$, and 
the $u$-type and $d$-type anti-quark distributions
$x\bar{U}$, $x\bar{D}$, where $x\bar{U} = x\bar{u}$ and 
$x\bar{D} = x\bar{d} +x\bar{s}$ at the chosen starting scale.
The generic form of the parameterisation for a PDF $f(x)$ is
\begin{equation}
 xf(x) = A x^{B} (1-x)^{C} (1 + D x + E x^2).
\label{eqn:pdf}
\end{equation}
For the gluon PDF,
an additional term of the form
$A_g'x^{B_g'}(1-x)^{C_g'}$ is subtracted\footnote{The parameter $C_g' = 25$
was fixed since the fit is not sensitive to this value, provided it is high 
enough ($C_g' > 15$) to ensure that the term does not contribute at
large $x$.}.

Not all the $D$ and $E$ parameters were required in the fit.
The so-called $\chi^2$ saturation method~\cite{HERAIcombi,HERAPDF20}
was used to reject redundant parameters.
Initially, all $D$ and $E$ parameters as well as $A_g'$  were set to 
zero.
Extra parameters were introduced 
one at a time until the $\chi^2$ of the fit could not be further
improved.
This resulted in a final parameterisation 
\begin{eqnarray}
\label{eq:xgpar}
xg(x) &=   & A_g x^{B_g} (1-x)^{C_g} - A_g' x^{B_g'} (1-x)^{C_g'}  ,  \\
xu_v(x) &=  & A_{u_v} x^{B_{u_v}}  (1-x)^{C_{u_v}}\left(1+E_{u_v}x^2 \right) , \\
\label{eq:xuvpar}
xd_v(x) &=  & A_{d_v} x^{B_{d_v}}  (1-x)^{C_{d_v}} , \\
\label{eq:xdvpar}
x\bar{U}(x) &=  & A_{\bar{U}} x^{B_{\bar{U}}} (1-x)^{C_{\bar{U}}}\left(1+D_{\bar{U}}x\right) , \\
\label{eq:xubarpar}
x\bar{D}(x) &= & A_{\bar{D}} x^{B_{\bar{D}}} (1-x)^{C_{\bar{D}}} .
\label{eq:xdbarpar}
\end{eqnarray}
The normalisation parameters, $A_g, A_{u_v}, A_{d_v}$, were constrained 
by the quark-number and momentum sum rules. 
The $B$ parameters, $B_{\bar{U}}$ and $B_{\bar{D}}$, were set equal,
resulting in a single $B$ parameter for the sea distributions. 

The strange-quark distribution was expressed 
as an $x$-independent fraction, $f_s$, of the $d$-type sea, 
$x\bar{s}= f_s x\bar{D}$ at the starting scale $\mu_{\rm f0}$.
The value $f_s=0.4$ 
was chosen to be a compromise between the suppressed 
strange sea seen in neutrino-induced di-muon 
production~\cite{Martin:2009iq,Nadolsky:2008zw} and  
the unsuppressed strange sea seen by the ATLAS 
collaboration~\cite{atlasstrange}. 
The further constraint 
$A_{\bar{U}}=A_{\bar{D}} (1-f_s)$, together with the requirement  
$B_{\bar{U}}=B_{\bar{D}}$,  ensured that 
$x\bar{u} \rightarrow x\bar{d}$ as $x \rightarrow 0$.

The final parameterisation together with the constraints
became the basis of the $14$-parameter fit which
was used throughout the analysis. The parameterisation is
identical to the parameterisation used previously for the analysis of the
inclusive data~\cite{HERAPDF20}.

\subsection{Model and parameterisation uncertainties}
\label{sec:sysunc}

Model and parameterisation uncertainties on the PDFs
were evaluated by using fits with modified input assumptions.
The central values of the model parameters and their variations
are summarised in Table~\ref{tab:model}.
The uncertainties on the PDFs obtained from
variations of $M_c$, $M_b$, $f_s$ and $Q^2_{\rm min}$ were 
added in quadrature, separately for positive and negative uncertainties,
and represent the model uncertainty. 

The symmetrised uncertainty obtained from the downward variation
of $\mu^2_{\rm f0}$ from 1.9\,GeV to 1.6\,GeV, see also Section\,\ref{sec:mcmb},
was taken as a parameterisation uncertainty.
In addition, a variation of the number of terms in
the polynomial $(1 + D x + E x^2)$ was  
considered for each of the parton distributions listed in
Eqs.~({\ref{eq:xgpar}) -- (\ref{eq:xdbarpar}).
For this, all 15-parameter fits which have one more non-zero free  
$D$ or $E$ parameter were considered as possible variants and the resulting
PDFs compared to the PDF from the 14-parameter central fit.
The only visible change in the shapes of the PDFs was observed for 
the addition of a $D_{u_v}$ parameter. 
The maximal deviation of the fit at each $x$ value was considered
an uncertainty, forming an envelope around the central fit.

\subsection{Optimisation of $M_c$ and $M_b$}
\label{sec:mcmb}

The RTOPT scheme used to calculate predictions
for the inclusive data requires 
the charm- and beauty-mass parameters, $M_c$ and $M_b$, as input.
The optimal values of these parameters were reevaluated using
the previously established procedure~\cite{HERAIcombi,HERAPDF20},
applied to the new combined HERA data on heavy quarks~\cite{combi-cb}
together with the combined inclusive data~\cite{HERAPDF20}.
The procedure comprises multiple pQCD fits with varying choices of the
$M_c$ and $M_b$ parameters.
The parameter values resulting in the lowest $\chi^2$ values of the fit
were chosen.
This was done both at NNLO and NLO to provide consistent sets
of $M_c$ and $M_b$ for future pQCD analyses.
The uncertainties of the mass parameters
were determined by fitting the $\chi^2$
values with a quadratic function and finding the
mass-parameter values corresponding to $\Delta\chi^2 = 1$.

At NNLO (NLO), the fits for the optimisation were performed
with fixed values of 
$\alpha_s=0.1155$\,\footnote{A cross-check was performed with the
fixed value of $\alpha_s=0.118$ and no significant difference in 
the resulting $M_c$ and $M_b$ values was observed.} 
($\alpha_s=0.118$)\,\footnote{The value 0.118 was used in the pQCD analysis
of heavy-quark data~\cite{combi-cb}.}.
As a first iteration at NNLO (NLO), the mass
parameter values used for HERAPDF2.0 NNLO (NLO) were used
as fixed points, so that $M_c$ was varied with fixed $M_b = 4.5$\,GeV
(4.5GeV) and $M_b$ was varied with fixed $M_c = 1.43$\,GeV (1.47\,GeV).
In every iteration to determine $M_b$ ($M_c$),
the mass-parameter value for $M_c$ ($M_b$)
as obtained from the
previous iteration was  used as a new fixed point.
The iterations were terminated once values stable to within 0.1\,\% for
$M_c$ and $M_b$ were obtained.
The final $\chi^2$ scans at NNLO are shown in
  Figs.~\ref{fig:mass-scans}\,a) and~\ref{fig:mass-scans}\,c) and at NLO in
  Figs.~\ref{fig:mass-scans}\,b) and~\ref{fig:mass-scans}\,d). 
  The resulting values at NNLO
  are $M_c = 1.41 \pm 0.04$\,GeV and $M_b=4.20 \pm 0.10$\,GeV,
  compatible with the values determined for HERAPDF2.0 NNLO, with slightly
  reduced uncertainties.
  The values at NLO are
  $M_c = 1.46 \pm 0.04$\,GeV and $M_b=4.30 \pm 0.10$\,GeV.
  The minimum in $\chi^2$ for the parameter $M_c$ at NNLO is observed 
  close to the technical limit of 
  the fitting procedure, $\mu_{\rm f0} < M_c$.
The model uncertainty due to $M_b$ was obtained
by varying $M_b$ by its one-standard-deviation uncertainty. 
The same procedure was not possible 
for $M_c$ because the downward variation created a conflict
with $\mu_{\rm f0}$, which  has to be less than $M_c$ in the RTOPT scheme,
in order that charm can be generated perturbatively.
Thus, only an upward variation of $M_c$ was considered and the
resulting uncertainty on the PDFs was symmetrised.
In addition, this requirement of $\mu_{\rm f0} < M_c$
created a conflict with the variation of $\mu_{\rm f0}^2$.
The normal procedure would have included
an upward variation of $\mu_{\rm f0}^2$ to $2.2~$GeV$^2$ but $\mu_{\rm f0}$
would have become larger than the upper boundary of the uncertainty interval
of $M_c$\,\footnote{In previous HERAPDF analyses, the 
uncertainty on $M_c$ was large enough to accommodate
the upward $\mu_{\rm f0}^2$ variation.}. 
Thus, $\mu_{\rm f0}^2$ was only varied downwards to $1.6~$GeV$^2$, and, again, the
resulting uncertainty on the PDFs was symmetrised.
The continued suitability of the chosen central parameterisation was verified 
for the new settings for $M_c$ and $M_b$ using the $\chi^2$ saturation method
as described in Section~\ref{sec:modpar}.

\subsection{Hadronisation uncertainties}
For the jet-data analysis, it was also necessary to consider  
the effect of the uncertainties on hadronisation corrections.
These, as determined for the original publications,
were reviewed for this analysis. The H1 uncertainties were used
as published; those for the ZEUS
data were
increased\footnote{This increase was necessary for technical reasons.}
to the maximum value quoted in the publications, 2\,\%.
This change resulted in
no significant difference to any of the results presented here.

In the HERAPDF2.0Jets NLO analysis~\cite{HERAPDF20},
hadronisation uncertainties were
applied using the offset method, i.e.\ performing separate fits with the
hadronisation corrections set to their maximal and minimal values.
This resulted in a hadronisation uncertainty on $\asmz$ of
$\pm 0.0012$~\cite{HERAPDF20}.
The current procedure improves upon this by including 
the uncertainties on the hadronisation corrections at the same level
as the other systematic uncertainties.
Thus, their contribution became part of the overall 
experimental (fit) uncertainties.
They were treated as 
50\,\% correlated and 50\,\% uncorrelated between bins and data sets. 
For fits with fixed $\asmz$, their contribution was negligible.
For fits with free $\asmz$, their contribution to the 
experimental uncertainty on $\asmz$ was $\pm 0.0006$.
This represents a significant reduction of the influence
of the hadronisation uncertainties compared to previous analyses.

The total uncertainties on the PDFs were obtained by adding the
experimental (fit), the model and the parameterisation uncertainties
in quadrature.

\section{HERAPDF2.0Jets NNLO -- results}
\subsection{Simultaneous determination of \boldmath{$\asmz$} and PDFs}
\label{sec:as}
In pQCD fits to inclusive DIS data alone, the gluon PDF is only determined 
via the DGLAP equations, using the observed scaling violations.
This results in a strong correlation between the shape of the 
gluon distribution and the value of $\asmz$. 
Data on jet-production cross sections provide an independent
constraint on the gluon distribution and
are also directly sensitive to $\asmz$. 
Thus, such data are essential for an accurate simultaneous determination
of $\asmz$ and the gluon distribution.

When determining $\asmz$, it is necessary to consider so-called
``scale uncertainties'', which
serve as a proxy
for the uncertainties due to the unknown higher-order contributions
in the perturbation expansion.
These uncertainties
were evaluated by varying the renormalisation and factorisation 
scales by a factor of two,
both separately and simultaneously\footnote{This procedure is often called the
9-point variation, where the nine variations are
$(0.5\mu_{\rm r},0.5 \mu_{\rm f})$, $(0.5\mu_{\rm r},1.0 \mu_{\rm f})$, $(0.5\mu_{\rm r},2.0 \mu_{\rm f})$,
$(1.0\mu_{\rm r},0.5 \mu_{\rm f})$, $(1.0\mu_{\rm r},1.0 \mu_{\rm f})$, $(1.0\mu_{\rm r},2.0 \mu_{\rm f})$,
$(2.0\mu_{\rm r},0.5 \mu_{\rm f})$, $(2.0\mu_{\rm r},1.0 \mu_{\rm f})$, $(2.0\mu_{\rm r},2.0 \mu_{\rm f})$.}.
The maximum positive and negative deviations of the result
were assigned as the scale uncertainties on  $\asmz$.
These were observed for the variations
$(2.0\mu_{\rm r},1.0 \mu_{\rm f})$ and  $(0.5\mu_{\rm r},1.0 \mu_{\rm f})$,
respectively.

The HERAPDF2.0Jets NNLO fit with free $\asmz$ resulted in
\begin{eqnarray}
\label{eq:asfit}  
\asmz =0.1156 \pm 0.0011~{\rm (exp)}~~
              ^{+0.0001}_{-0.0002}~{\rm (model+parameterisation)}
           ~~ \pm 0.0029~{\rm (scale)}~~,
\end{eqnarray}
where ``exp'' denotes
the experimental uncertainty, which was taken as the fit uncertainty,
including the contribution from hadronisation uncertainties.
The value of $\asmz$ and the size of the experimental uncertainty
were confirmed by 
a scan in $\asmz$, for which the resulting $\chi^2$
values are  shown in Fig.~\ref{fig:alphasscan2}.
The clear minimum observed in $\chi^2$ coincides with the value of $\asmz$
listed in Eq.~(\ref{eq:asfit}). The width of the minimum in $\chi^2$ confirms
the fit uncertainty.
The combined model and parameterisation uncertainty shown
in Fig.~\ref{fig:alphasscan2}
was determined by performing similar scans, for which the values of the
model parameters and the parameterisation were 
varied as described in Section~\ref{sec:modpar}.

Figure~\ref{fig:alphasscan2} also shows the scale uncertainty, which
dominates the total uncertainty.
The scale uncertainty as listed in Eq.~(\ref{eq:asfit}) 
was evaluated under the assumption 
of 100\,\% correlated uncertainties
between bins and data sets.
The previously published result at NLO~\cite{HERAPDF20}
had scale uncertainties calculated under
the assumption of 50\,\% correlated and 50\,\% uncorrelated uncertainties
between bins and data sets, owing to the inclusion
of heavy-quark and trijet data.
A strong motivation to determine
$\asmz$ at NNLO was the expectation of a substantial reduction in the
scale uncertainty.
Therefore, the analysis was repeated for these
assumptions in order to compare the NNLO to the NLO scale
uncertainties.
The re-evaluated NNLO scale uncertainty of ($\pm 0.0022$) is indeed
significantly lower than the ($+0.0037,-0.0030$) previously
observed in the HERAPDF2.0Jets NLO analysis.

The HERAPDF2.0Jets NNLO fit with free $\asmz$  was based on
1363 data points and had a
$\chi^2/$degree of freedom\,(d.o.f.)~$= 1614/1348 = 1.197$. This can be compared
to the $\chi^2/$d.o.f.~$= 1363/1131 = 1.205$ for HERAPDF2.0 NNLO
based on inclusive data only~\cite{HERAPDF20}.
The similarity of the $\chi^2/$d.o.f. values indicates that 
the data on jet production do not introduce
any additional tension into the fit and are fully consistent
with the inclusive data.

The question of whether data at relatively low $Q^2$ bias the
determination of $\asmz$ arose within the context of the
HERAPDF2.0 analysis~\cite{HERAPDF20}.
Figure~\ref{fig:alphasscan3}\,a) shows the result of $\asmz$ scans with  
$Q^2_{\rm min}$ for the inclusive data
set to 3.5\,GeV$^2$, 10\,GeV$^2$ and 20\,GeV$^2$.
The positions of the minima are in good agreement, indicating that any anomalies
at low $Q^2$ are small.
Figure~\ref{fig:alphasscan3}\,b) shows the result of similar scans with only the
inclusive data used as input~\cite{HERAPDF20}. The inclusive data alone cannot
sufficiently constrain $\asmz$.

To verify that the use of the $A_g'$ term
in the gluon parameterisation does not bias the determination
of $\asmz$,
cross-checks were made with two modified gluon
parameterisations.
These are  $A_g'=0$ and $xg(x) = A_g x^{B_g} (1-x)^{C_g}$ as well as
the alternative gluon parameterisation, AG~\cite{HERAPDF20}, for which
$A_g'=0$ and $xg(x) = A_g x^{B_g} (1-x)^{C_g} (1 + D_g x)$.
A value of $\asmz = 0.1151 \pm 0.0010~{\rm (exp)}$ was obtained for both
modifications of the parameterisation, which is in agreement
with the result for the standard parameterisation.
The value of $D_g$ in the AG parameterisation was consistent with zero.
These results demonstrate that the present
$\asmz$ determination is not very sensitive to the details of the gluon
parameterisation.

Previous  determinations of $\asmz$ at NNLO
using jet data~\cite{H1lovesDaniel,nnlojetalphas}
used predetermined PDFs.
 These analyses were performed with a cut $\mu > 2M_b$,
 which is quite similar to the $\mu > 10.0$\,GeV cut used for this analysis.
 Thus, the scale uncertainties can be compared.
 The H1 result~\cite{H1lovesDaniel} is based on H1 data only
 and the quoted scale uncertainty
 is $\pm 0.0039$.
 The scale uncertainty published by NNLOjet~\cite{nnlojetalphas}
 using only H1 and ZEUS inclusive jet data is
 $\pm 0.0033$. 
 This can be compared to the $\pm 0.0029$ obtained for the analysis
 presented here.
The H1 collaboration also provided one simultaneous fit of $\asmz$ and PDFs
using a zero-mass variable-flavour-number scheme~\cite{H1lovesDaniel}. It was
based on H1 inclusive and jet data  with $Q^2_{\rm min} = 10$\,GeV$^2$.
For comparison, the analysis presented here was modified 
by also setting 
$Q^2_{\rm min} = 10$\,GeV$^2$.
The value of $\asmz$  published by H1 is
$\asmz = 0.1147 \pm 0.0011~{\rm (exp)}
  \pm 0.0002~{\rm (model)} 
  \pm 0.0003~{\rm (parameterisation)} \pm 0.0023~{\rm (scale)}$ 
while the current modified analysis resulted in
$\asmz = 0.1156 \pm 0.0011~{\rm (exp)}
\pm 0.0002~{\rm (model+parameterisation)} \pm 0.0021~{\rm (scale)}$.
These values agree within uncertainties.
Overall, the various determinations of $\asmz$ provide a very consistent
picture up to NNLO.


\subsection{The PDFs of HERAPDF2.0Jets NNLO obtained for fixed \boldmath{$\asmz$}}
\label{sec:PDF}

Fixed values of $\asmz = 0.1155$ and  $\asmz = 0.118$ 
were used for the determination of the two sets of PDFs
released from the HERAPDF2.0Jets NNLO analysis, see Appendix A.
The 
value of  $\asmz = 0.1155$
corresponds\footnote{After much analysis work had been done at the
  initial fit result of 0.1155, further theoretical work led to the
  final fit value drifting to 0.1156. In order to avoid a large amount of
  extra work, it was decided to continue using the value of
  0.1155 for the analysis presented in this section, in the  
  knowledge that such a tiny discrepancy could not make any difference
  to the conclusions.}  
to the determination of
$\asmz$ presented in Section~\ref{sec:as}.
The value of  $\asmz = 0.118 $ was the result of the HERAPDF2.0Jets
NLO analysis and was used for the HERAPDF2.0 analyses at NNLO 
based on inclusive data only~\cite{HERAPDF20}.
The PDFs of HERAPDF2.0Jets NNLO are shown
in Fig.~\ref{fig:as0-116-118}\,a) and~b) for
fixed $\asmz=0.1155$ and fixed $\asmz=0.118$, respectively,
at the scale $\mu_{\rm f}^2=10$\,GeV$^2$.
The uncertainties shown are the experimental (fit) uncertainties
as well as the model and parameterisation uncertainties defined
in Section~\ref{sec:sysunc}.
The introduction of the parameter $D_{u_v}$ as a variation dominates the
parameterisation uncertainty.

As the PDFs were derived with fixed $\asmz$ values, uncertainties
on the PDFs from varying the scales in the fit procedure were not considered,
since in this case, a quantification of the influence of
higher orders by varying the renormalisation and factorisation scales
in the fit becomes questionable.
Any variation  of the renormalisation scale effectively amounts,
in its numerical effect, to a modification of the value of $\asmz$,
since the  
compensation with the explicit scale-dependent terms
in the NLO and NNLO coefficients is incomplete.
If a fit is performed with a fixed value of $\asmz$, it might thus not
reach a local minimum, which is required
to estimate the 
influence of higher orders
by varying the scales.
Nevertheless, a cross-check with scale variations as described in
Section~\ref{sec:as} was made.
The impact on the resulting PDFs was found to be
negligible compared to the other uncertainties presented in
Fig.~\ref{fig:as0-116-118}.

A comparison between the PDFs obtained for $\asmz = 0.1155$ and $\asmz = 0.118$
is provided in Figs.~\ref{fig:as0-116vsas0-118} 
and~\ref{fig:as0-116vsas0-118-mz} for the scales $\mu_{\rm f}^2=$10\,GeV$^2$
and
$\mu_{\rm f}^2=M_Z^2$, respectively. Here, only total uncertainties are shown.
At the lower scale, 
a significant difference is observed between the gluon PDFs;
the gluon PDF for $\asmz = 0.1155$ is above the gluon PDF for
$\asmz = 0.118$ for $x$ less than $\approx 10^{-2}$.
This correlation between the value of $\asmz$ and the shape of
the gluon PDF is as expected from QCD evolution.
At the scale of $M_Z^2$, the differences
become negligible in the visible range of $x$.

A comparison of the PDFs obtained for $\asmz = 0.118$
by HERAPDF2.0Jets NNLO to
the PDFs of HERAPDF2.0 NNLO, based on inclusive data only,
is provided in Fig.~\ref{fig:as0-118vsherapdf2}. These two sets of PDFs
do not show any significant difference in the central values.
However, the HERAPDF2.0Jets  NNLO    
analysis results in a significant reduction of
the uncertainties on the gluon PDFs 
as shown in Fig.~\ref{fig:unc-as0-118vsherapdf2-0118} at the scale of
$\mu_{\rm f}^2 = 10\,$GeV$^2$ and in Fig.~\ref{fig:unc-as0-118vsherapdf2-MZ-0118} at the scale
of $\mu_{\rm f}^2 = M_Z^2$. The reduction in the uncertainties
for HERAPDF2.0Jets NNLO for  $\asmz = 0.1155$ compared to $\asmz = 0.118$
is shown in Figs.~\ref{fig:unc-as0-118vsherapdf2}
and~\ref{fig:unc-as0-118vsherapdf2-MZ}.
At high $x$ and $\mu_{\rm f}^2 = M_Z^2$, the parameterisaton uncertainties
become important, as
can be seen by comparing Figs.~\ref{fig:unc-as0-118vsherapdf2-MZ}\,b) and
~\ref{fig:unc-as0-118vsherapdf2-MZ}\,c).

The reduction in model and parameterisation uncertainty for $x < 10^{-3}$
for HERAPDF2.0Jets NNLO
compared to HERAPDF2.0 NNLO 
is mostly due to the improved procedure to estimate this uncertainty.
The reduced ranges of variation of $M_c$ and $M_b$
had little effect.
The major effect came from 
symmetrising the results of the variations of $\mu_{\rm f0}^2$ and $M_c^2$,
as discussed in Section~\ref{sec:mcmb}.
This removed a double counting of sources of uncertainty
that had been present in the orginal HERAPDF2.0 procedure.
On the other hand, the reduction of
experimental as well as model and parameterisation uncertainties
for $x > 10^{-3}$ is due to the influence of the jet data. 
This is also demonstrated in Fig.~\ref{fig:mark1}, which shows ratios of
the uncertainties with respect to the total uncertainties of HERAPDF2.0 NNLO
based on inclusive data only.
Shown are the experimental,
the experimental plus model, and the experimental plus parameterisation
uncertainties.
Other selected ratio plots are provided in Appendix B.

\subsection{Comparisons of HERAPDF2.0Jets NNLO predictions to jet data}
\label{sec:comp:fit:jets} 

Comparisons of the predictions based on the PDFs of HERAPDF2.0Jets NNLO with
fixed $\asmz = 0.1155$
to the data on jet production used as input to the fit 
are shown in Figs.~\ref{fig:h1old-jet-data-lowQ2}
to
\ref{fig:zeus-jet-data-dijets}.
Each figure presents a direct comparison of the cross sections and
the respective ratios.

The uncertainties on the NNLO predictions as
calculated by NNLOJET
were taken into account in all HERAPDF2.0Jets NNLO fits.
The predictions based on the PDFs of HERAPDF2.0Jets NNLO were computed
using the assumption of massless 
jets, i.e.\ the transverse energy, $E_{\rm T}$, and the transverse momentum
of a jet, $p_{\rm T}$, were assumed to be equivalent. 
For the inclusive-jet analyses, each jet $p_{\rm T}$ entered
the cross-section calculation separately.
For dijet 
analyses, the average 
of the transverse momenta of the two jets,
$\langle p_{\rm T}\rangle_2$, was used.
The factorisation and renormalisation scales were set
accordingly
for calculating predictions.
Scale uncertainties were not considered~\cite{nnlojet2} for
the comparisons to data.
The predictions 
based on the PDFs of HERAPDF2.0Jets NNLO describe the data
on jet production well, demonstrating consistency
of the inclusive and the jet-production data sets that were
used in the current analysis. 

\section{Summary}

The HERA DIS data set on inclusive $ep$ scattering as published by the 
H1 and ZEUS collaborations~\cite{HERAPDF20}, 
together with selected data on jet production, published separately by 
the two collaborations, have been used as input to a pQCD analysis
at NNLO.

An analysis was performed where
$\asmz$ and the PDFs were fitted simultaneously.
This resulted in a value of 
$\asmz = 0.1156 \pm 0.0011~{\rm (exp)} ^{+0.0001} _{-0.0002}~
{\rm (model+parameterisation)}~\pm 0.0029~{\rm (scale)}$.
This result for $\asmz$ is compatible with the world average~\cite{PDG18} 
and is competitive in comparison with other determinations at NNLO.
The scale uncertainties were calculated under the assumption
of fully correlated uncertainties between bins and data sets.
They would decrease to $\pm 0.0022$ under the assumption 
of 50\,\% correlated and 50\,\% uncorrelated uncertainties, which
is the value that can be directly compared to the previously
published~\cite{HERAPDF20} scale uncertainties
of ($+0.0037,-0.0030$) observed in the HERAPDF2.0Jets NLO analysis. 

Two sets of PDFs were determined for HERAPDF2.0Jets NNLO for fixed
$\asmz=0.1155$ and $\asmz=0.118$. They are available to the
community~\cite{fullcorr}.
Comparisons between the PDFs of HERAPDF2.0Jets NNLO
obtained for the two values of $\asmz$ 
were shown, as well as comparisons to HERAPDF2.0 NNLO, for which
jet data were not used as input to the fit.
The PDFs of HERAPDF2.0Jets NNLO and HERAPDF2.0 NNLO are consistent over
the whole kinematic range.
This also demonstrates the consistency of the jet data and the inclusive data
at NNLO level.
The switch from NLO to NNLO led to a lower value of $\asmz$.
The inclusion of the jet data
reduced the uncertainty on the gluon PDF.
Predictions based on the PDFs of HERAPDF2.0Jets NNLO
give an excellent description of
the jet-production data used as input.

The PDFs of HERAPDF2.0Jets NNLO complete the HERAPDF2.0 ensemble of
parton distribution functions. This ensemble of PDFs, extracted from
HERA data alone, presents a self-consistent picture in the framework of pQCD
and is one of the major legacies of HERA.

\section{Acknowledgements}

We are grateful to the HERA machine group whose outstanding
efforts have made the success of H1 and ZEUS possible.
We appreciate the contributions to the construction, 
maintenance and operation of the H1 and ZEUS detectors of 
many people who are not listed as authors.
We thank our funding agencies for financial 
support, the DESY technical staff for continuous assistance and the 
DESY directorate for their support and for the hospitality they 
extended to the non-DESY members of the collaborations. 
We would like to give credit to all partners contributing to the 
EGI computing infrastructure for their support.
We acknowledge the support of the IPPP Associateship program
for this project. One of the authors, A.\,Cooper-Sarkar, would like
to thank the Leverhulme Trust for their support.

\clearpage
\bibliography{desy-21-206}

\clearpage

\begin{table}
\vskip 2cm
  \begin{center}
\begin{scriptsize}
\begin{tabular}{|lr|rr|c|c|c|c|c|c|c|}
\hline
Data set & taken~~~~ &\multicolumn{2}{ c|}{$Q^2 [$GeV$^2$] range}&${\cal L}$ & $e^+/e^-$ & $\sqrt{s}$ & Norma-& All & Used & Ref.\\
\multicolumn{2}{|r|}{from ~~~~ to~~~}          & from   & to     & pb$^{-1}$ &       & GeV    & lised  & points & points & \\
\hline
 H1 HERA\,I normalised jets & 1999 -- 2000 & $150$  &$15000$ & $65.4$  &$e^+p$   & $319$  & yes & 24 & 24 &\cite{h1highq2oldjets} \\
 H1 HERA\,I jets at low  $Q^2$    & 1999 -- 2000 &   $5$  &  $100$ & $43.5$  &$e^+p$   & $319$  & no  & 28 & 20 &\cite{h1lowq2jets} \\
 H1 normalised inclusive jets at high $Q^2$  & 2003 -- 2007 & $150$  &$15000$ & $351$  &$e^+p$/$e^-p$   & $319$  & yes & 30 & 30 &\cite{h1highq2newjets,h1lowq2newjets} \\
 H1 normalised dijets at high $Q^2$  & 2003 -- 2007 & $150$  &$15000$ & $351$  &$e^+p$/$e^-p$   & $319$  & yes & 24 & 24 &\cite{h1highq2newjets}\\
 H1 normalised inclusive jets at low $Q^2$  & 2005 -- 2007 & $5.5$    &   $80$ & $290$  &$e^+p$/$e^-p$   & $319$  & yes & 48 & 37 & \cite{h1lowq2newjets}\\
 H1 normalised dijets at low $Q^2$     & 2005 -- 2007  & $5.5$    &   $80$ & $290$  &$e^+p$/$e^-p$   & $319$  & yes & 48 & 37 &\cite{h1lowq2newjets}\\
\hline
ZEUS inclusive jets  & 1996 -- 1997 & $125$  &$10000$ & $38.6$  &$e^+p$    & $301$ & no  & 30 & 30 &\cite{zeus9697jets}  \\
ZEUS dijets ~~~~~~~~~~~~~~~~~~ 1998 --2000 \& & 2004 -- 2007 & $125$  &$20000$ & $374$&$e^+p$/$e^-p$& $318$ & no  & 22 & 16 &\cite{zeusdijets} \\
\hline
\end{tabular}
\end{scriptsize}
\end{center}
\caption{\label{tab:jet-data}
The jet-production data sets from H1 and ZEUS  
used for the HERAPDF2.0Jets NNLO fits. The term normalised indicates that
these cross sections are normalised to the respective neutral current
inclusive cross sections. }
\end{table}

\begin{table}[h]
\vskip 3cm
\centerline{
\begin{tabular}{|ll|c|c|c|}
\hline
\multicolumn{2}{|c|}{Parameter} &
\multicolumn{1}{ c|}{Central value} &
\multicolumn{1}{ c|}{Downwards variation} &
\multicolumn{1}{ c|}{Upwards variation}  \\
\hline
$Q^2_{\rm min}$ & [GeV$^2$]& $3.5\phantom{0}$& $2.5\phantom{0^*}$ &
                                          $5.0\phantom{0^*}$   \\
\hline
$f_s$  &  & $0.4\phantom{0}$  & $0.3\phantom{0^*}$ & $0.5\phantom{0^*}$  \\
\hline
$M_c$ & [GeV]  & $1.41$  & $1.37^*$ & $1.45\phantom{^*}$   \\
$M_b$ & [GeV]  & $4.20$  & $4.10\phantom{^*}$ & $4.30\phantom{^*}$  \\
\hline
$\mu^2_{f0}$& [GeV$^2$] & $1.9\phantom{0}$  & $1.6\phantom{0^*}$  & $2.2^*\phantom{0}$  \\
\hline
\end{tabular}}
\vskip 0.4cm
\caption{Central values of model input parameters
  and their one-sigma variations. It was not possible to implement the
  variations marked $^*$ because $\mu_{\rm f0} < M_c$ is required,
  see Section~\ref{sec:mcmb}. In these cases,
  the uncertainty on the PDF obtained from the other variation was symmetrised.
}
\label{tab:model}
\end{table}

\clearpage

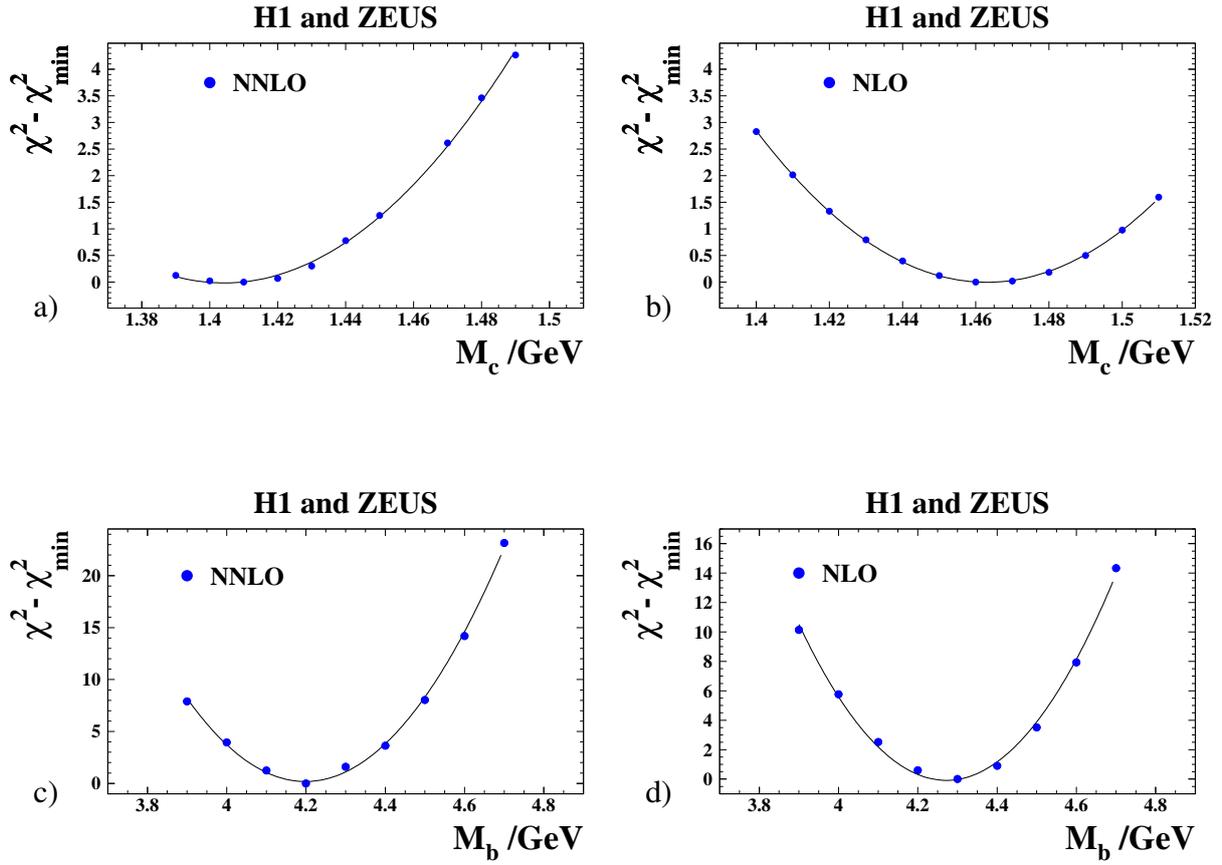
\begin{figure}
  \centering
  \setlength{\unitlength}{0.1\textwidth}
  \begin{picture} (9,8)
    \put(-.5,4.0){\includegraphics[width=0.55\textwidth]{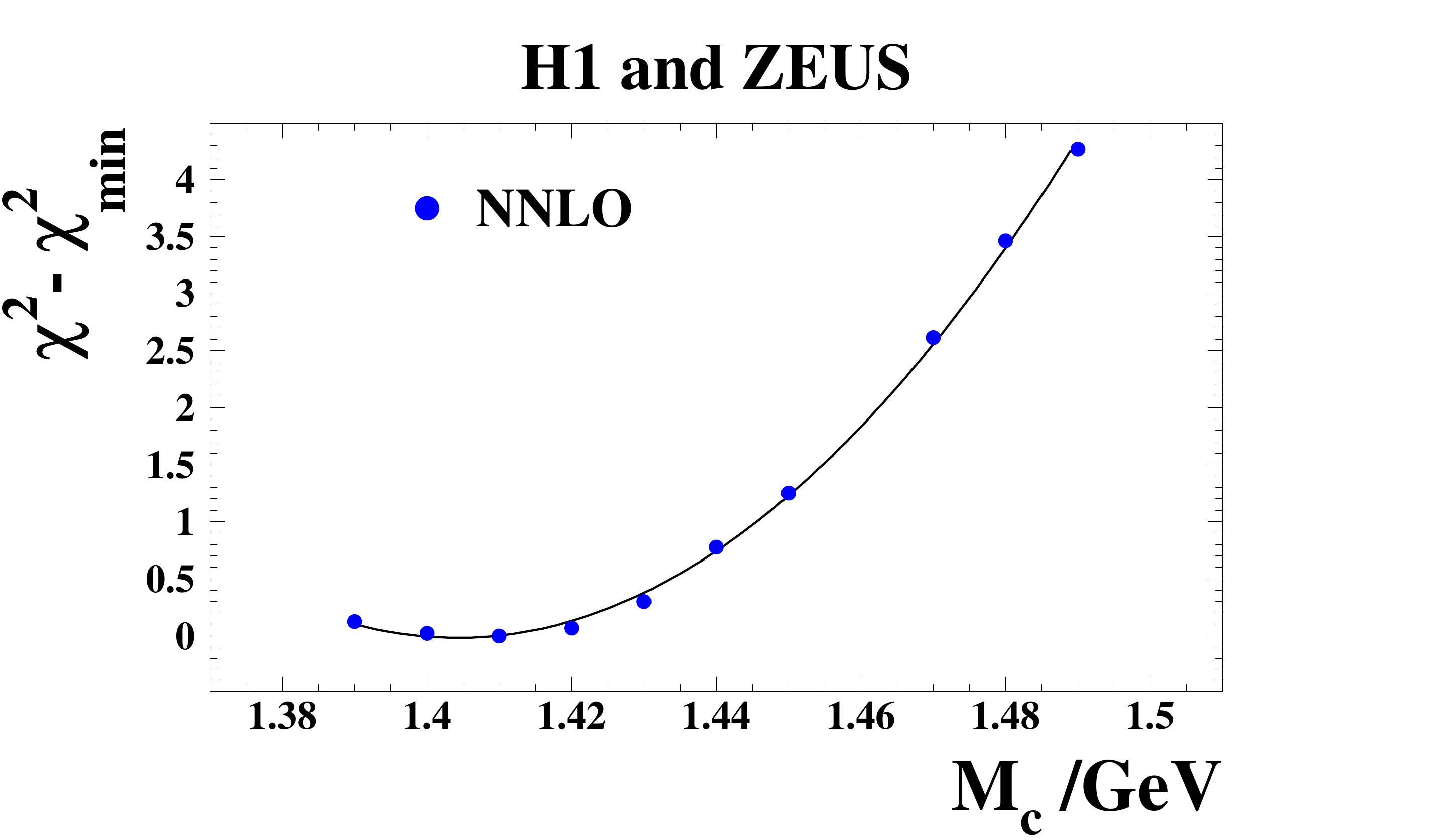}}
    \put(4.5,4.0){\includegraphics[width=0.55\textwidth]{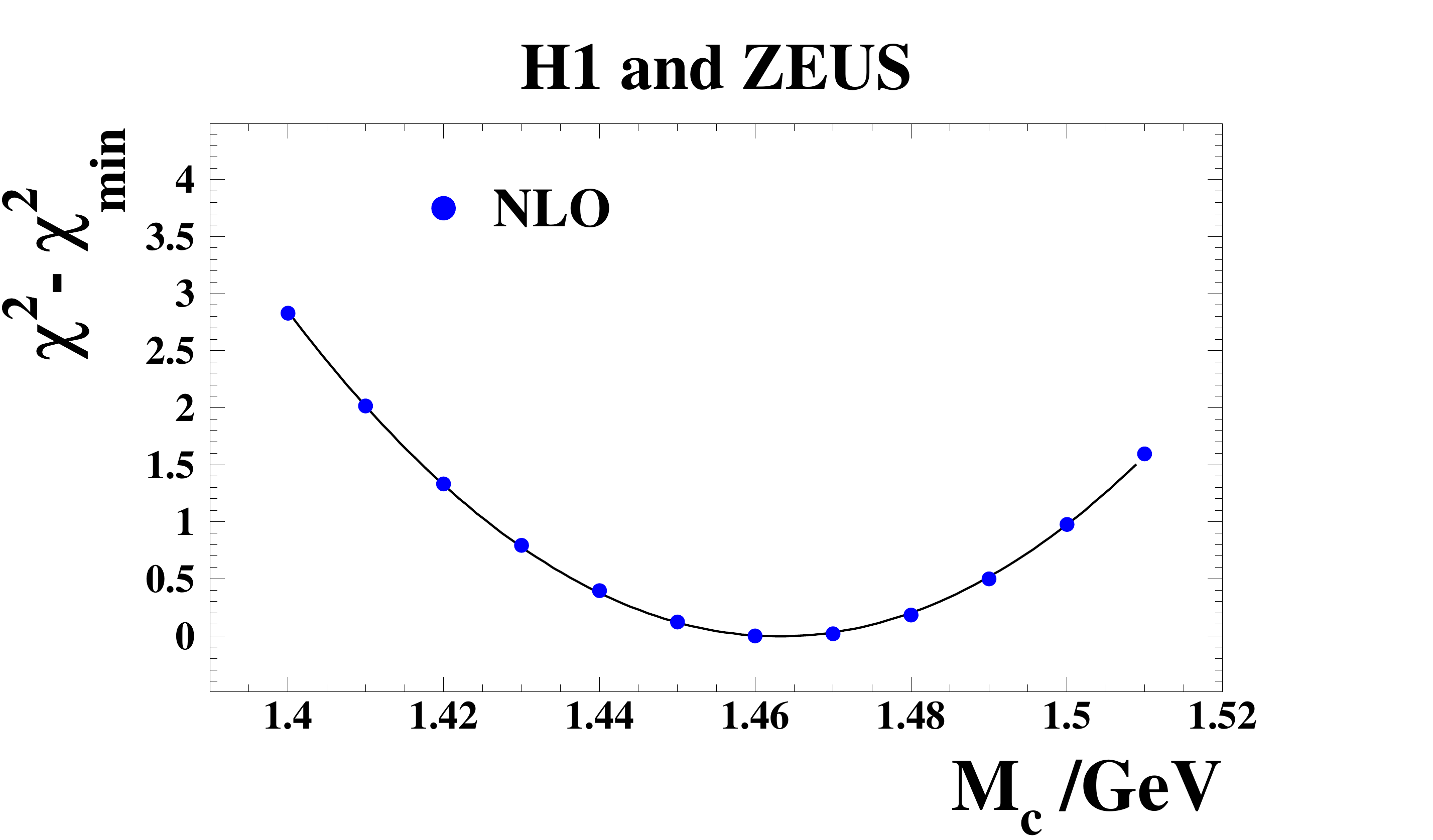}}
  \put(-.5,0.0){\includegraphics[width=0.55\textwidth]{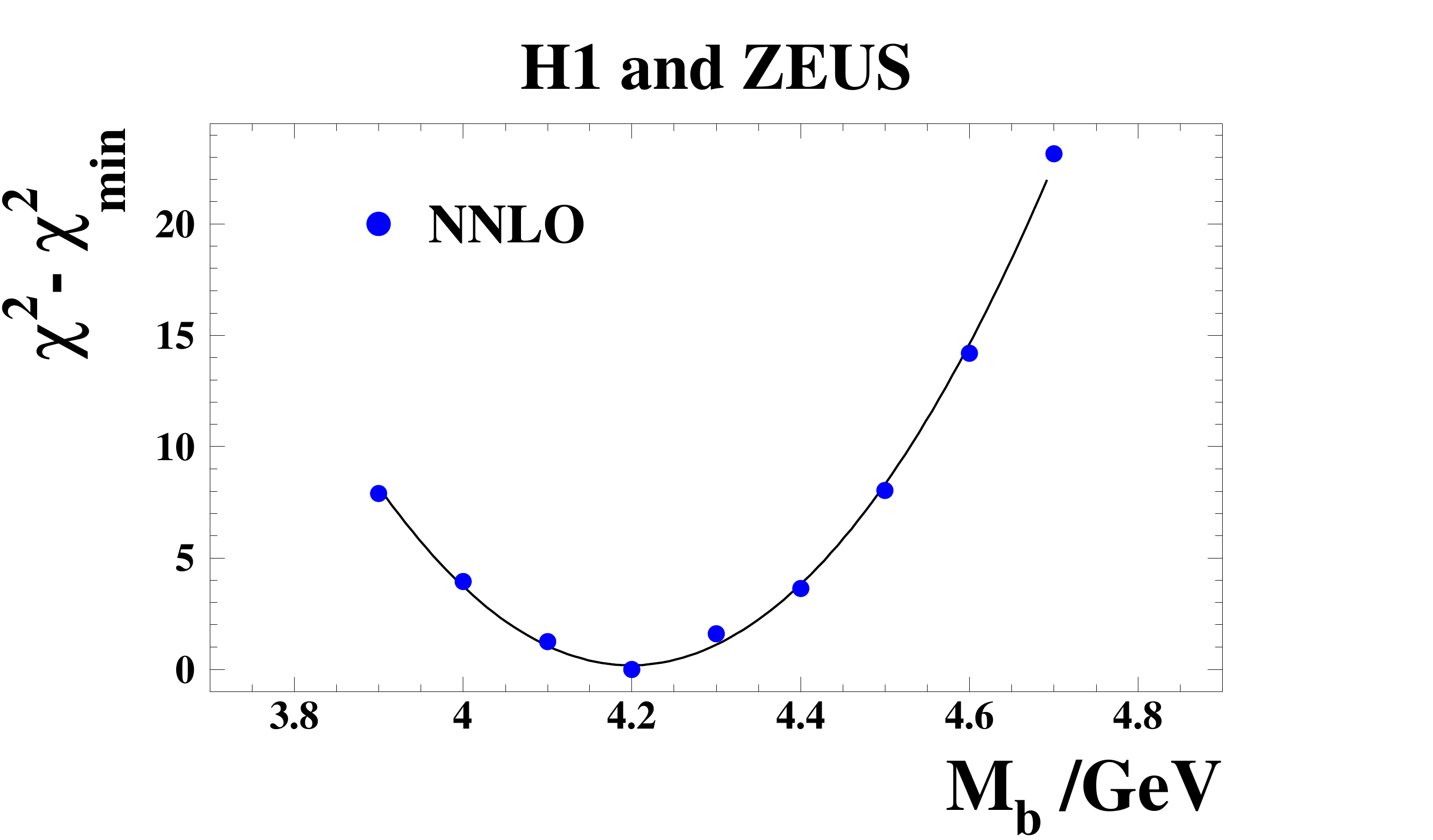}}
  \put(4.5,0.0){\includegraphics[width=0.55\textwidth]{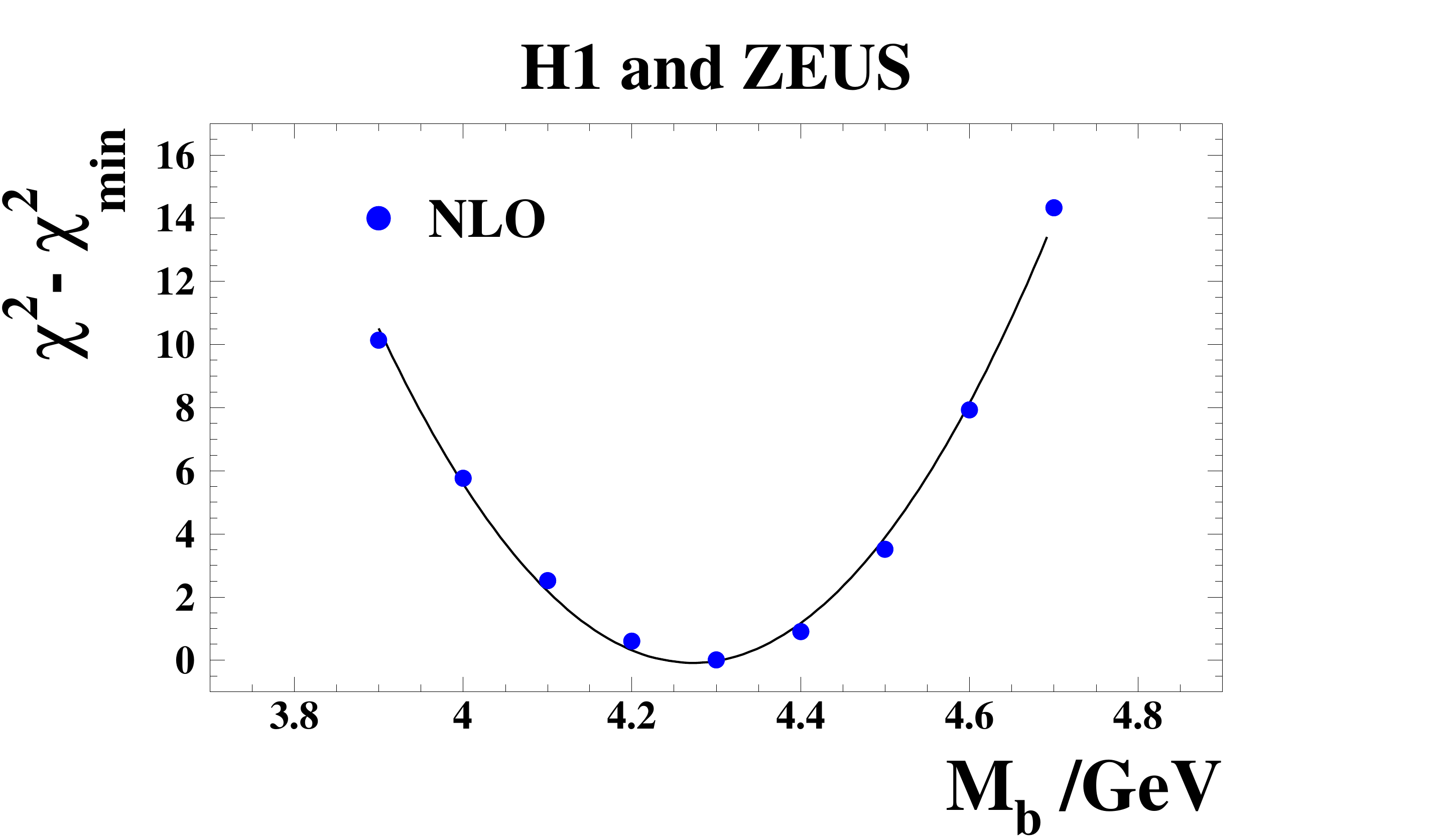}}
  \put (-.3,4.5) {a)}
  \put (4.7,4.5) {b)}
  \put (-.3,0.5) {c)}
  \put (4.7,0.5) {d)}
  \end{picture}
  \vskip 0.7cm
  \caption {Difference between $\chi^2$ and  $\chi^2_{\rm min}$ versus
    a) $M_c$ for $M_b = 4.2$\,GeV at NNLO with $\asmz=0.1155$,
    b) $M_c$ for $M_b = 4.3$\,GeV at NLO  with $\asmz=0.118$,
    c) $M_b$ with $M_c=1.41\,$GeV at NNLO with $\asmz=0.1155$,
    d) $M_b$ with $M_c=1.46\,$GeV at NLO with $\asmz=0.118$.
}
\label{fig:mass-scans}
\end{figure}

\clearpage

\begin{figure}
  \centering
  \setlength{\unitlength}{0.1\textwidth}
  \begin{picture} (9,8)
  \put(0,0){\includegraphics[width=0.9\textwidth]{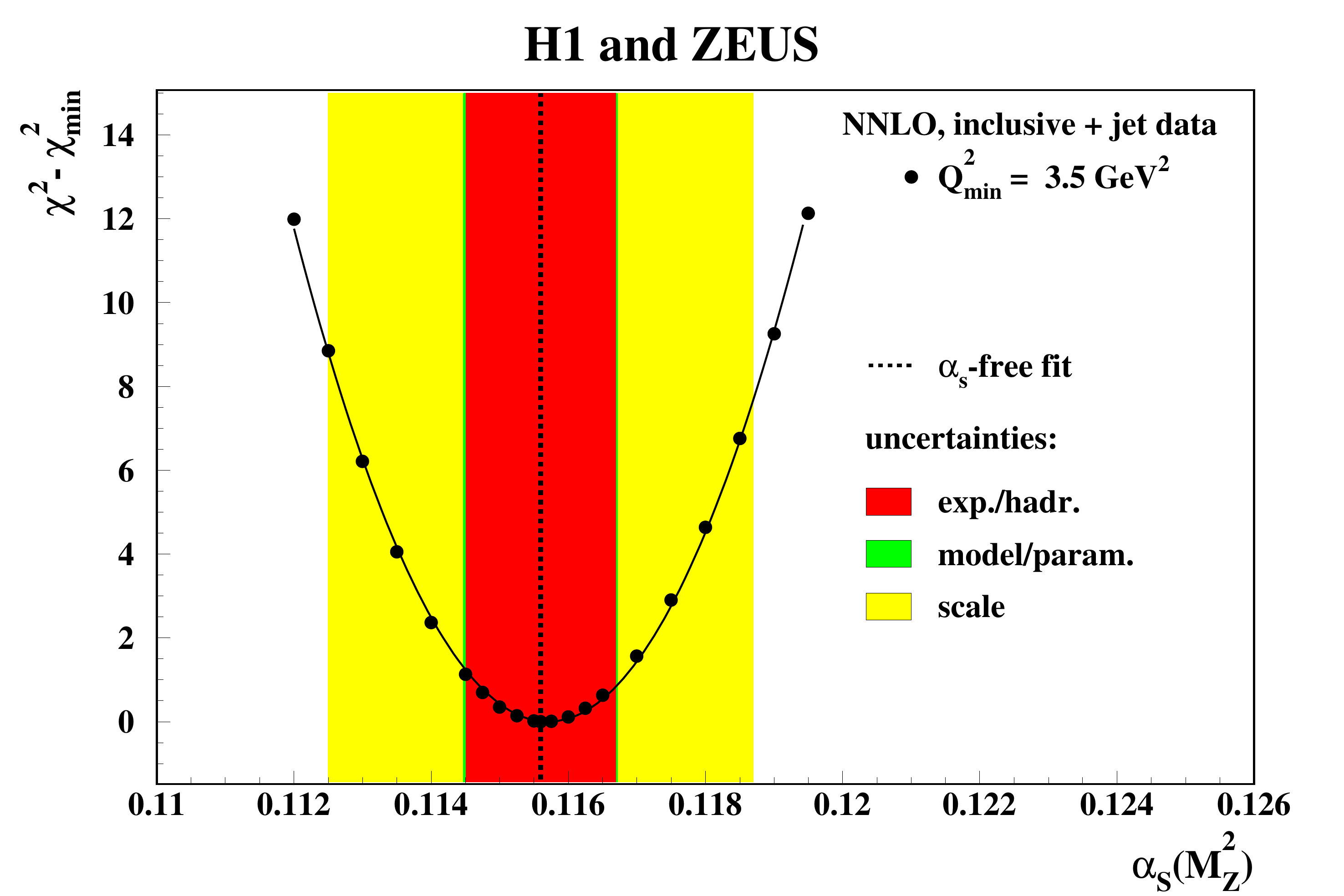}}
  \end{picture}
  \caption {Difference between $\chi^2$ and  $\chi^2_{\rm min}$ versus
    $\asmz$ for
    HERAPDF2.0Jets NNLO fits with fixed $\asmz$.
    The result and all uncertainties determined for
    the HERAPDF2.0Jets NNLO fit
    with free $\asmz$ are also shown, added in quadrature.
}
\label{fig:alphasscan2}
\end{figure}

\clearpage
\begin{figure}
  \centering
  \setlength{\unitlength}{0.1\textwidth}
  \begin{picture} (9,8)
  \put(0.0,0.0){\includegraphics[width=0.9\textwidth]{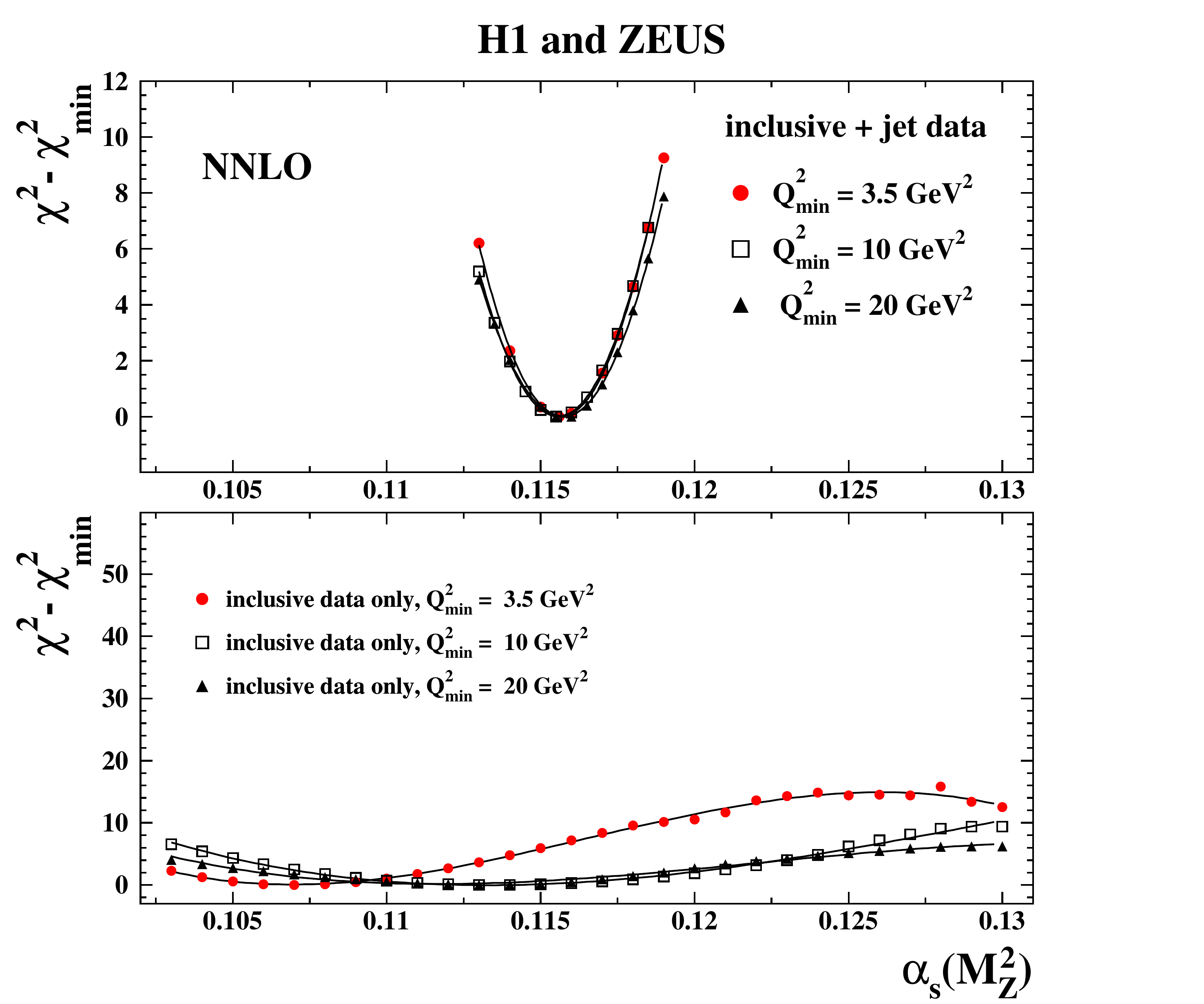}}
  \put (0.2,3.9) {a)}
  \put (0.2,0.7) {b)}
  \end{picture}
  \caption {Difference between $\chi^2$ and  $\chi^2_{\rm min}$ versus
    $\asmz$ for
    a) HERAPDF2.0Jets NNLO fits with fixed $\asmz$ with the standard
    $Q^2_{\rm min}$ for the inclusive data of 3.5\,GeV$^2$ and
    $Q^2_{\rm min}$ set to 10\,GeV$^2$ and 20\,GeV$^2$.
    b) For comparison, the situation for fits to only inclusive data,
    HERAPDF2.0 NNLO, is shown, taken from~\cite{HERAPDF20}.
}
\label{fig:alphasscan3}
\end{figure}
\clearpage

\begin{figure}[tbp]
  \centering
  \setlength{\unitlength}{0.1\textwidth}
  \begin{picture} (7,13)
    \put (0,6.7) {\includegraphics[width=0.7\textwidth]{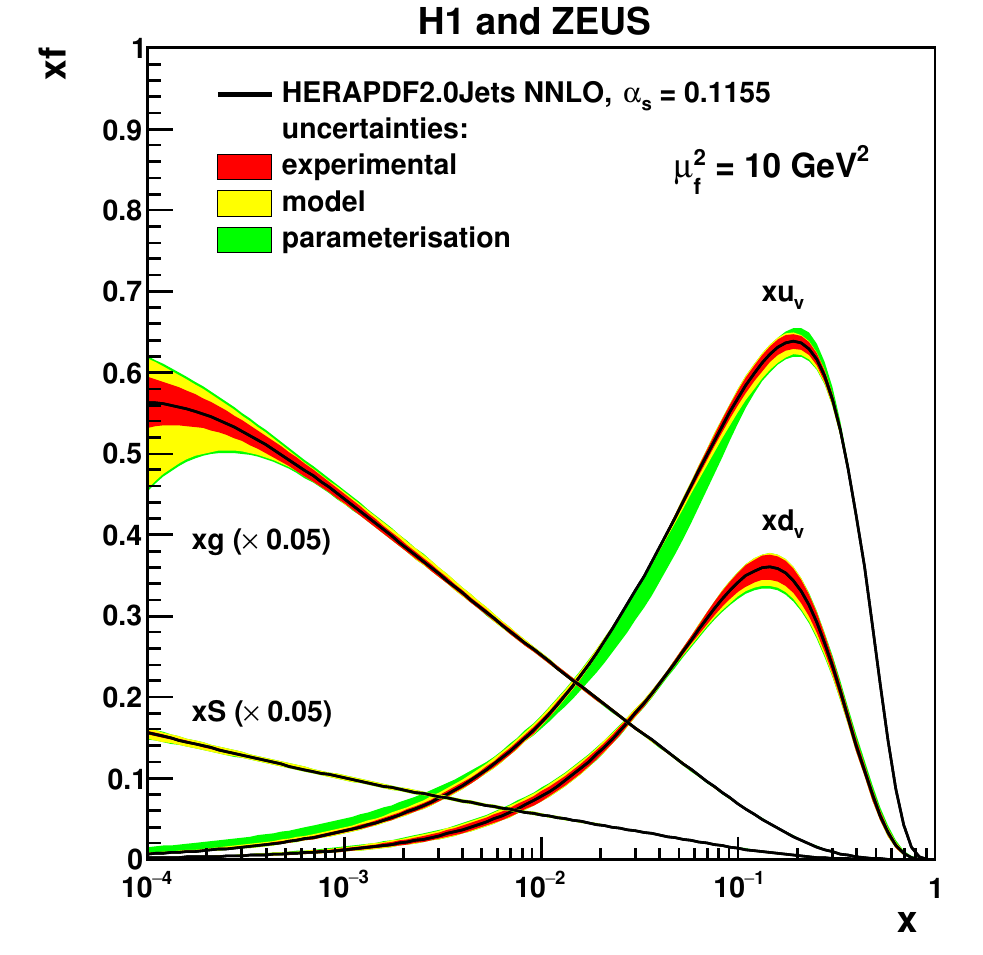}}
    \put (0,0)  {\includegraphics[width=0.7\textwidth]{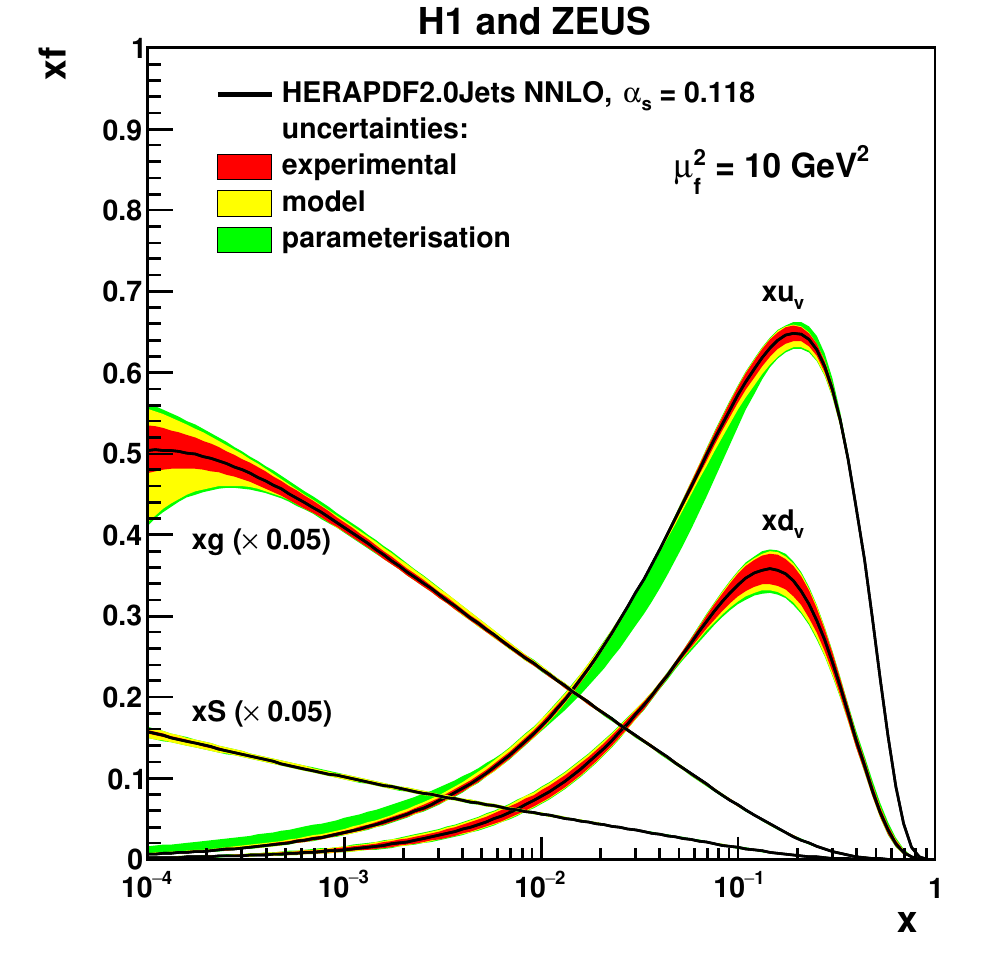}}
    \put (0,7.2) {a)}
    \put (0,0.5) {b)}
   \end{picture}
\caption { 
The parton distribution functions 
$xu_v$, $xd_v$, $xg$ and $xS=x(\bar{U}+\bar{D})$ of
HERAPDF2.0Jets NNLO, with a) $\asmz$ fixed to 0.1155 and
b) $\asmz$ fixed to 0.118,
at the scale $\mu_{\rm f}^2 = 10$\,GeV$^{2}$.
The uncertainties are shown as differently shaded bands.
}
\label{fig:as0-116-118}
\end{figure}

\clearpage

\begin{figure}[tbp]
  \centering
  \vskip -6cm
  \setlength{\unitlength}{0.1\textwidth}
  \begin{picture} (12,12)
   \put(-1.0,0.0){\includegraphics[width=1.2\textwidth]{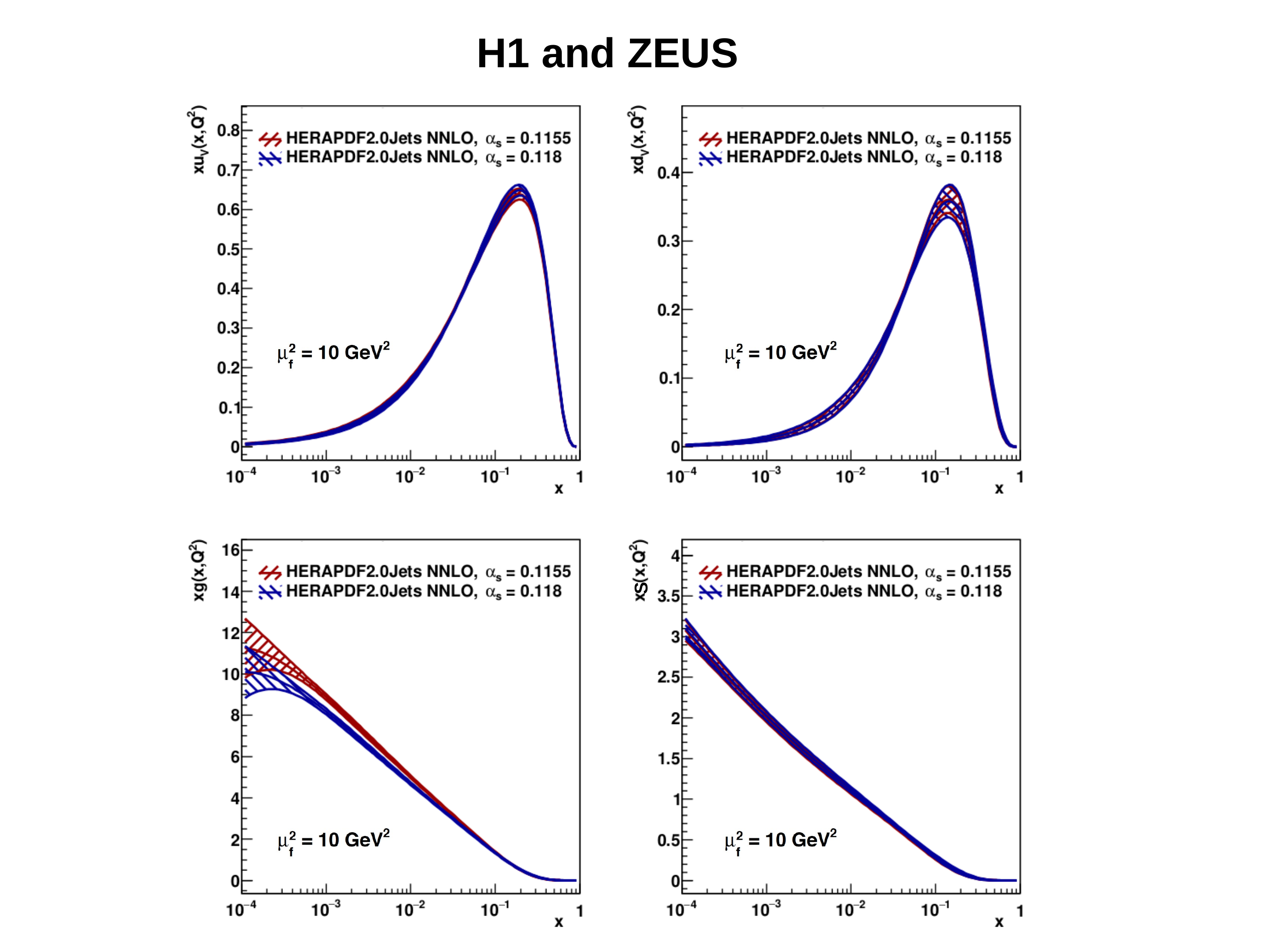}}    
  \put (0.4,5.5) {a)}
  \put (5.2,5.5) {b)}
  \put (0.4,0.7) {c)}
  \put (5.2,0.7) {d)}
  \end{picture}
\vspace{-0.5cm} 
\caption {  
Comparison of the parton distribution functions 
a) $xu_v$, b) $xd_v$, c) $xg$ and d) $xS=x(\bar{U}+\bar{D})$ of 
HERAPDF2.0Jets NNLO with fixed $\asmz = 0.1155$ and $\asmz = 0.118$,
at the scale $\mu_{\rm f}^{2} = 10\,$GeV$^{2}$.
The total uncertainties are shown as differently hatched bands.
}
\label{fig:as0-116vsas0-118}
\end{figure}
\clearpage

\begin{figure}[tbp]
  \centering
  \vskip -6cm  
  \setlength{\unitlength}{0.1\textwidth}
  \begin{picture} (12,12)
    \put(-0.6,0.0){\includegraphics[width=1.17\textwidth]{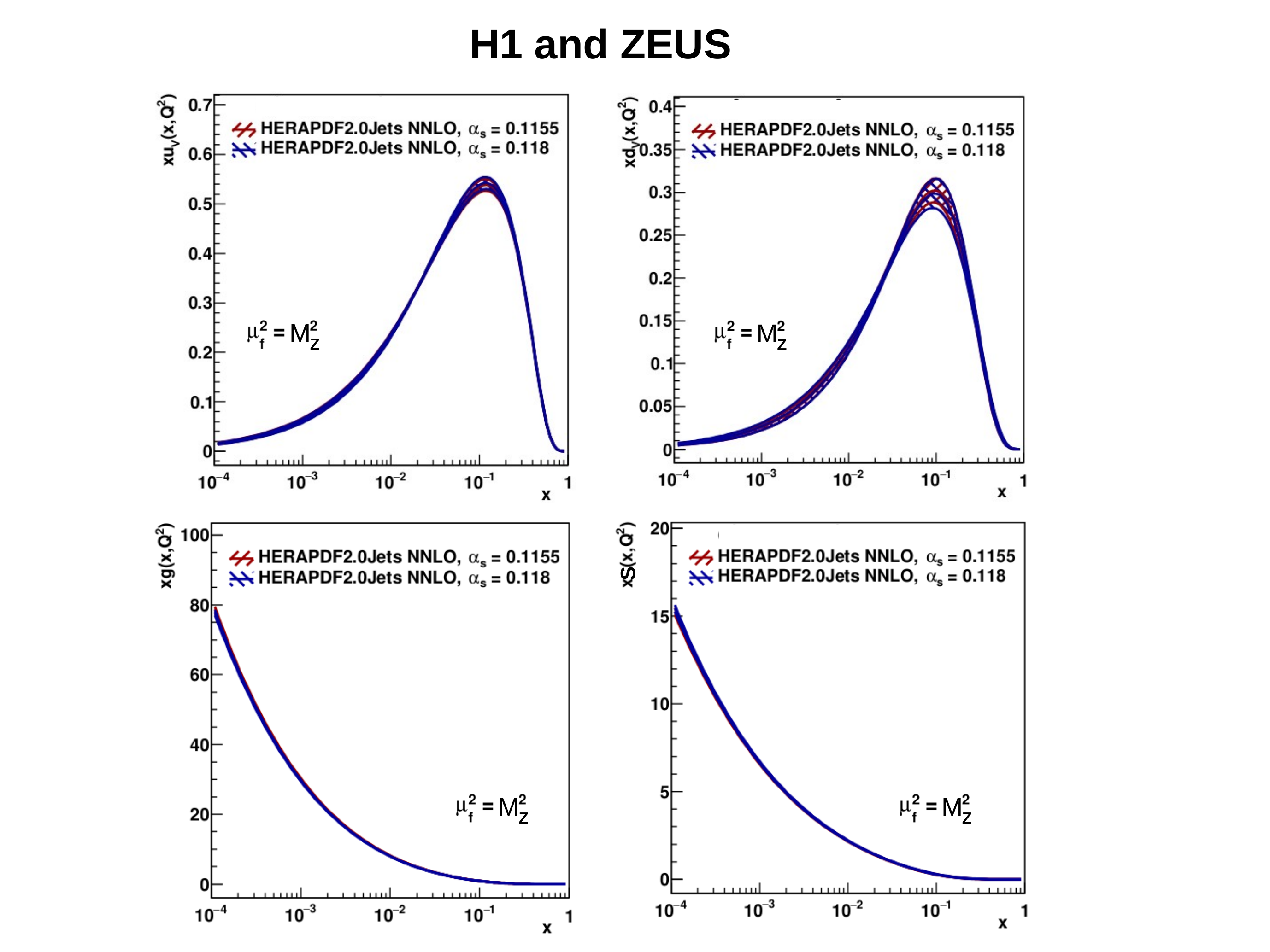}}
  \put (0.4,5.5) {a)}
  \put (5.2,5.5) {b)}
  \put (0.4,0.7) {c)}
  \put (5.2,0.7) {d)}
  \end{picture}
\vspace{-0.5cm} 
\caption { 
Comparison of the parton distribution functions 
a) $xu_v$, b) $xd_v$, c) $xg$ and d) $xS=x(\bar{U}+\bar{D})$ of 
HERAPDF2.0Jets NNLO with fixed $\asmz = 0.1155$ and $\asmz = 0.118$,
at the scale $\mu_{\rm f}^{2} = M_Z^2$ with $M_Z=91.19$\,GeV~\cite{PDG18}.
The total uncertainties are shown as differently hatched bands.
}
\label{fig:as0-116vsas0-118-mz}
\end{figure}

\clearpage

\begin{figure}[tbp]
  \centering
  \setlength{\unitlength}{0.1\textwidth}
  \includegraphics[width=1.0\textwidth]{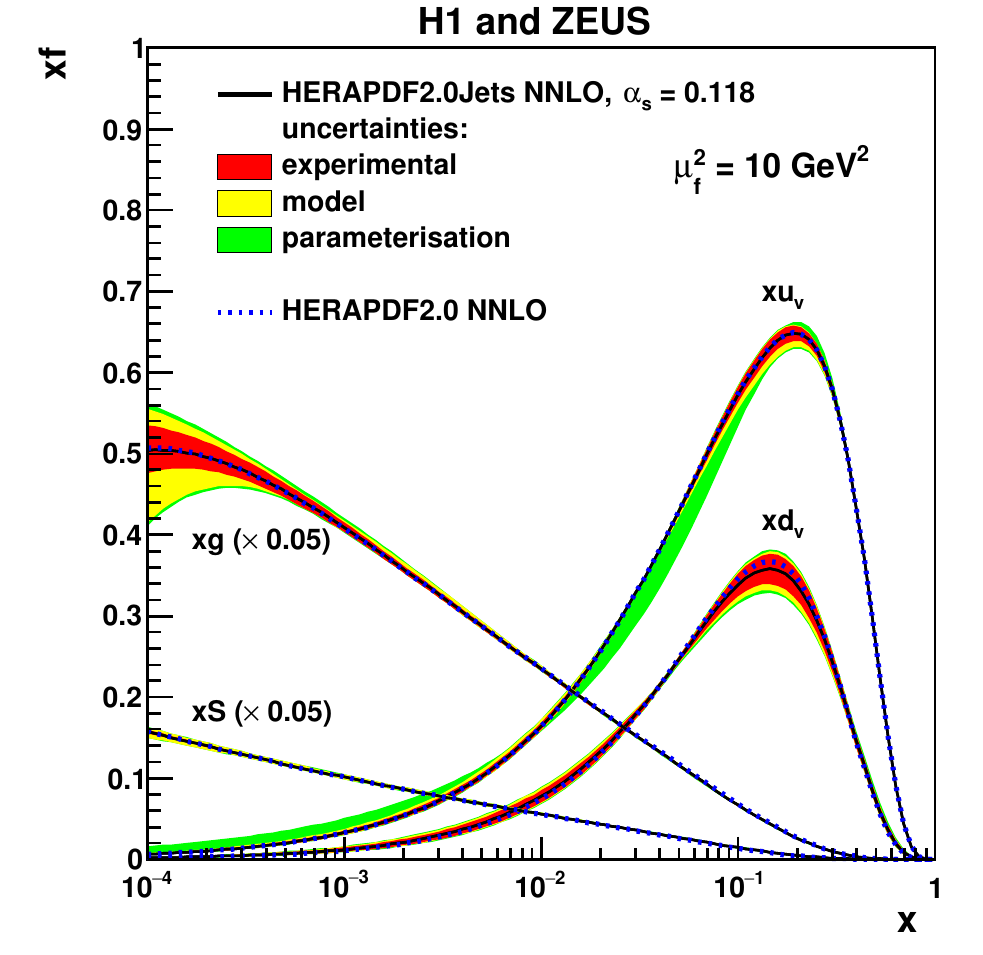}
  \caption {
     Comparison of the parton distribution functions 
     $xu_v$, $xd_v$, $xg$ and $xS=x(\bar{U}+\bar{D})$ of
     HERAPDF2.0Jets NNLO
     with HERAPDF2.0 NNLO based
     on inclusive data only, both with fixed $\asmz = 0.118$,
     at the scale $\mu_{\rm f}^{2} = 10\,$GeV$^{2}$.
     The uncertainties of HERAPDF2.0Jets NNLO
     are shown as differently shaded bands
     and the central value of HERAPDF2.0 NNLO is shown as a dotted line.
}
\label{fig:as0-118vsherapdf2}
\end{figure}


\clearpage

\begin{figure}
  \centering
  \vskip -6cm
  \setlength{\unitlength}{0.1\textwidth}
  \begin{picture} (12,12)
  \put(-0.4,0.0){\includegraphics[width=1.2\textwidth]{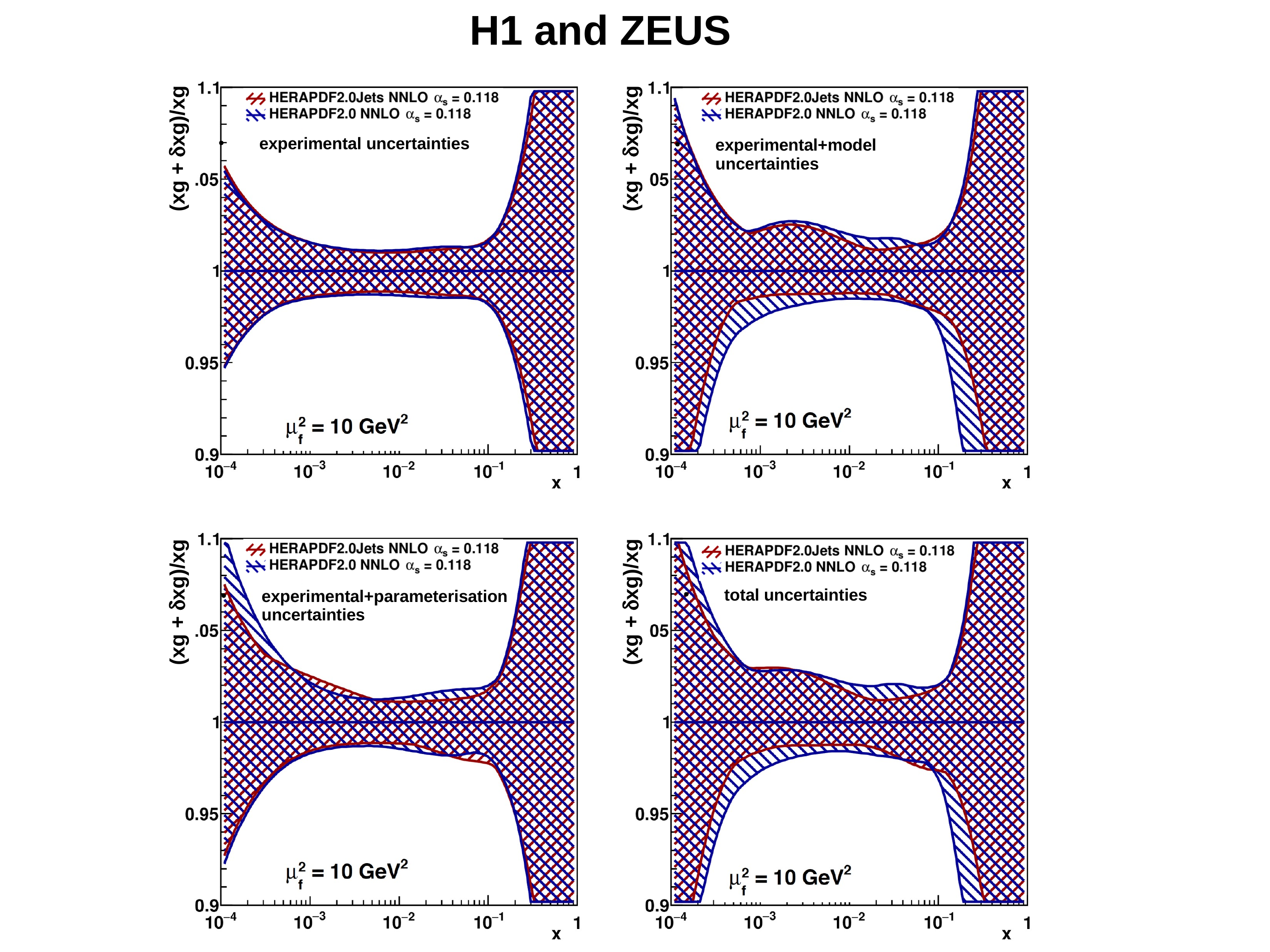}} 
  \put (0.5,5.3) {a)}
  \put (5.2,5.3) {b)}
  \put (0.5,0.7) {c)}
  \put (5.2,0.7) {d)}
  \end{picture}
  \caption{Comparison of the normalised uncertainties on the gluon PDFs
    of HERAPDF2.0Jets NNLO
    and HERAPDF2.0 NNLO, both for $\asmz = 0.118$
    and at the scale  $\mu_{\rm f}^{2} = 10\,$GeV$^{2}$,
    for
    a) experimental (fit),
    b) experimental plus model,
    c) experimental plus parameterisation,
    d) total
    uncertainties. 
    The uncertainties on both gluon PDFs are shown as
    differently hatched bands.
}
\label{fig:unc-as0-118vsherapdf2-0118}
\end{figure}

\clearpage

\begin{figure}
  \centering
  \vskip -6cm
  \setlength{\unitlength}{0.1\textwidth}
  \begin{picture} (12,12)
  \put(-0.4,0.0){\includegraphics[width=1.2\textwidth]{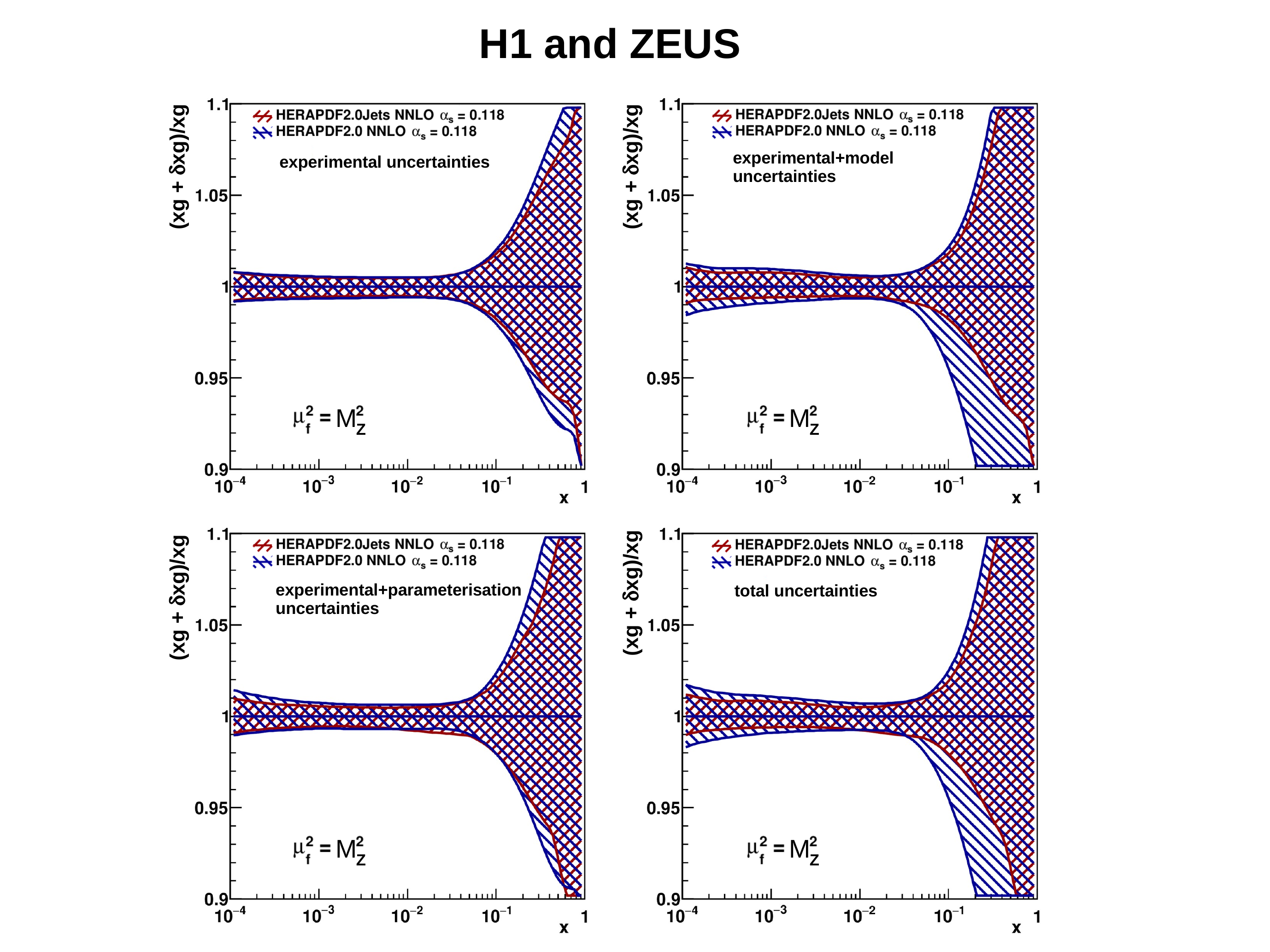}}  
  \put (0.5,5.3) {a)}
  \put (5.2,5.3) {b)}
  \put (0.5,0.7) {c)}
  \put (5.2,0.7) {d)}
  \end{picture}
  \caption{Comparison of the normalised uncertainties on the gluon PDFs
    of HERAPDF2.0Jets NNLO
    and HERAPDF2.0 NNLO, both for $\asmz = 0.118$
    and at the scale  $\mu_{\rm f}^{2} = M_Z^{2}$,
    for
    a) experimental (fit),
    b) experimental plus model,
    c) experimental plus parameterisation,
    d) total
    uncertainties. 
    The uncertainties on both gluon PDFs are shown as
    differently hatched bands.
}
\label{fig:unc-as0-118vsherapdf2-MZ-0118}
\end{figure}

\begin{figure}
  \centering
  \vskip -6cm
  \setlength{\unitlength}{0.1\textwidth}
  \begin{picture} (12,12)
  \put(-0.4,0.0){\includegraphics[width=1.2\textwidth]{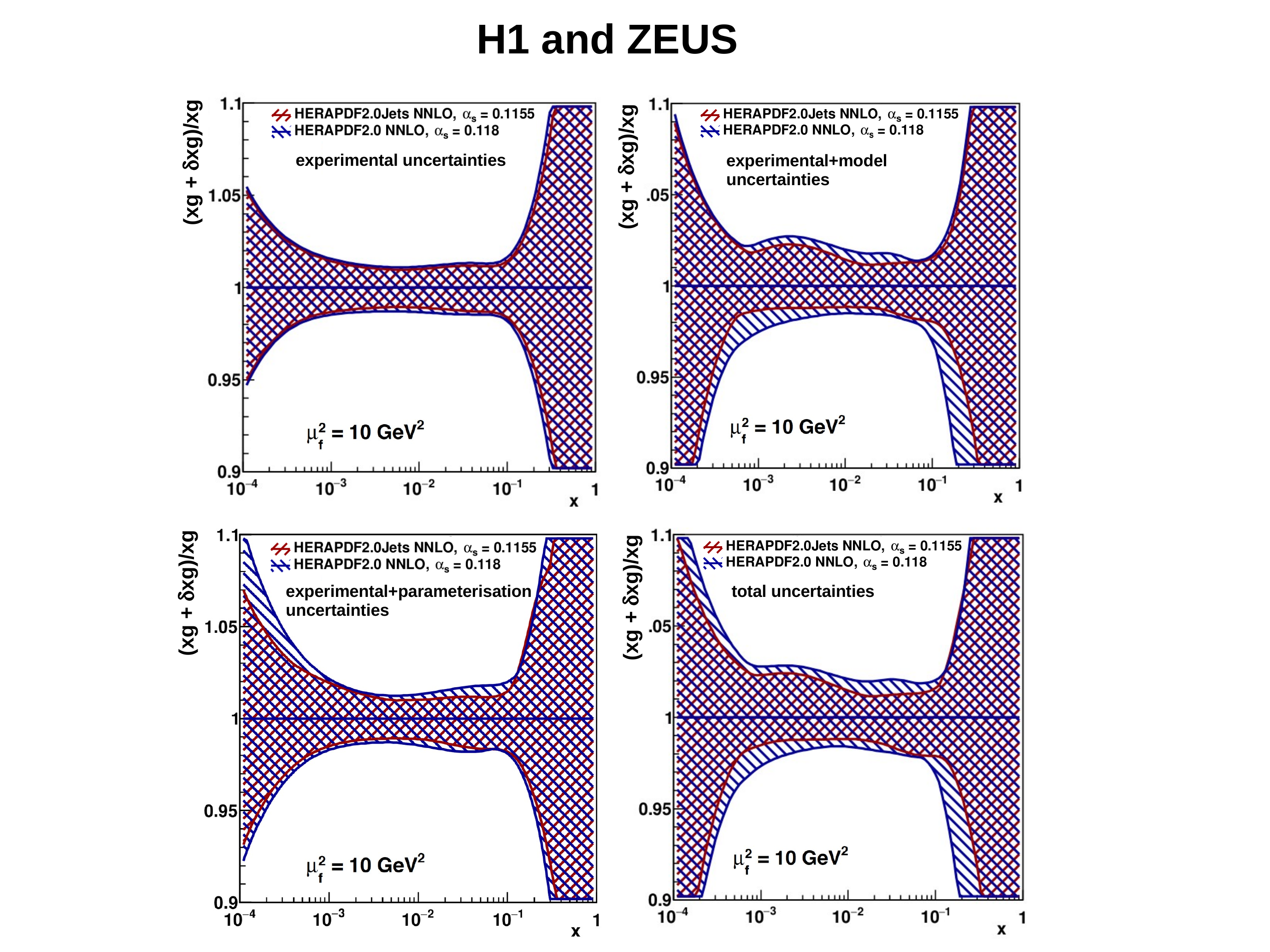}}  
  \put (0.5,5.3) {a)}
  \put (5.2,5.3) {b)}
  \put (0.5,0.7) {c)}
  \put (5.2,0.7) {d)}
  \end{picture}
  \caption{Comparison of the normalised uncertainties on the gluon PDFs
    of HERAPDF2.0Jets NNLO for $\asmz = 0.1155$
    and HERAPDF2.0 NNLO for $\asmz = 0.118$, 
    both at the scale  $\mu_{\rm f}^{2} = 10\,$GeV$^{2}$,
    for
    a) experimental (fit),
    b) experimental plus model,
    c) experimental plus parameterisation,
    d) total
   uncertainties. 
    The uncertainties on both gluon PDFs are shown as
    differently hatched bands.
}
\label{fig:unc-as0-118vsherapdf2}
\end{figure}

\clearpage

\begin{figure}
  \centering
  \vskip -6cm
  \setlength{\unitlength}{0.1\textwidth}
  \begin{picture} (12,12)
  \put(-0.4,0.0){\includegraphics[width=1.2\textwidth]{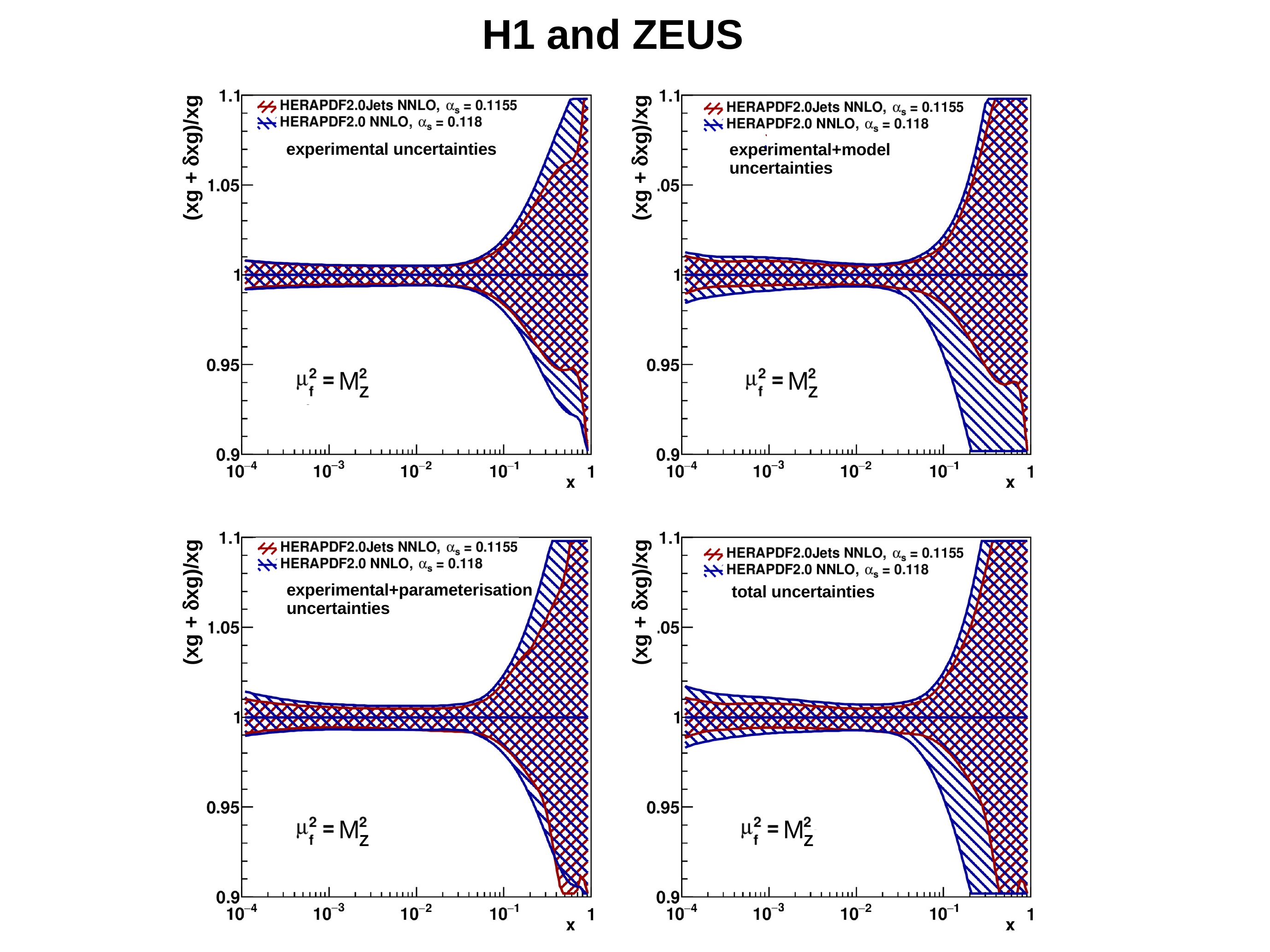}}  
  \put (0.5,5.3) {a)}
  \put (5.2,5.3) {b)}
  \put (0.5,0.7) {c)}
  \put (5.2,0.7) {d)}
  \end{picture}
  \caption{Comparison of the normalised uncertainties on the gluon PDFs
    of HERAPDF2.0Jets NNLO for $\asmz = 0.1155$
    and HERAPDF2.0 NNLO for $\asmz = 0.118$,
    both at the scale  $\mu_{\rm f}^{2} = M_Z^{2}$,
    for
    a) experimental, i.e.\ fit,
    b) experimental plus model,
    c) experimental plus parameterisation,
    a) total
    uncertainties. 
    The uncertainties on both gluon PDFs are shown as
    differently hatched bands.
}
\label{fig:unc-as0-118vsherapdf2-MZ}
\end{figure}

\clearpage

\begin{figure}
  \centering
  \vskip -3cm
  \setlength{\unitlength}{0.1\textwidth}
  \begin{picture} (12,12)
  \put(0,0.0){\includegraphics[width=1.0\textwidth]{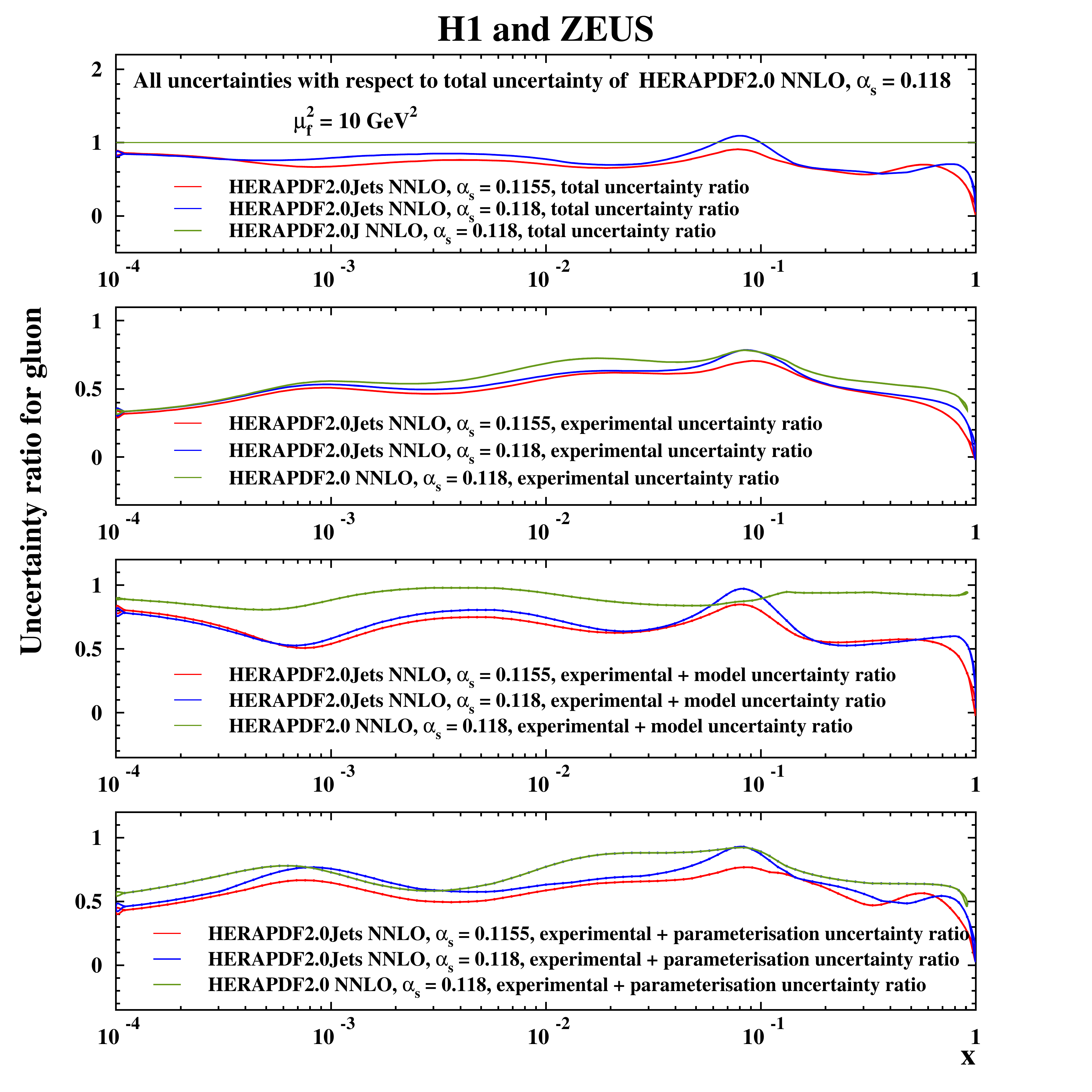}}  
  \put (0.6,7.6) {a)}
  \put (0.6,5.3) {b)}
  \put (0.6,3.0) {c)}
  \put (0.6,0.7) {d)}
  \end{picture}
  \caption{Ratios of uncertainties relative to the total
    uncertainties of HERAPDF2.0 NNLO for $\asmz=0.118$
    a) total,
    b) experimental,
    c) experimental plus model,
    d) experimental plus parameterisation
    uncertainties  for HERAPDF2.0Jets NNLO
    for $\asmz=0.118$ and $\asmz=0.1155$,
    all at the scale $\mu_{\rm f}^{2} = 10$\,GeV$^{2}$.
}
\label{fig:mark1}
\end{figure}
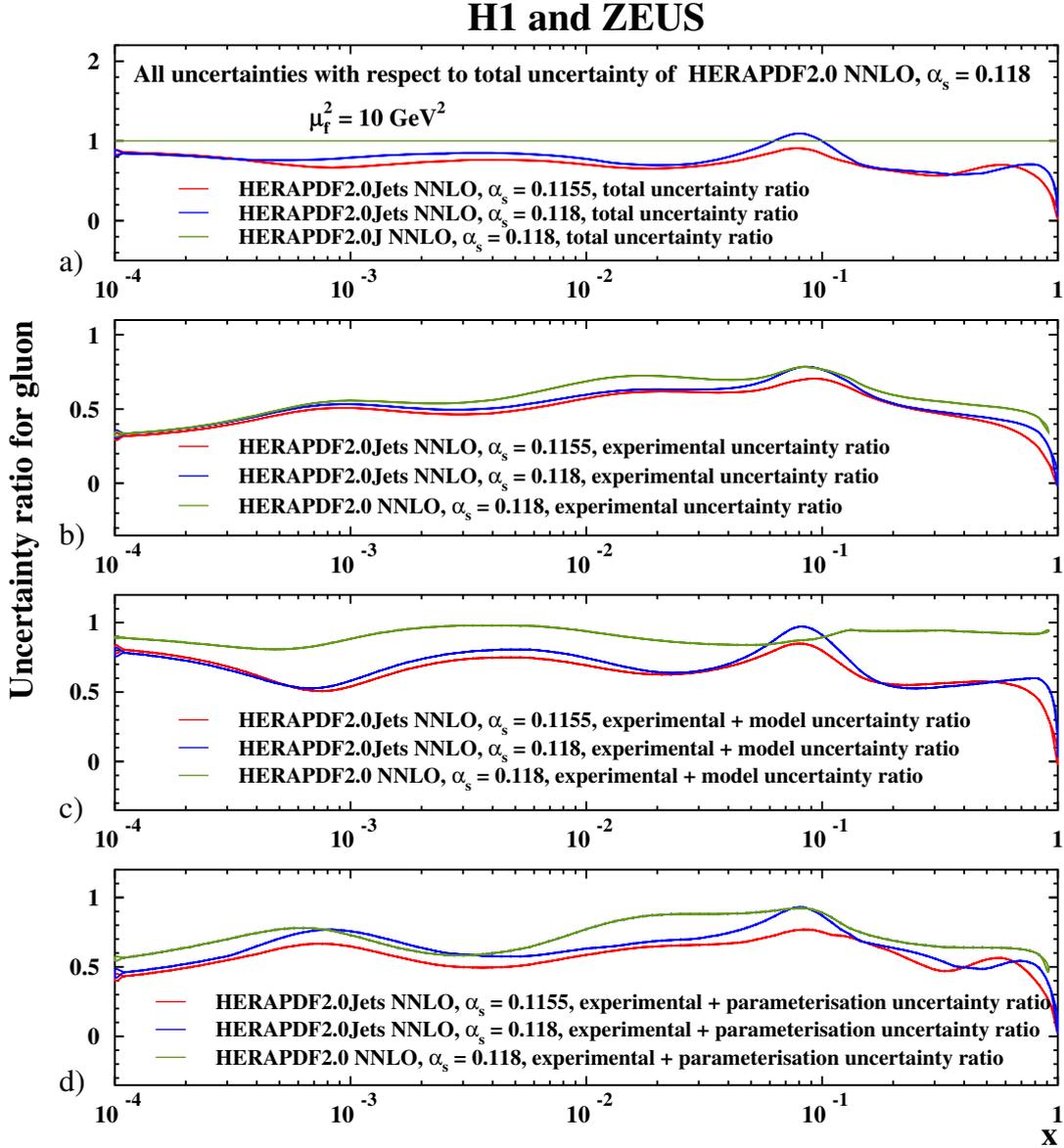

\clearpage


\begin{figure}
  \centering
  \setlength{\unitlength}{0.1\textwidth}
  \begin{picture} (9,12)
  \put(0,7){\includegraphics[width=0.9\textwidth]{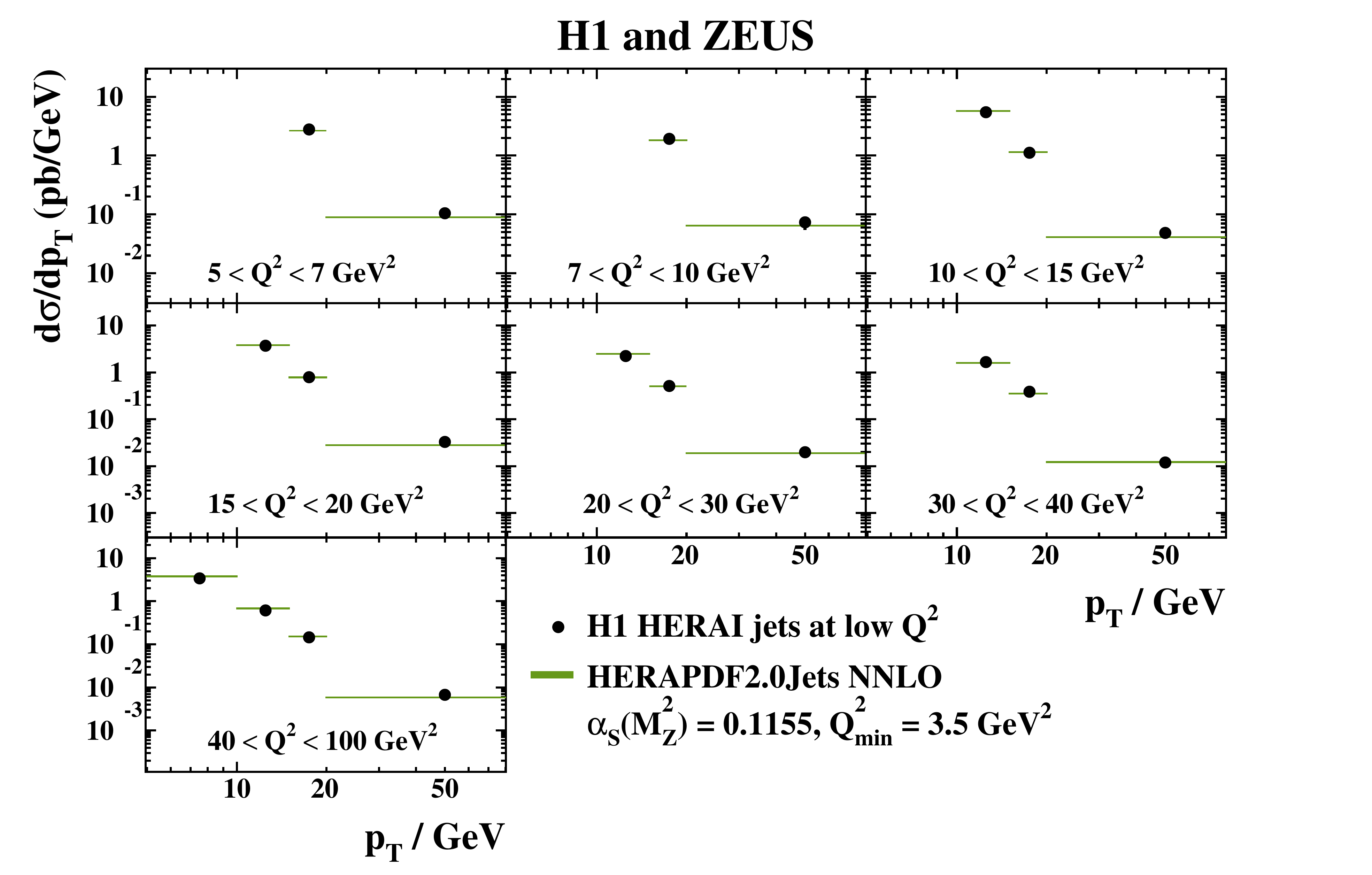}}
  \put(0,1){\includegraphics[width=0.9\textwidth]{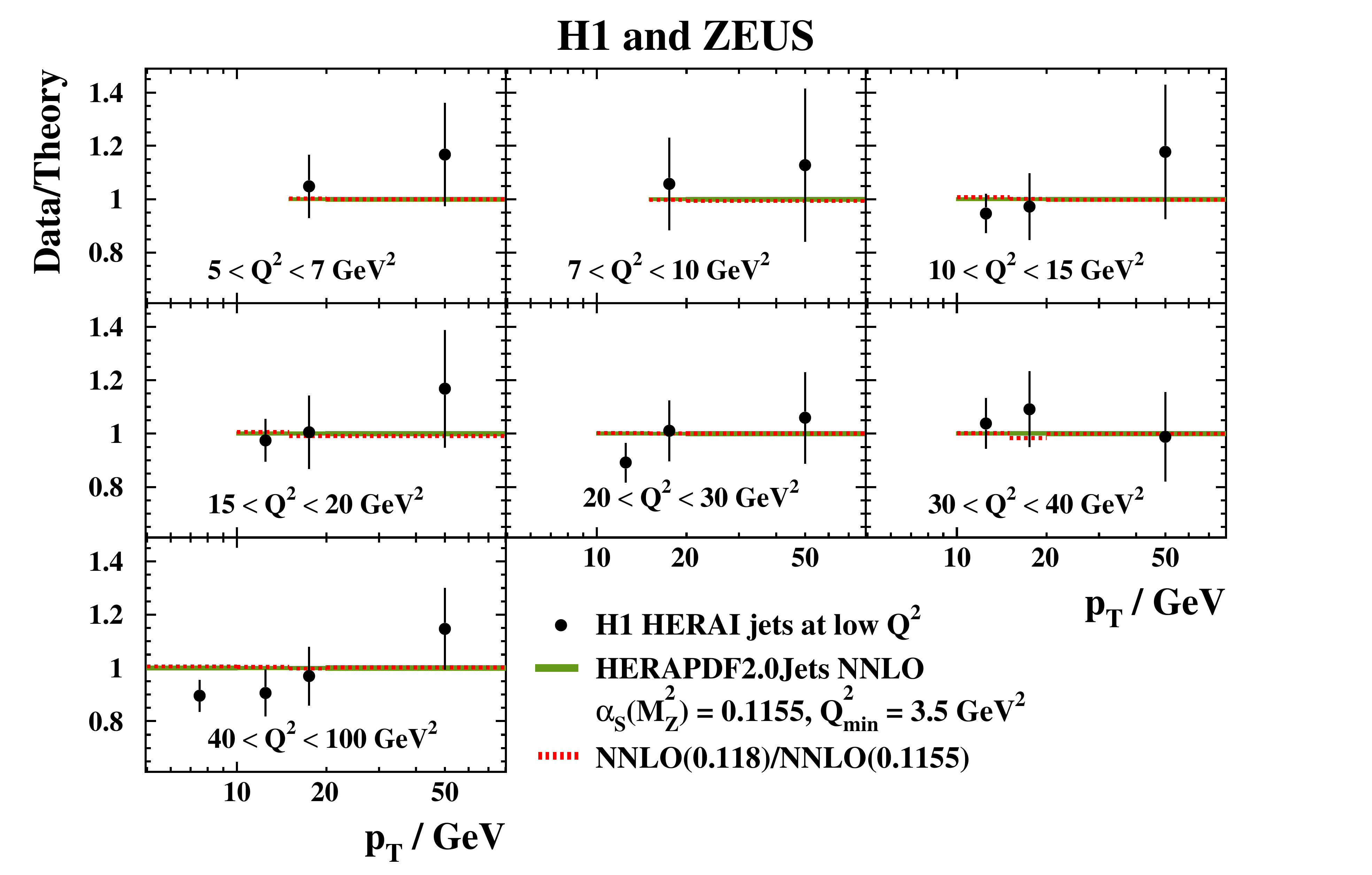}}
  \put (0.1,7.2) {a)}
  \put (0.1,1.2) {b)}
  \end{picture}
 \vskip -1cm 
\caption{
  a) Differential jet-cross-section predictions,
   ${\rm d}\sigma/{\rm d}p_{\rm T}$, based on HERAPDF2.0Jets NNLO
   with $\asmz = 0.1155$
   in bins of $Q^2$ between 5 and 100\,GeV$^2$ compared to H1
   data~\cite{h1lowq2jets}.
   Only data used in the fit are shown.
  b) Measured cross sections divided by
   predictions based on HERAPDF2.0Jets NNLO.
   The bands represent the total uncertainties on the predictions
   excluding scale uncertainties;
   the bands are so narrow that they mostly appear as lines.
   Error bars indicate the full uncertainties on the data and are smaller
   than the symbols in a).
   In b), the ratio of predictions based on HERAPDF2.0Jets NNLO with
   $\asmz = 0.118$ and $\asmz = 0.1155$ is also shown.
}
\label{fig:h1old-jet-data-lowQ2}
\end{figure}

\clearpage

\begin{figure}
  \centering
  \setlength{\unitlength}{0.1\textwidth}
  \begin{picture} (9,12)
  \put(0,7){\includegraphics[width=0.9\textwidth]{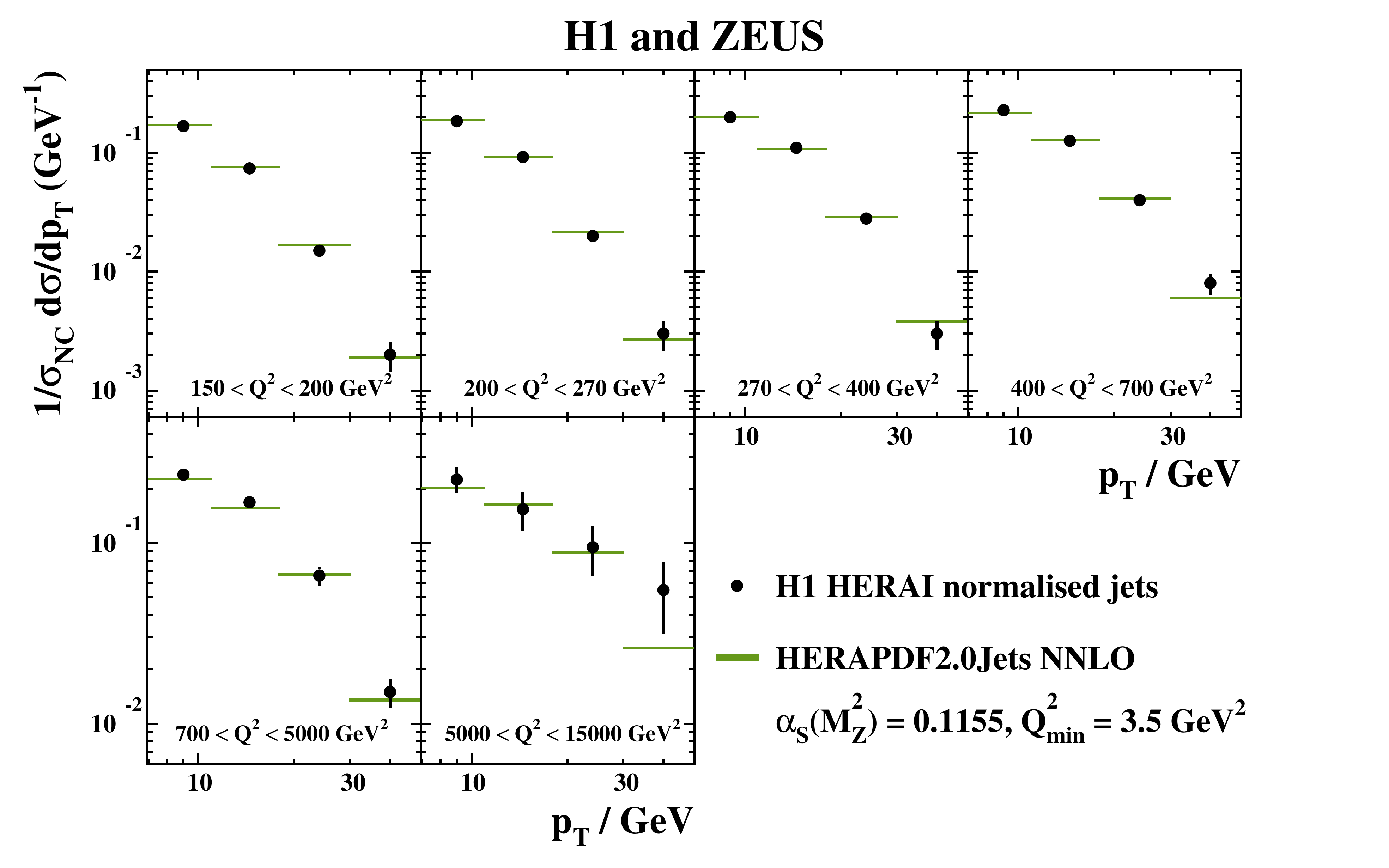}}
  \put(0,1){\includegraphics[width=0.9\textwidth]{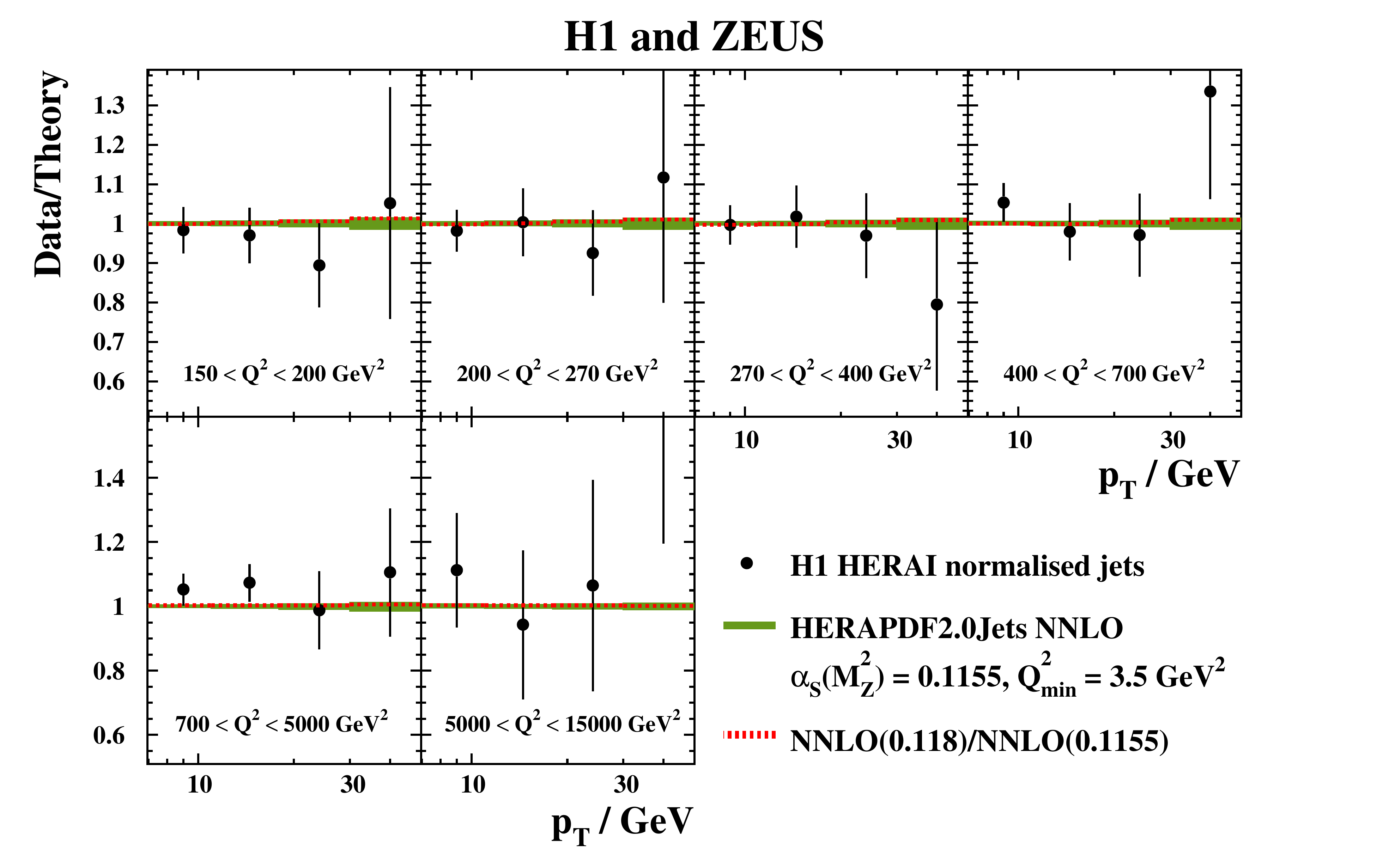}}
  \put (0.1,7.2) {a)}
  \put (0.1,1.2) {b)}
  \end{picture}
 \vskip -1cm 
\caption{
  a) Differential jet-cross-section predictions,
   ${\rm d}\sigma/{\rm d}p_{\rm T}$, based on HERAPDF2.0Jets NNLO 
   with $\asmz = 0.1155$
   in bins of $Q^2$ between 150 and 15000\,GeV$^2$ compared to H1 data
   normalised to neutral current (NC) inclusive cross
   sections~\cite{h1highq2oldjets}.
   Only data used in the fit are shown.
  b) Measured normalised cross sections divided by
   predictions based on HERAPDF2.0Jets NNLO.
   The bands represent the total uncertainties on the predictions
   excluding scale uncertainties; 
   the bands are so narrow that they mostly appear as lines.
   Error bars indicate the full uncertainties on the data and are smaller
   than the symbols for some bins in a).
   In b), the ratio of predictions based on HERAPDF2.0Jets NNLO with
   $\asmz = 0.118$ and $\asmz = 0.1155$ is also shown.
}
\label{fig:h1old-jet-data-highQ2}
\end{figure}
\clearpage


\begin{figure}
  \centering
  \setlength{\unitlength}{0.1\textwidth}
  \vskip 1.5 cm
  \begin{picture} (9,11)
  \put(0,7){\includegraphics[width=0.9\textwidth]{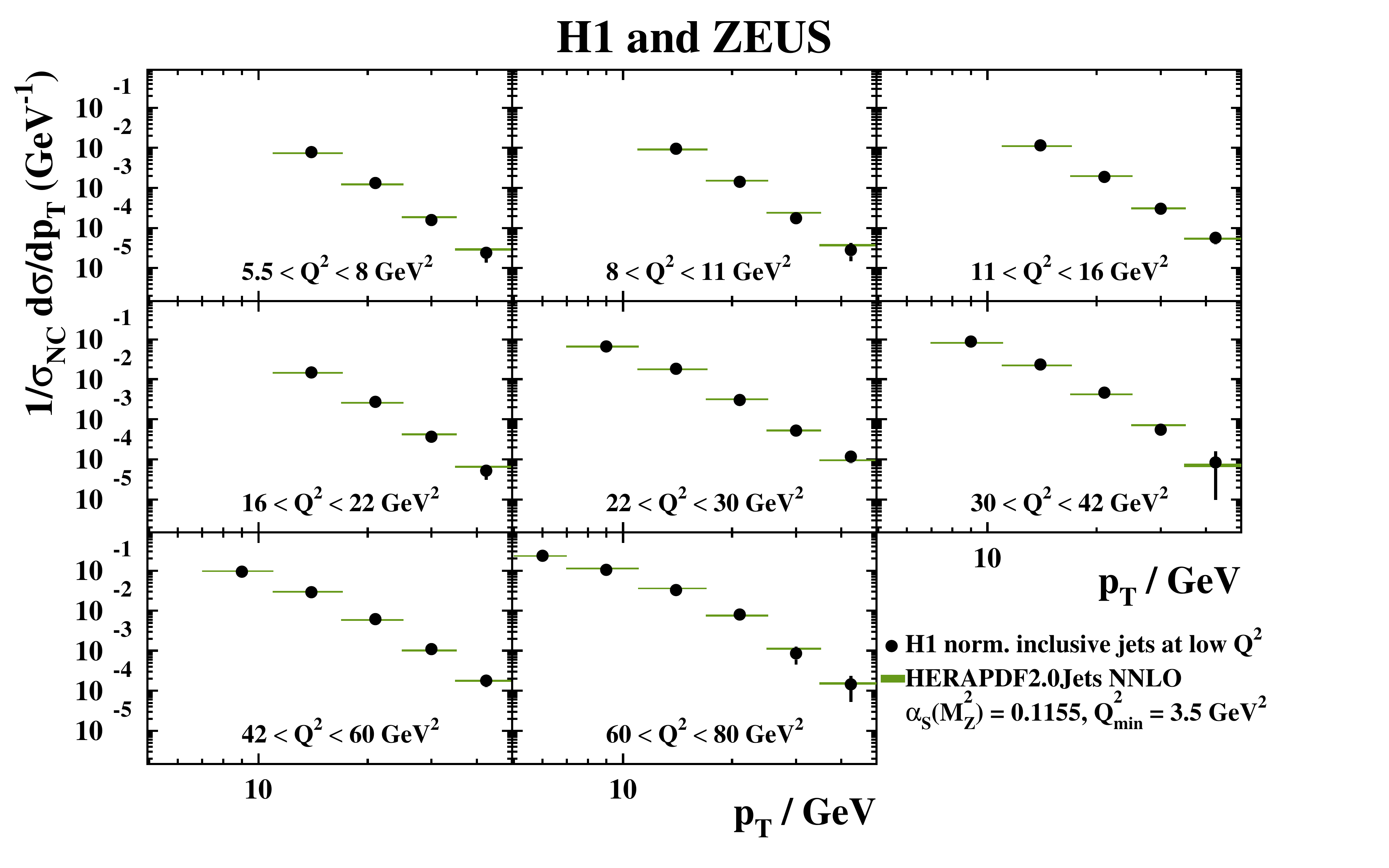}}
  \put(0,1){\includegraphics[width=0.9\textwidth]{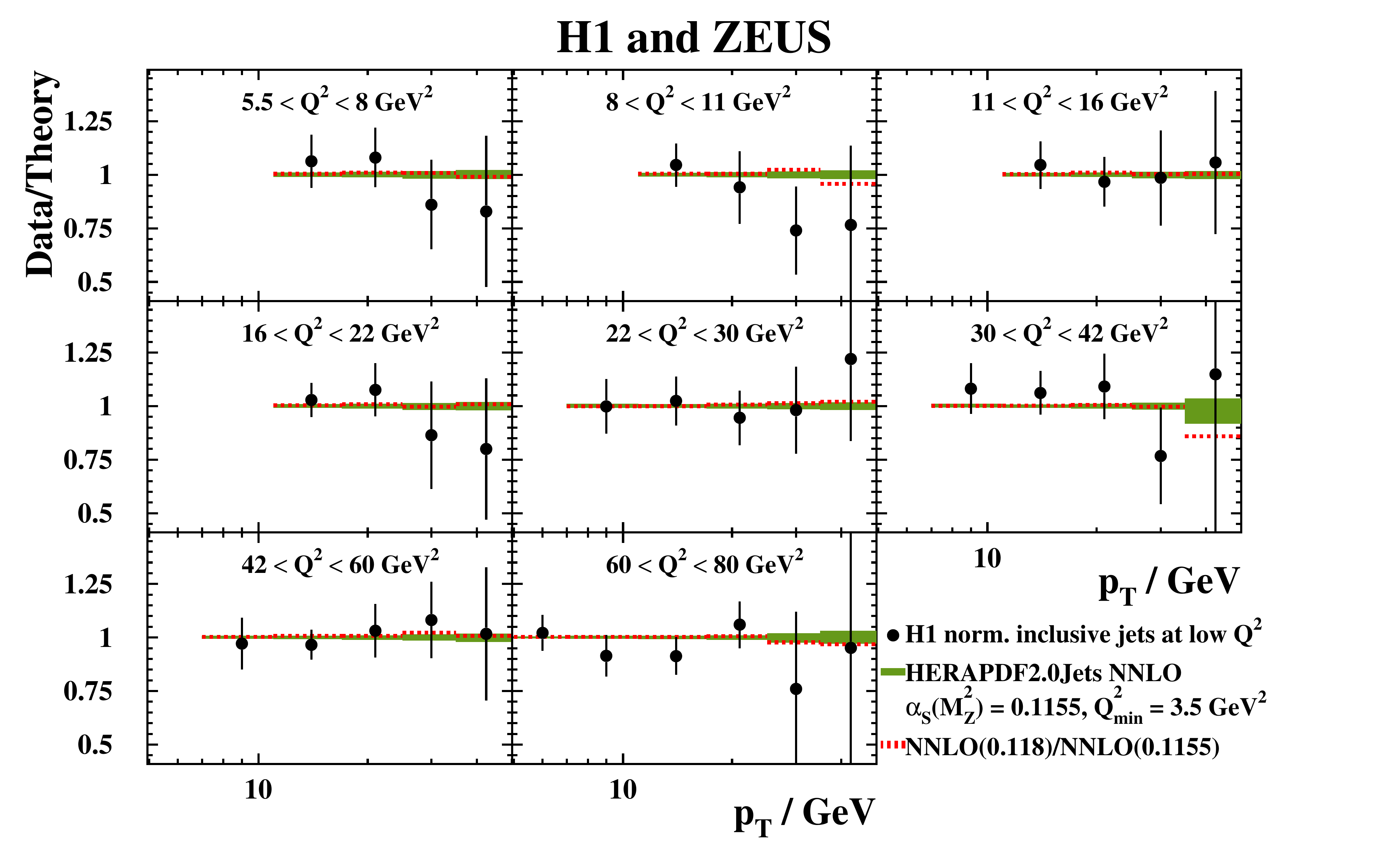}}
  \put (0.1,7.2) {a)}
  \put (0.1,1.2) {b)}
  \end{picture}
  \vskip -1.0 cm
  \caption{
  a) Differential jet-cross-section predictions,
   ${\rm d}\sigma/{\rm d}p_{\rm T}$, based on HERAPDF2.0Jets NNLO 
   with $\asmz = 0.1155$
   in bins of $Q^2$ between 5.5 and 80\,GeV$^2$ compared to H1 data
   normalised to neutral current (NC) inclusive cross
   sections~\cite{h1lowq2newjets}.
   Only data used in the fit are shown.
  b) Measured normalised cross sections divided by
   predictions based on HERAPDF2.0Jets NNLO.
   The bands represent the total uncertainties on the predictions
   excluding scale uncertainties; 
   the bands are so narrow that they mostly appear as lines.
   Error bars indicate the full uncertainties on the data and are mostly
   smaller than the symbols in a).
   In b), the ratio of predictions based on HERAPDF2.0Jets NNLO with
   $\asmz = 0.118$ and $\asmz = 0.1155$ is also shown.
}
\label{fig:h1-jet-data-low-Q2}
\end{figure}
\clearpage

\begin{figure}
  \centering
  \setlength{\unitlength}{0.1\textwidth}
  \vskip 1.5cm
  \begin{picture} (9,11)
  \put(0,7){\includegraphics[width=0.9\textwidth]{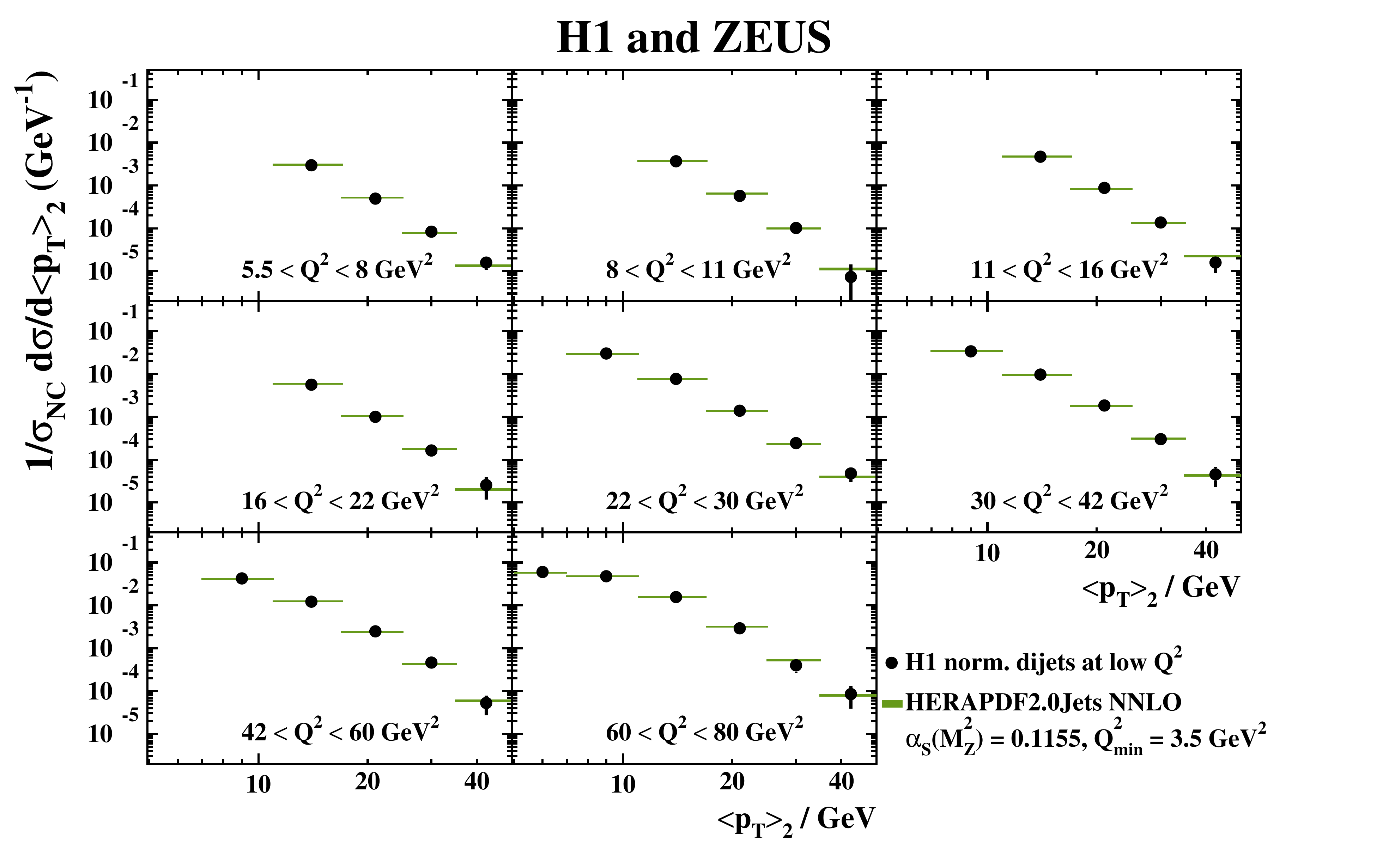}}
  \put(0,1){\includegraphics[width=0.9\textwidth]{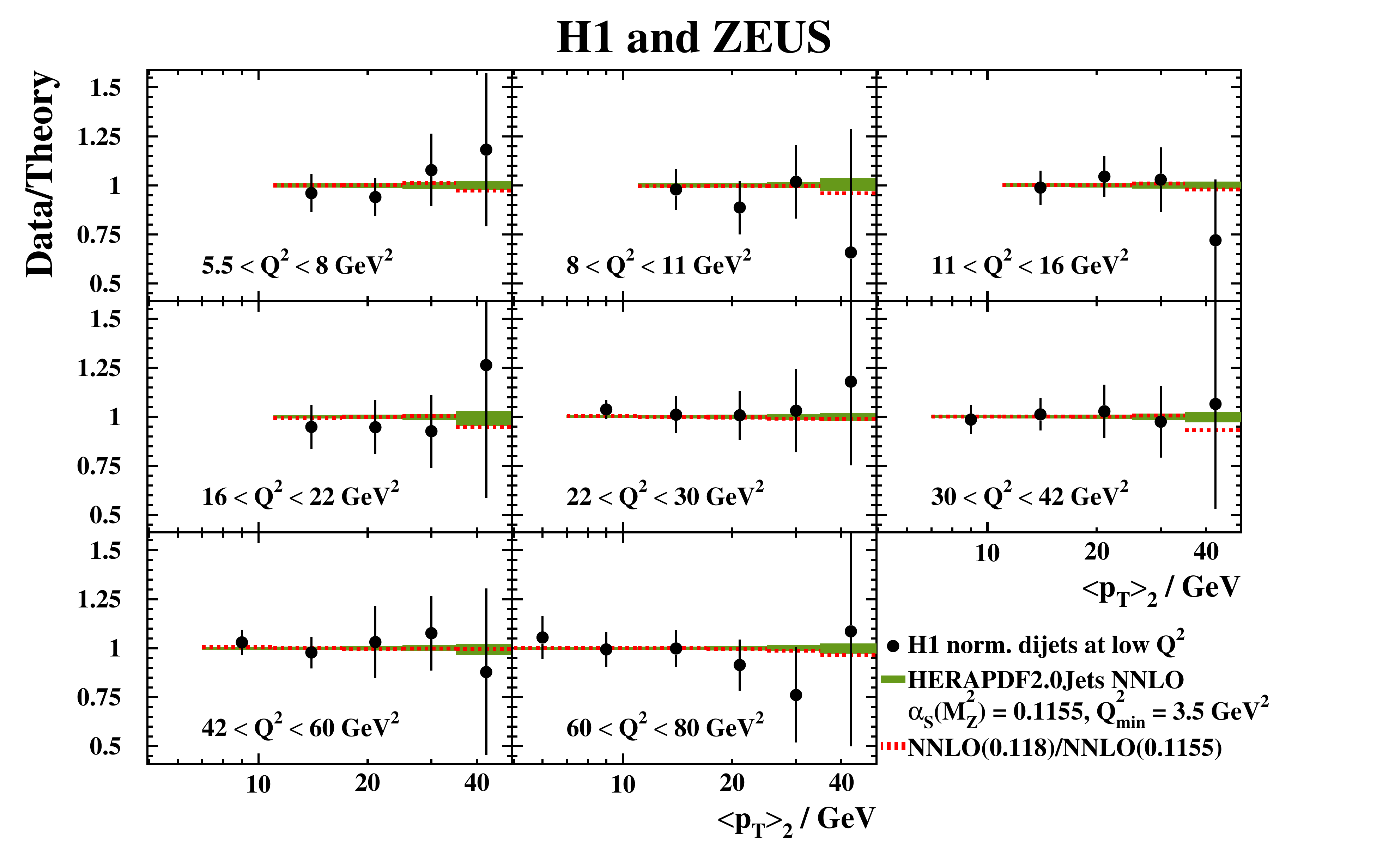}}
  \put (0.1,7.2) {a)}
  \put (0.1,1.2) {b)}
  \end{picture}
  \vskip -1.0 cm
  \caption{
   a) Differential dijet-cross-section predictions,
   ${\rm d}\sigma/{\rm d}\langle p_{\rm T}\rangle_2$, 
   based on HERAPDF2.0Jets NNLO 
   with $\asmz = 0.1155$
   in bins of $Q^2$ between 5.5 and 80\,GeV$^2$ compared to H1 data
   normalised to neutral current (NC) inclusive cross
   sections~\cite{h1lowq2newjets}.
   The variable $\langle p_{\rm T}\rangle_2$ 
   denotes the average $p_{\rm T}$ of the two jets. 
   Only data used in the fit are shown.
  b) Measured dijet cross sections divided by
   predictions based on HERAPDF2.0Jets NNLO.
   The bands represent the total uncertainties on the predictions
   excluding scale uncertainties; 
   the bands are so narrow that they mostly appear as lines.
   Error bars indicate the full uncertainties on the data and are mostly
   smaller than the symbols in a).
   In b), the ratio of predictions based on HERAPDF2.0Jets NNLO with
   $\asmz = 0.118$ and $\asmz = 0.1155$ is also shown.
}
\label{fig:h1-jet-data-low-Q2-dijets}
\end{figure}
\clearpage

\begin{figure}
  \centering
  \setlength{\unitlength}{0.1\textwidth}
  \vskip 1cm
  \begin{picture} (9,11)
    \put(0,7){\includegraphics[width=0.9\textwidth]{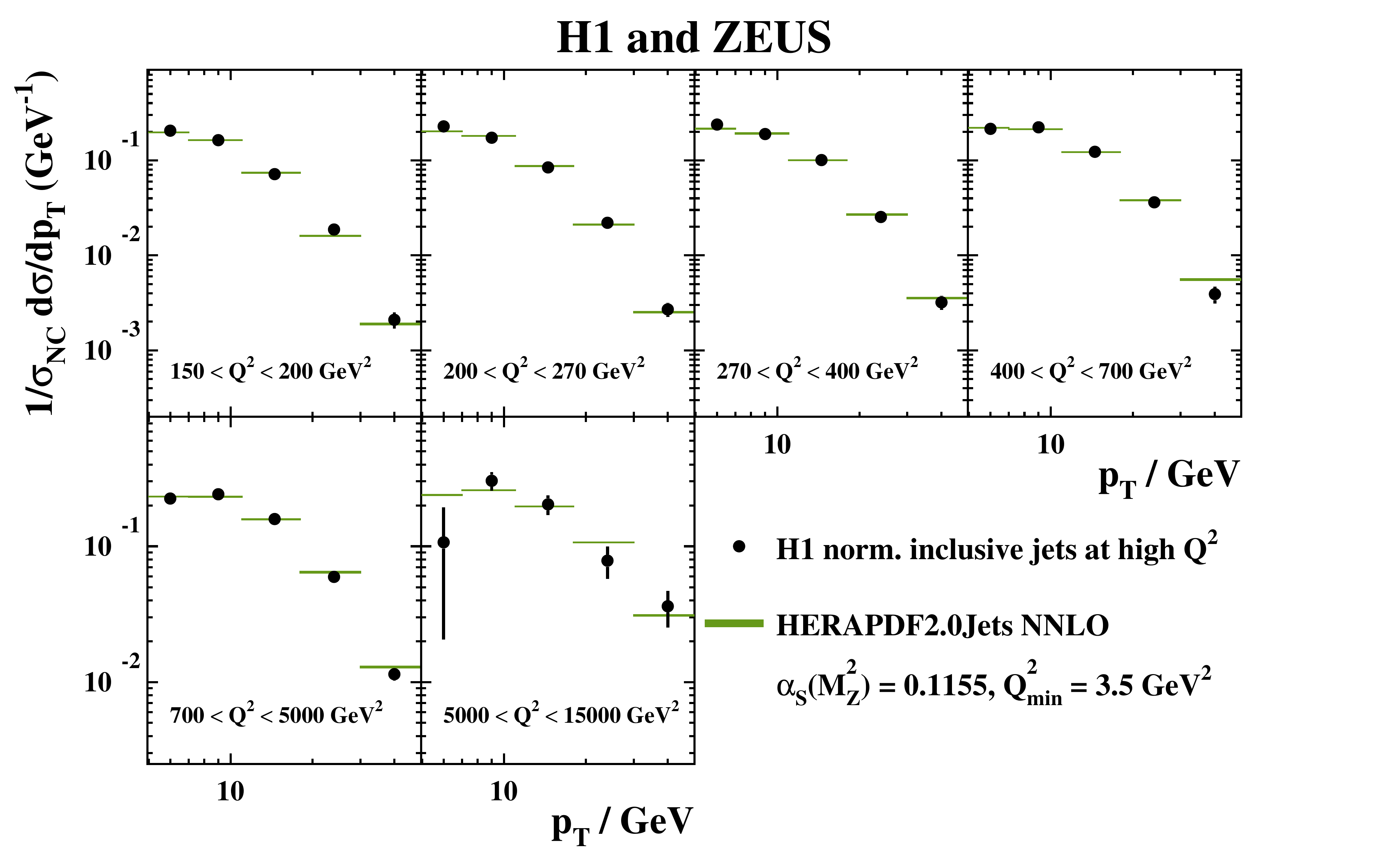}}
    \put(0,1){\includegraphics[width=0.9\textwidth]{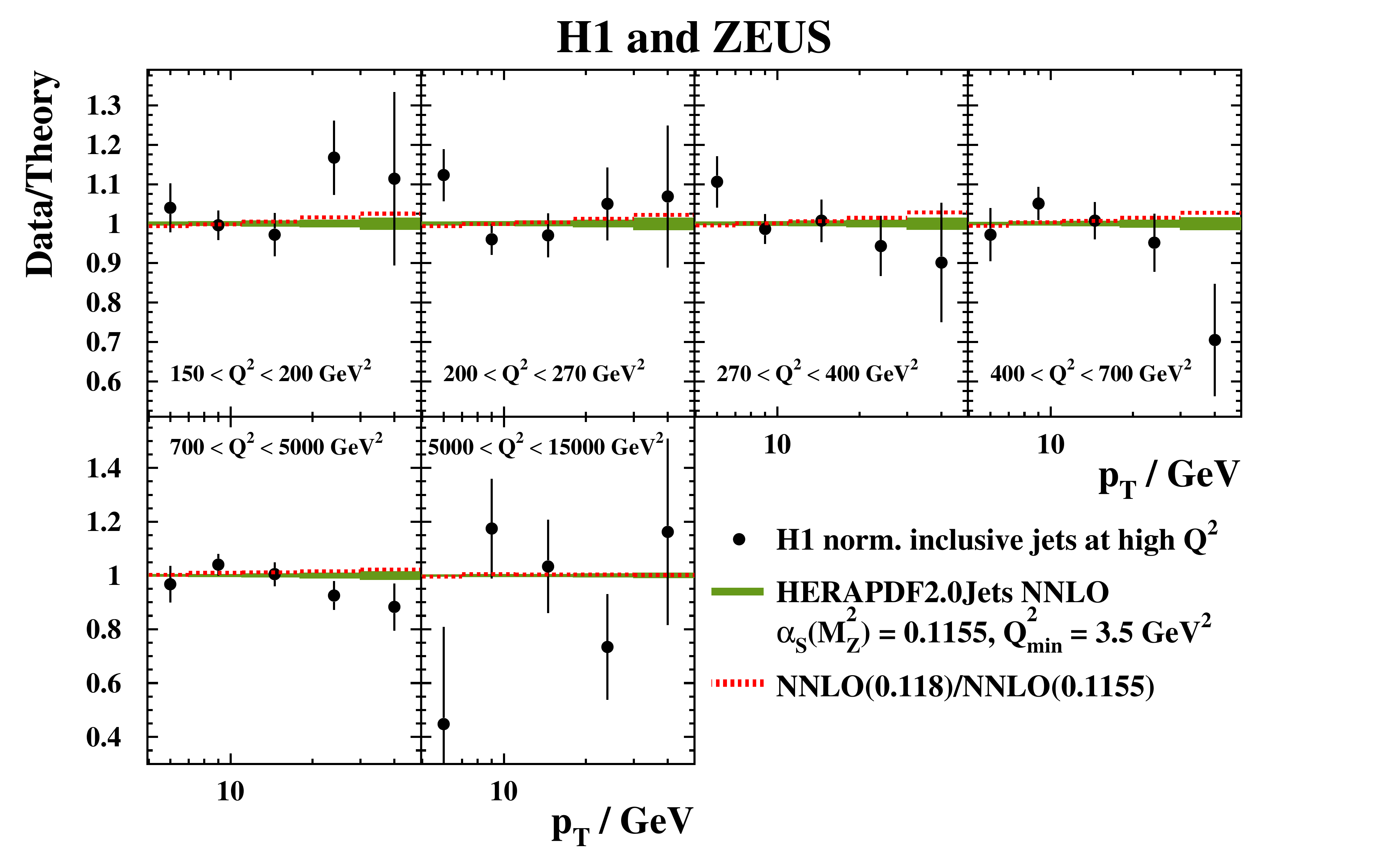}}
  \put (0.1,7.2) {a)}
  \put (0.1,1.2) {b)}
  \end{picture}
  \vskip -1.0cm
  \caption{
   a) Differential jet-cross-section predictions,
   ${\rm d}\sigma/{\rm d}p_{\rm T}$, based on HERAPDF2.0Jets NNLO 
   with $\asmz = 0.1155$
   in bins of $Q^2$ between 150 and 15000\,GeV$^2$ compared to H1 data
   normalised to neutral current (NC) inclusive cross
   sections~normalised to neutral current (NC) inclusive cross
   sections~\cite{h1highq2newjets,h1lowq2newjets}.
   Only data used in the fit are shown.
  b) Measured normalised cross sections divided by
   predictions based on HERAPDF2.0Jets NNLO.
   The bands represent the total uncertainties on the predictions
   excluding scale uncertainties;
   the bands are so narrow that they mostly appear as lines.
   Error bars indicate the full uncertainties on the data
   and are smaller
   than the symbols for most bins in a). 
   In b), the ratio of predictions based on HERAPDF2.0Jets NNLO with
   $\asmz = 0.118$ and $\asmz = 0.1155$ is also shown.
}
\label{fig:h1-jet-data-highQ2}
\end{figure}
\clearpage

\begin{figure}
  \centering
  \setlength{\unitlength}{0.1\textwidth}
  \vskip 1cm
  \begin{picture} (9,11)
    \put(0,7){\includegraphics[width=0.9\textwidth]{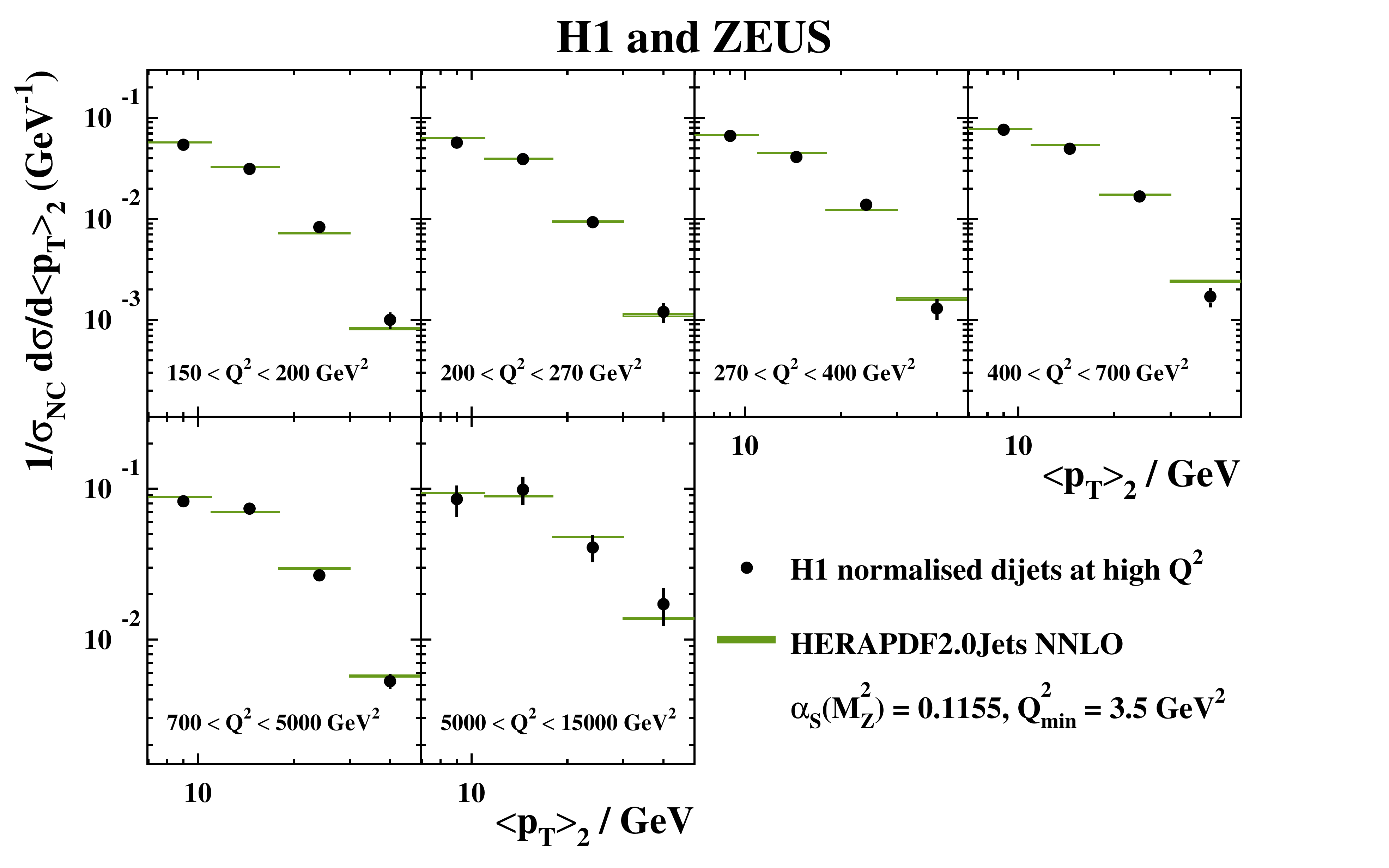}}
    \put(0,1){\includegraphics[width=0.9\textwidth]{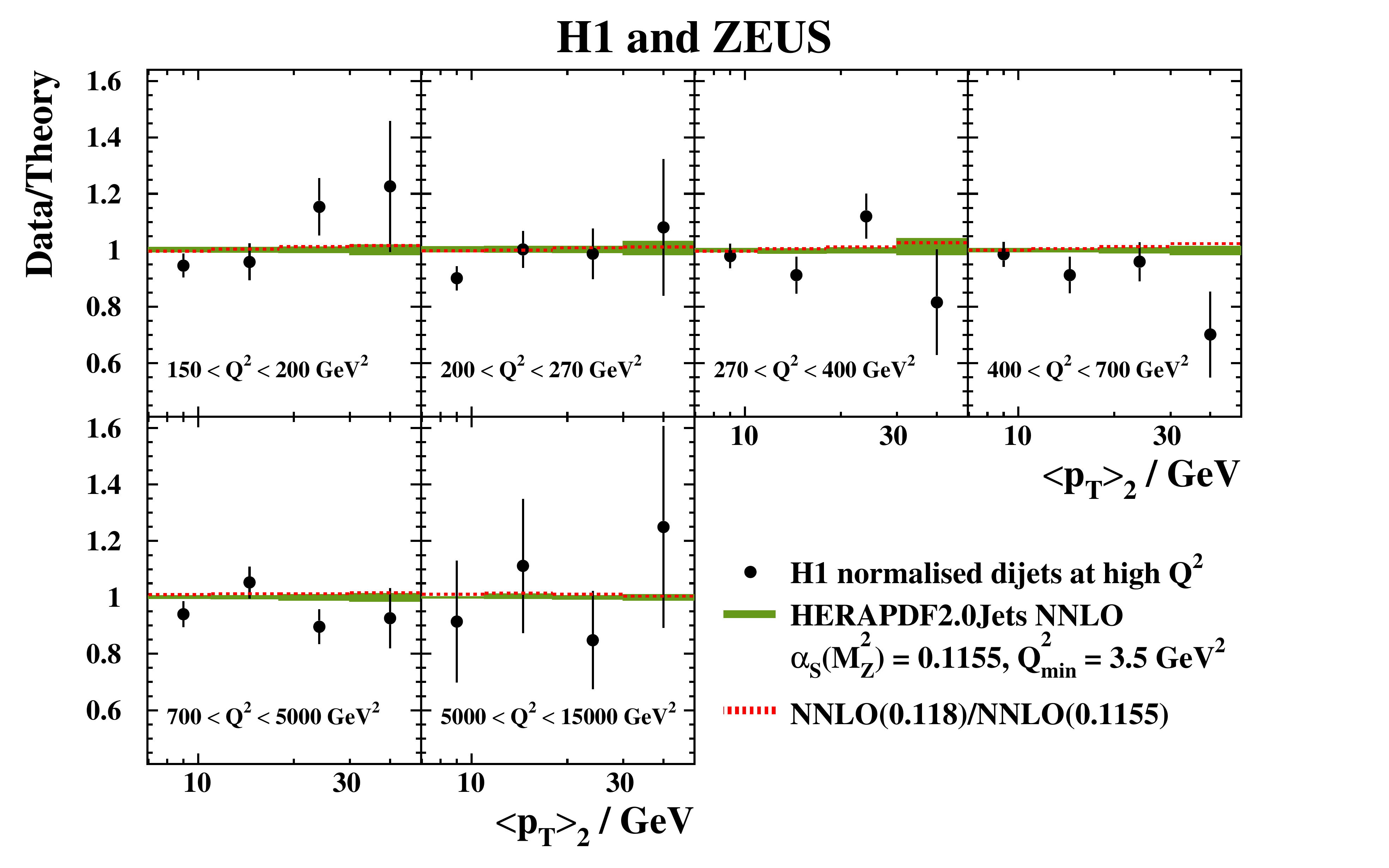}}
  \put (0.1,7.2) {a)}
  \put (0.1,1.2) {b)}
  \end{picture}
  \vskip -1.0 cm
  \caption{
   a) Differential dijet-cross-section predictions,
   ${\rm d}\sigma/{\rm d}\langle p_{\rm T}\rangle_2$, 
   based on HERAPDF2.0Jets NNLO 
   with $\asmz = 0.1155$
   in bins of $Q^2$ between 150 and 15000\,GeV$^2$ compared to H1 data
   normalized to neutral current (NC) cross
   sections~\cite{h1highq2newjets}.
   The variable $\langle p_{\rm T}\rangle_2$ 
   denotes the average $p_{\rm T}$ of the two jets. 
   Only data used in the fit are shown.
  b) Measured dijet cross sections divided by
   predictions based on HERAPDF2.0Jets NNLO.
   The bands represent the total uncertainties on the predictions
   excluding scale uncertainties; 
   the bands are so narrow that they mostly appear as lines.
   Error bars indicate the full uncertainties on the data and are mostly
   smaller than the symbols in a).
   In b), the ratio of predictions based on HERAPDF2.0Jets NNLO with
   $\asmz = 0.118$ and $\asmz = 0.1155$ is also shown.
}
\label{fig:h1-jet-data-highQ2-dijets}
\end{figure}
\clearpage


\begin{figure}
  \centering
  \setlength{\unitlength}{0.1\textwidth}
  \begin{picture} (9,11)
 \put(0,7){\includegraphics[width=0.9\textwidth]{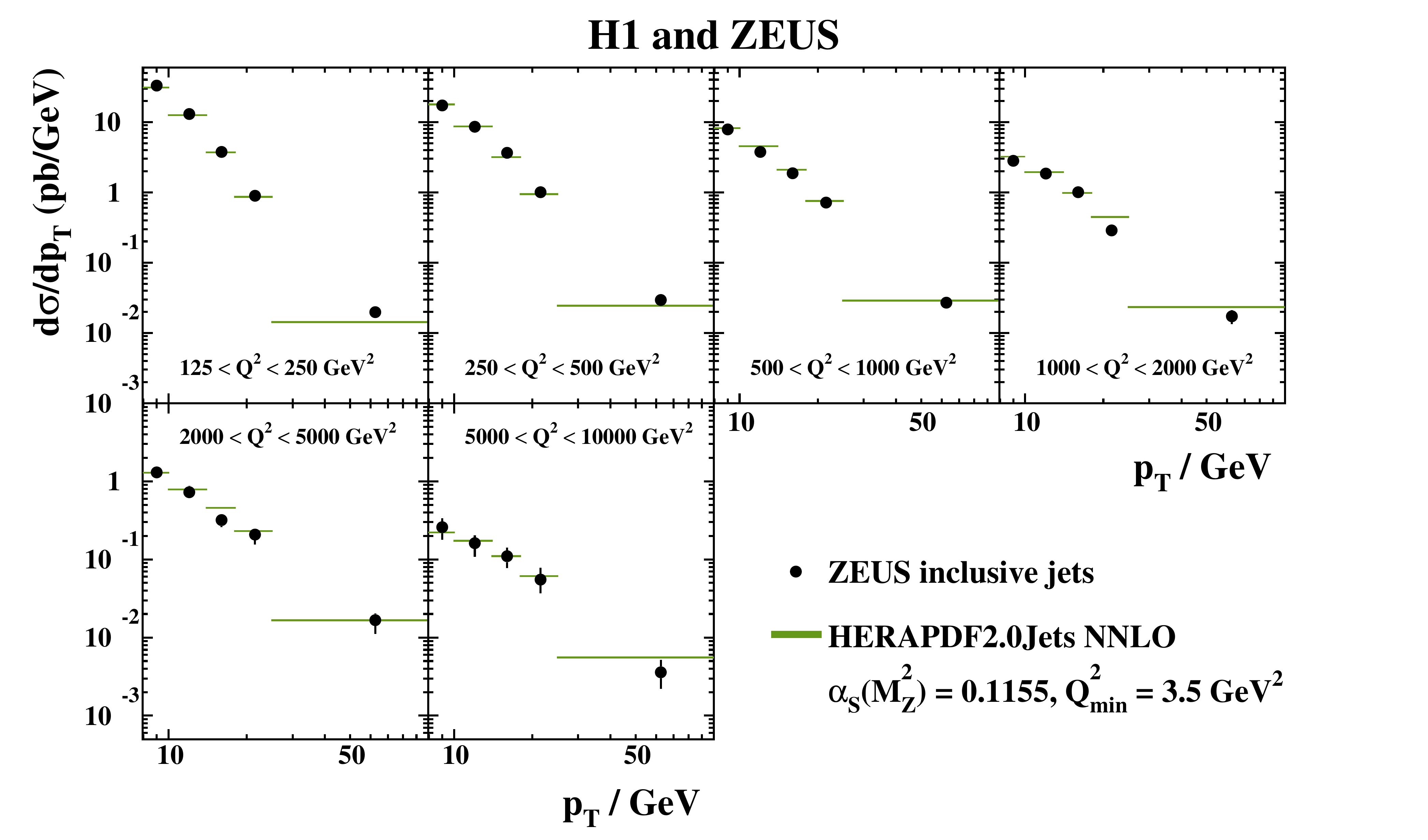}}
 \put(0,1){\includegraphics[width=0.9\textwidth]{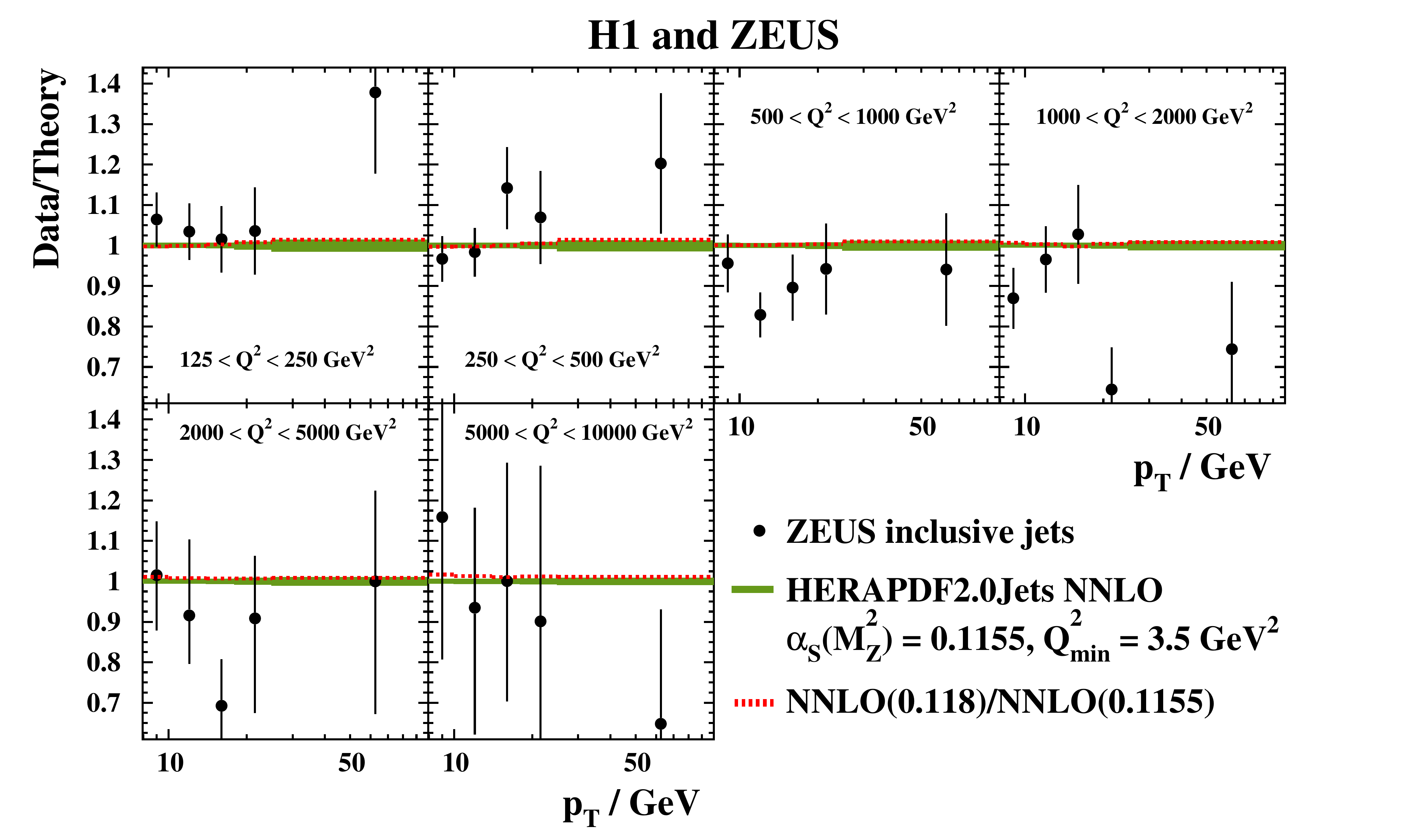}}
  \put (0.1,7.2) {a)}
  \put (0.1,1.2) {b)}
  \end{picture}
   \vskip -1.0 cm
   \caption{
   a) Differential jet-cross-section predictions,
   ${\rm d}\sigma/{\rm d}p_{\rm T}$, based on HERAPDF2.0Jets NNLO 
   with $\asmz = 0.1155$
   in bins of $Q^2$ between 125 and 10000\,GeV$^2$ compared to ZEUS
   data~\cite{zeus9697jets}.
   Only data used in the fit are shown.
  b) Measured cross sections divided by
   predictions based on HERAPDF2.0Jets NNLO.
   The bands represent the total uncertainties on the predictions
   excluding scale uncertainties; 
   the bands are so narrow that they mostly appear as lines.
   Error bars indicate the full uncertainties
   on the data and are smaller
   than the symbols for most bins in a). 
   In b), the ratio of predictions based on HERAPDF2.0Jets NNLO with
   $\asmz = 0.118$ and $\asmz = 0.1155$ is also shown.
}
\label{fig:zeus-jet-data}
\end{figure}
\clearpage

\begin{figure}
  \centering
  \setlength{\unitlength}{0.1\textwidth}
  \begin{picture} (9,11)
 \put(0,7){\includegraphics[width=0.9\textwidth]{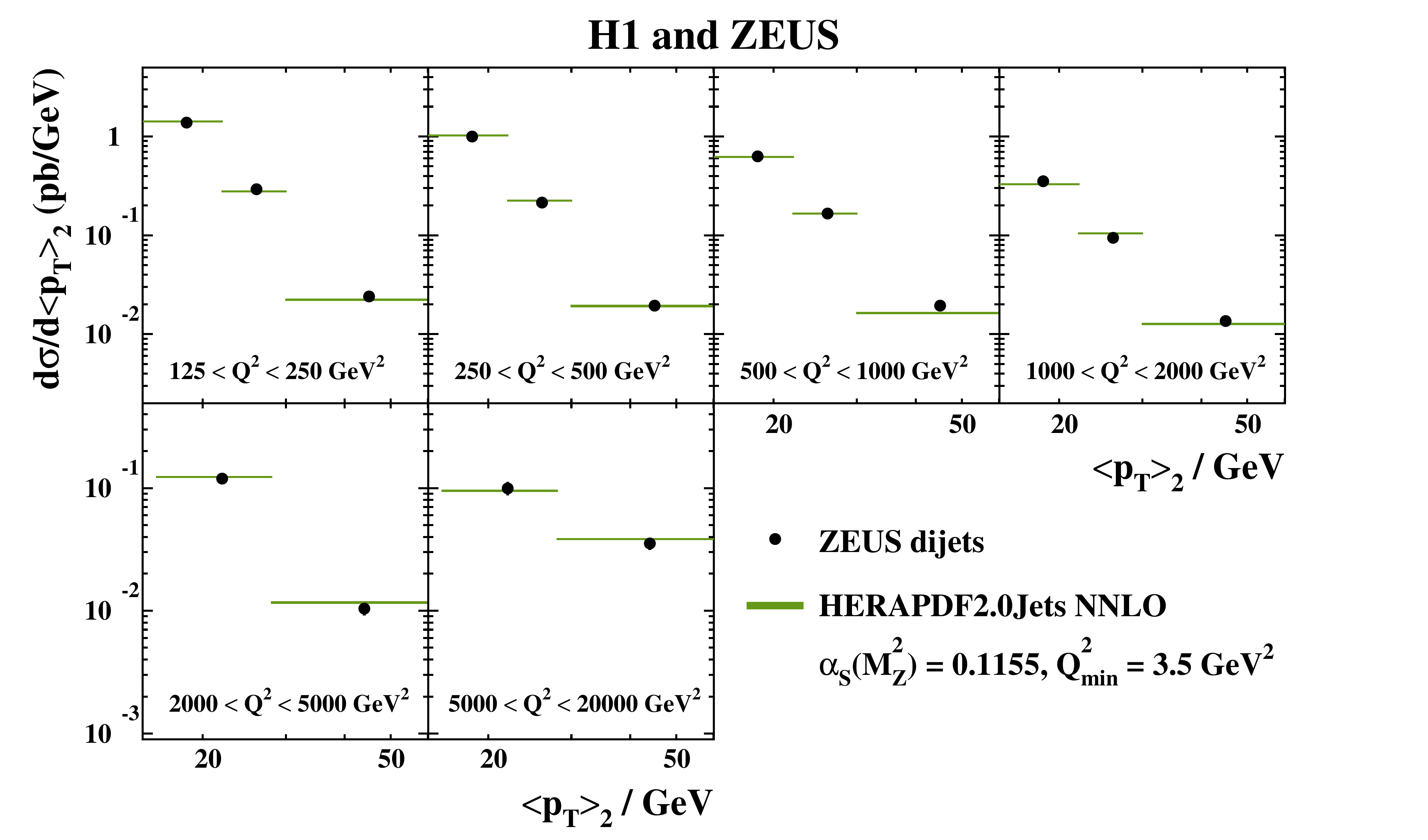}}
 \put(0,1){\includegraphics[width=0.9\textwidth]{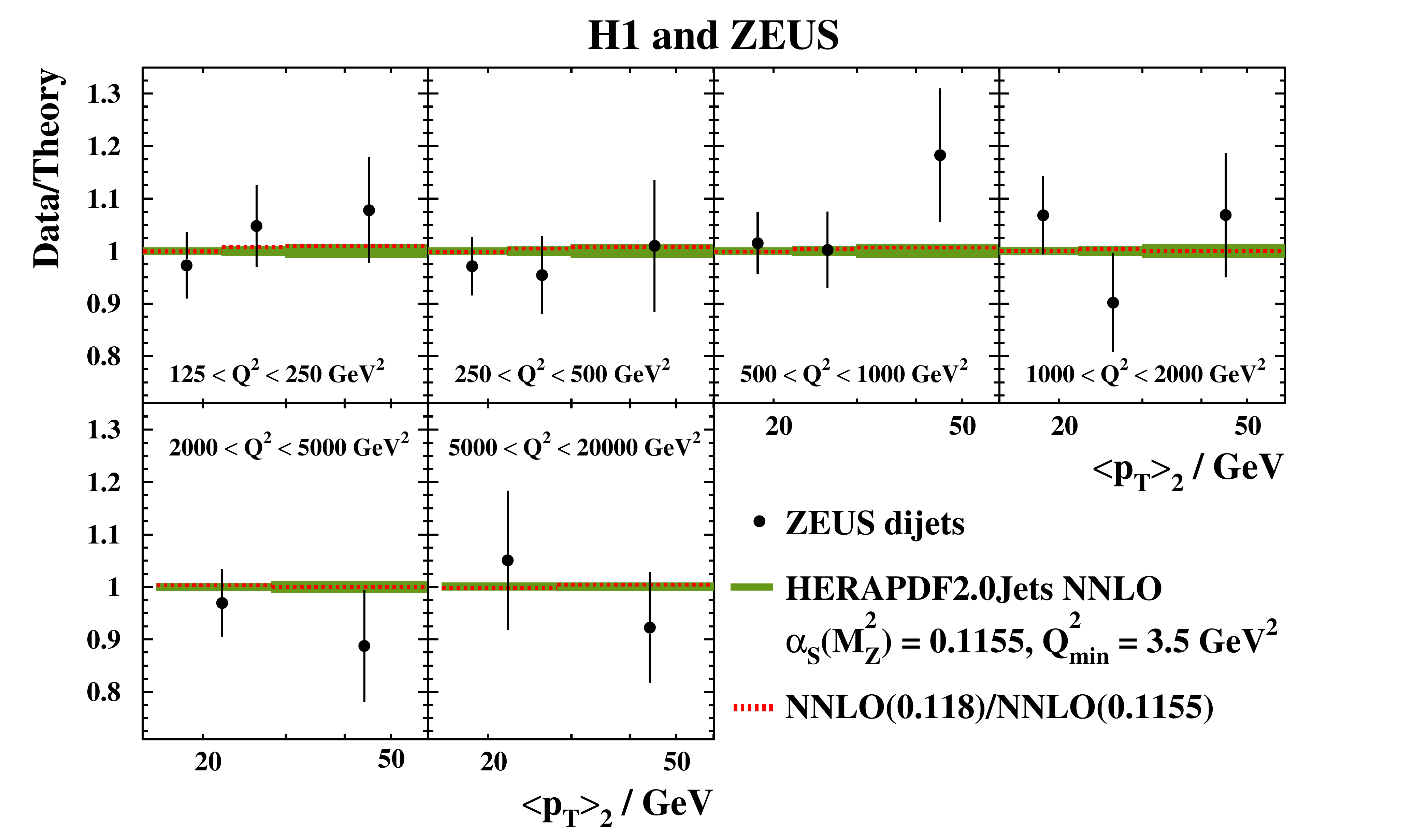}}
  \put (0.1,7.2) {a)}
  \put (0.1,1.2) {b)}
  \end{picture}
   \vskip -1.0 cm
   \caption {
  a) Differential dijet-cross-section predictions,
   ${\rm d}\sigma/{\rm d}\langle p_{\rm T}\rangle_2$, 
   based on HERAPDF2.0Jets NNLO 
   with $\asmz = 0.1155$
   in bins of $Q^2$ between 125 and 20000\,GeV$^2$ compared to ZEUS
   data~\cite{zeusdijets}.
   The variable $\langle p_{\rm T}\rangle_2$ 
   denotes the average $p_{\rm T}$ of the two jets. 
   Only data used in the fit are shown.
  b) Measured dijet cross sections divided by
   predictions based on HERAPDF2.0Jets NNLO.
   The bands represent the total uncertainties on the predictions
   excluding scale uncertainties; 
   the bands are so narrow that they mostly appear as lines.
   Error bars indicate the full uncertainties on the data and are
   smaller than the symbols in a).
   In b), the ratio of predictions based on HERAPDF2.0Jets NNLO with
   $\asmz = 0.118$ and $\asmz = 0.1155$ is also shown.
}
\label{fig:zeus-jet-data-dijets}
\end{figure}
\clearpage

{\huge \bf Appendix A:}

\vskip 0.5cm
{\Large \bf PDF sets released } 

The following two sets of PDFs are released~\cite{fullcorr} and available on
LHAPDF:

(https://lhapdf.hepforge.org/pdfsets.html).

\begin{itemize}

\item HERAPDF2.0Jets NNLO
  \begin{itemize}
  \item based on the combination of inclusive data
        from the H1 and ZEUS collaborations
        and selected data on jet production;
  \item with $Q^2_{\rm min}=3.5\,$GeV$^2$;
  \item using the RTOPT variable-flavour-number scheme;    
  \begin{itemize}
    \item with fixed value of $\asmz = 0.1155$;
    \item with fixed value of $\asmz = 0.118$;
  \end{itemize}
  \item 14 eigenvector pairs give Hessian experimental (fit) 
        uncertainties including hadronisation uncertainties;
  \item grids of 14 variations are released
        to describe the model and parameterisation uncertainties.
  \end{itemize}
\end{itemize}
\clearpage

{\huge \bf Appendix B:}

\vskip 0.5cm
{\Large \bf Additional ratio plots on gluon PDF uncertainties}

\begin{figure} [h]
  \centering
  \vskip -3cm
  \setlength{\unitlength}{0.1\textwidth}
  \begin{picture} (12,12)
  \put(0,0.0){\includegraphics[width=1.0\textwidth]{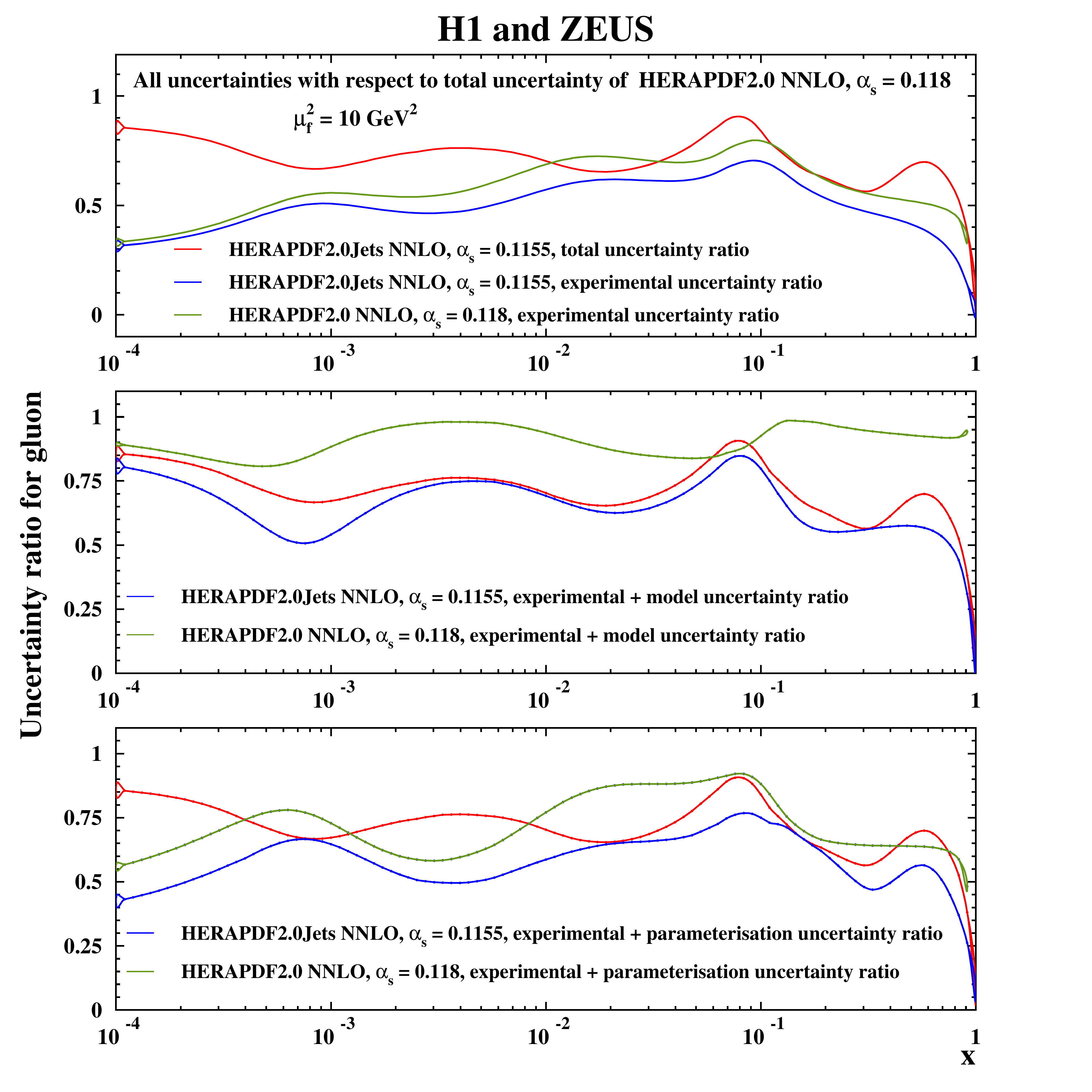}} 
  \put (0.5,6.7) {a)}
  \put (0.5,3.6) {b)}
  \put (0.5,0.5) {c)}
  \end{picture}
  \caption{Ratios of uncertainties relative to the total
    uncertainties of HERAPDF2.0 NNLO with $\asmz=0.118$ for
    the total uncertainty of  HERAPDF2.0Jets NNLO with $\asmz=0.1155$ and the
    a) experimental,
    b) experimental plus model, 
    c) experimental plus parameterisation
    uncertainty of HERAPDF2.0Jets NNLO with $\asmz=0.1155$
    as well as HERAPDF2.0 NNLO with $\asmz=0.118$
    at the scale $\mu_{\rm f}^{2} =10$\,GeV$^{2}$.
}
\label{fig:mark2}
\end{figure}

\clearpage

\begin{figure}
  \centering
  \vskip -3cm
  \setlength{\unitlength}{0.1\textwidth}
  \begin{picture} (12,12)
  \put(0,0.0){\includegraphics[width=1.0\textwidth]{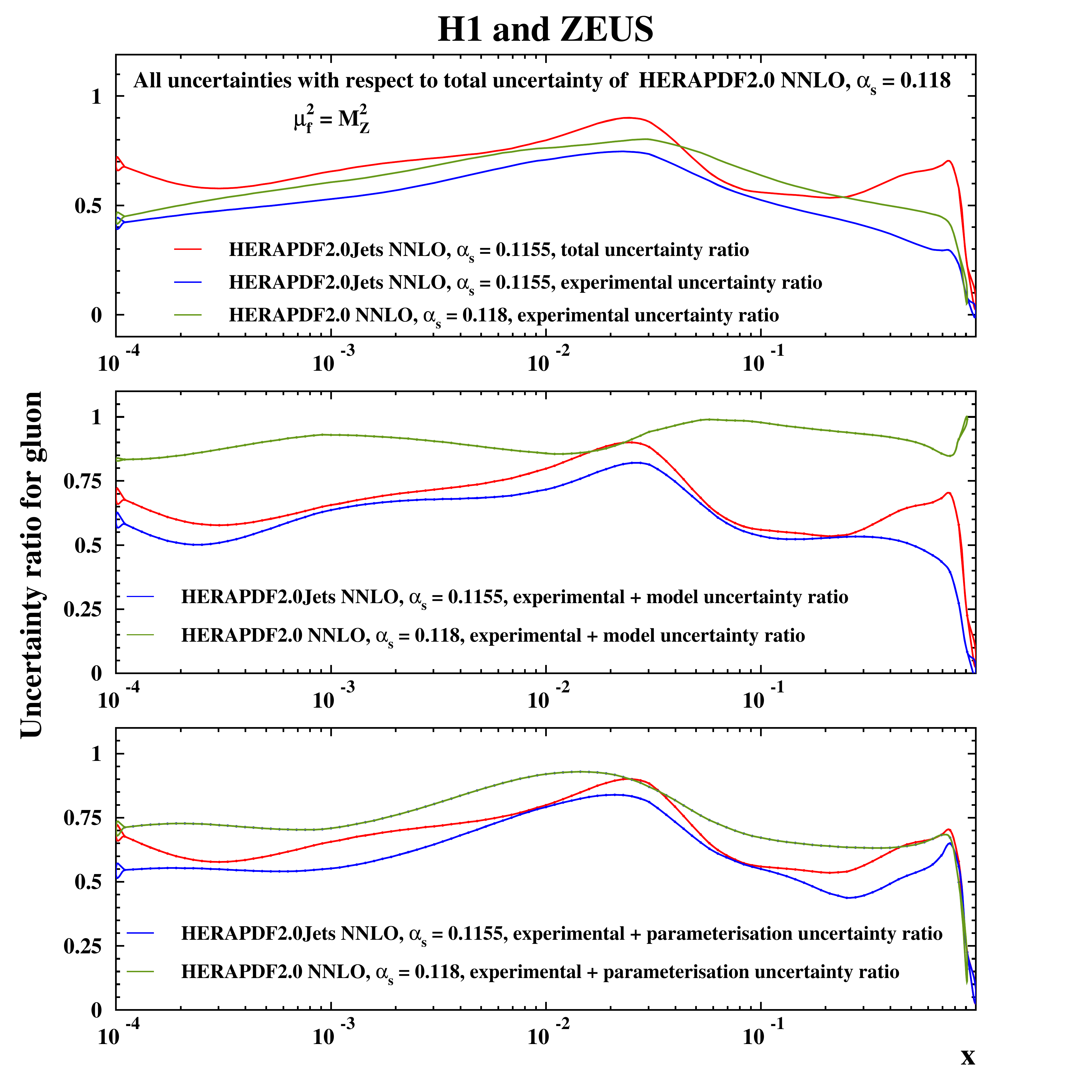}}  
  \put (0.5,6.7) {a)}
  \put (0.5,3.6) {b)}
  \put (0.5,0.5) {c)}
  \end{picture}
  \caption{Ratios of uncertainties relative to the total
    uncertainties of HERAPDF2.0 NNLO with $\asmz=0.118$ for
    the total uncertainty of  HERAPDF2.0Jets NNLO with $\asmz=0.1155$ and the
    a) experimental,
    b) experimental plus model, 
    c) experimental plus parameterisation
    uncertainty of HERAPDF2.0Jets NNLO with $\asmz=0.1155$
    as well as HERAPDF2.0 NNLO with $\asmz=0.118$
    at the scale $\mu_{\rm f}^{2} = M_Z^{2}$.
}
\label{fig:mark3}
\end{figure}
\clearpage

\begin{figure}
  \centering
  \vskip -3cm
  \setlength{\unitlength}{0.1\textwidth}
  \begin{picture} (12,12)
  \put(0,0.0){\includegraphics[width=1.0\textwidth]{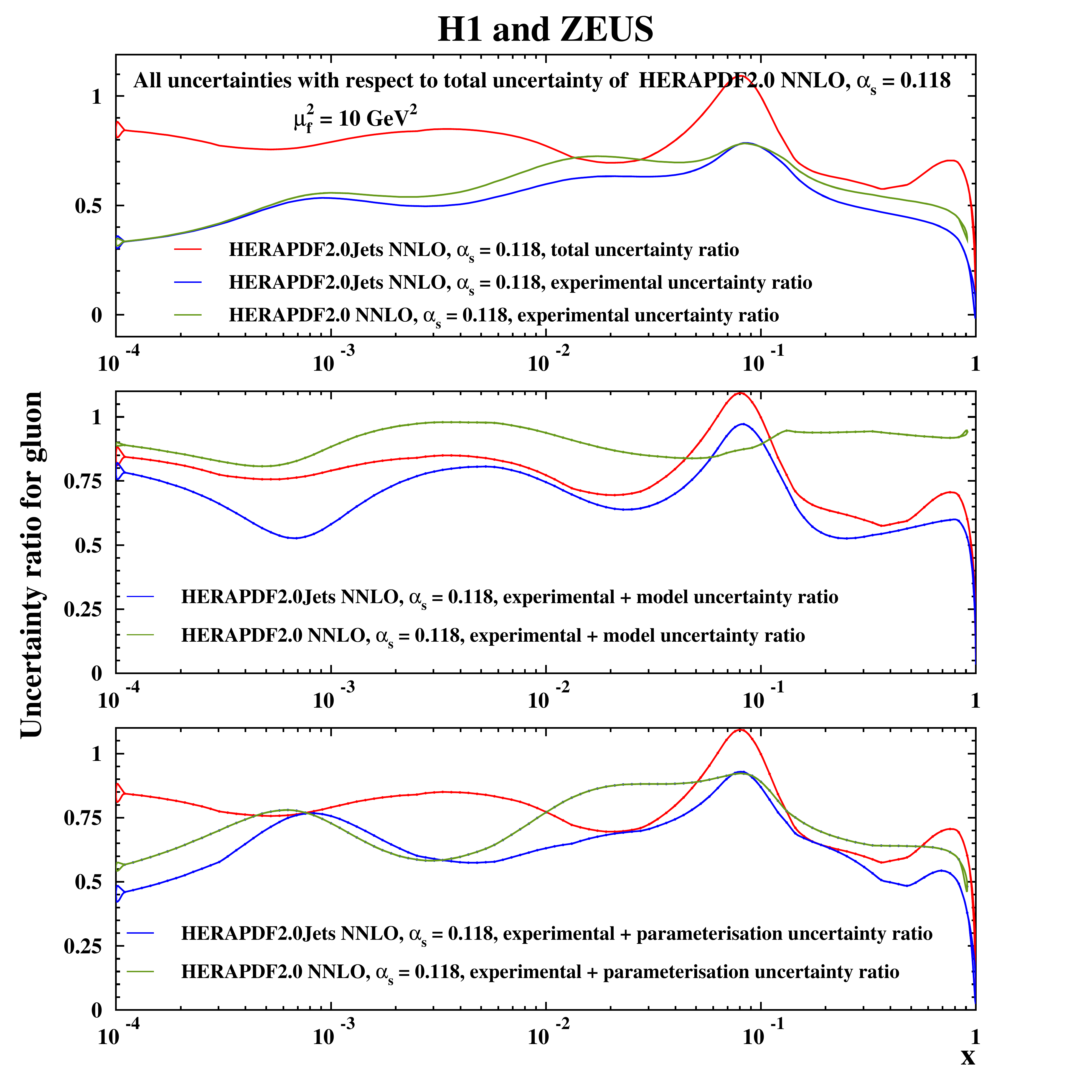}}  
  \put (0.5,6.7) {a)}
  \put (0.5,3.6) {b)}
  \put (0.5,0.5) {c)}
  \end{picture}
  \caption{Ratios of uncertainties relative to the total
    uncertainties of HERAPDF2.0 NNLO with $\asmz=0.118$ for
    the total uncertainty of  HERAPDF2.0Jets NNLO with $\asmz=0.118$ and the
    a) experimental,
    b) experimental plus model, 
    c) experimental plus parameterisation
    uncertainty of HERAPDF2.0Jets NNLO with $\asmz=0.118$
    as well as HERAPDF2.0 NNLO with $\asmz=0.118$
    at the scale $\mu_{\rm f}^{2} =10$\,GeV$^{2}$.
}
\label{fig:mark4}
\end{figure}
\clearpage

\begin{figure}
  \centering
  \vskip -3cm
  \setlength{\unitlength}{0.1\textwidth}
  \begin{picture} (12,12)
  \put(0,0.0){\includegraphics[width=1.0\textwidth]{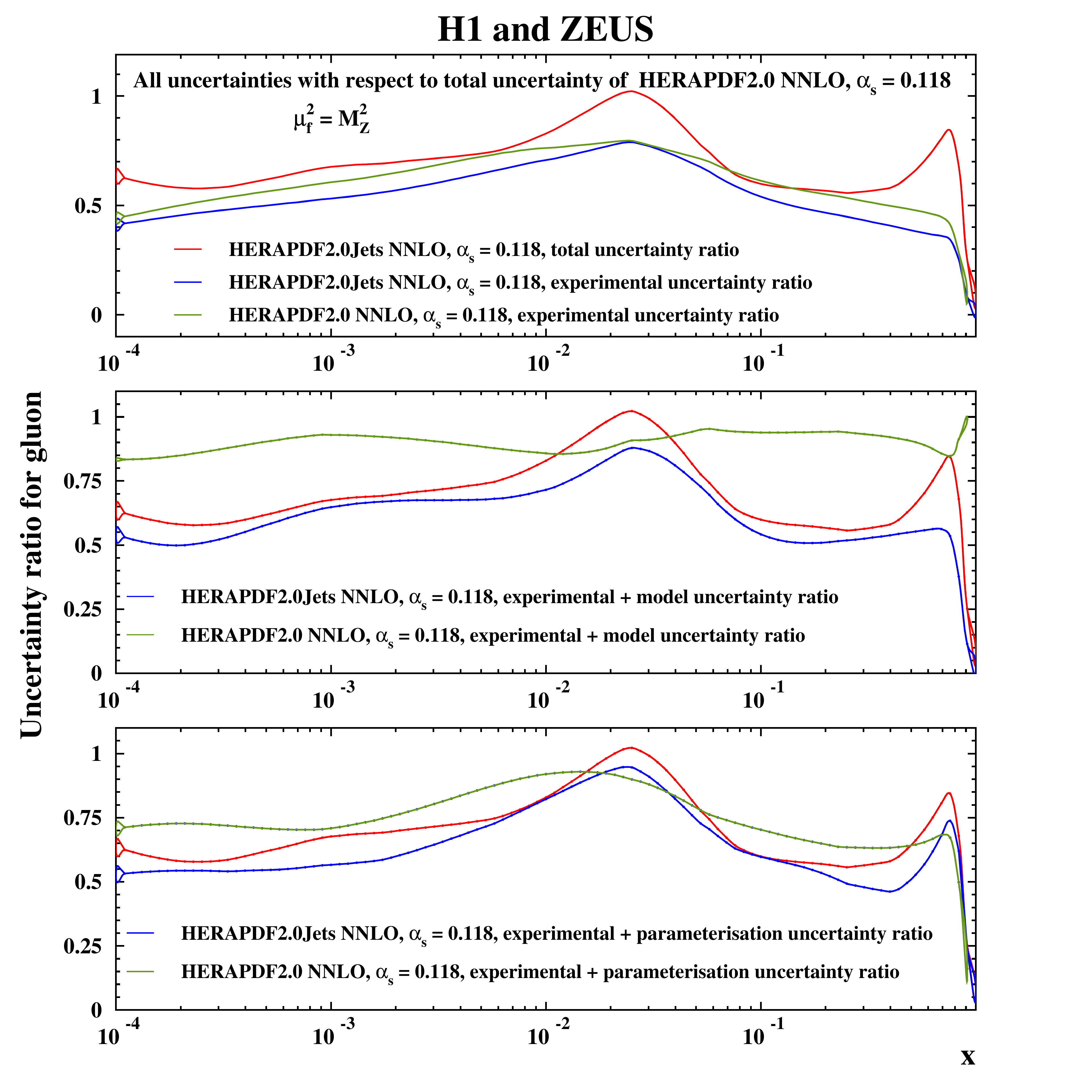}}  
  \put (0.5,6.7) {a)}
  \put (0.5,3.6) {b)}
  \put (0.5,0.5) {c)}
  \end{picture}
  \caption{Ratios of uncertainties relative to the total
    uncertainties of HERAPDF2.0 NNLO with $\asmz=0.118$ for
    the total uncertainty of  HERAPDF2.0Jets NNLO with $\asmz=0.118$ and the
    a) experimental,
    b) experimental plus model, 
    c) experimental plus parameterisation
    uncertainty of HERAPDF2.0Jets NNLO with $\asmz=0.118$
    as well as HERAPDF2.0 NNLO with $\asmz=0.118$
    at the scale $\mu_{\rm f}^{2} = M_Z^{2}$.
}
\label{fig:mark5}
\end{figure}
\clearpage

\end{document}